\documentclass[useAMS,usenatbib]{mn2e}

\usepackage{amsmath,amssymb}
\usepackage{aas_macros}
\usepackage[pdftex]{graphicx}

\newcommand{\neutrino}[1]{{#1}_{\nu}}
\newcommand{\modelname}[1]{\texttt{#1}}
\newcommand{\subpanel}[1]{\textit{#1}}
\newcommand{\region}[1]{\textit{#1}}

\newcommand{\msol}{M_{\odot}}
\newcommand{\msun}{M_{\sun}}

\newcommand{\Msol}{M_{\odot}}

\newcommand{\inv}[1]{\frac{1}{#1}}

\newcommand{\zehn}[1]{10^{#1}}
\newcommand{\zehnh}[2]{{#1} \times 10^{#2}}

\newcommand{\pns}{proto neutron star}
\newcommand{\Alfven}{Alfv{\'e}n}
\newcommand{\Alfvenic}{Alfv{\'e}nic}

\newcommand{\ms}{\textrm{ms}}
\newcommand{\km}{\textrm{km}}

\newcommand{\cms}{\textrm{cm s}^{-1}}
\newcommand{\erg}{\textrm{erg}}

\newcommand{\gccm}{\textrm{g\,cm}^{-3}}

\newcommand{\Gauss}{\textrm{G}}
\newcommand{\grad}{^{\circ}}

\newcommand{\MeV}{\textrm{MeV}}

\newcommand{\nth}[1]{${#1}^{\mathrm{th}}$}

\renewcommand{\eqref}[1]{Eq.\,(\ref{#1})}
\newcommand{\figref}[1]{Fig.\,\ref{#1}}
\newcommand{\tabref}[1]{Tab.\,\ref{#1}}
\newcommand{\secref}[1]{Sect.\,\ref{#1}}

\begin{document}

\title[Non-rotating core collapse with magnetic fields]{
  Magnetic field amplification and magnetically supported explosions
  of collapsing, non-rotating stellar cores
}

\author[M.~Obergaulinger et al.]{
  M.~Obergaulinger$^1$, H.-Th.~Janka$^{2}$, M.A.~Aloy Tor{\'a}s$^1$
  \\
  $^1$ Departament d{\'{}}Astronomia i Astrof{\'i}sica, Universitat de
  Val{\`e}ncia, \\ Edifici d{\'{}}Investigaci{\'o} Jeroni Munyoz, C/
  Dr.~Moliner, 50, E-46100 Burjassot (Val{\`e}ncia), Spain 
  \\ $^2$ Max-Planck-Institut f{\"u}r Astrophysik,
  Karl-Schwarzschild-Str. 1, D-85748 Garching, Bavaria, Germany
}

\maketitle

\begin{abstract}
  We study the amplification of magnetic fields in the collapse and
  the post-bounce evolution of the core of a non-rotating star of 15
  $\msun$ in axisymmetry.  To this end, we solve the coupled equations
  of magnetohydrodynamics and neutrino transport in the two-moment
  approximation.  The pre-collapse magnetic field is strongly
  amplified by compression in the infall.  Initial fields of the order
  of $10^{10} \, \mathrm{G}$ translate into proto-neutron star fields
  similar to the ones observed in pulsars, while stronger initial
  fields yield magnetar-like final field strengths.  After core
  bounce, the field is advected through the hydrodynamically unstable
  neutrino-heating layer, where non-radial flows due to convection and
  the standing accretion shock instability amplify the field further.
  Consequently, the resulting amplification factor of order five is
  the result of the number of small-eddy turnovers taking place within
  the time scale of advection through the post-shock layer.  Due to
  this limit, most of our models do not reach equipartition between
  kinetic and magnetic energy and, consequently, evolve similarly to
  the non-magnetic case, exploding after about 800 ms when a single or
  few high-entropy bubbles persist over several dynamical time scales.
  In the model with the strongest initial field we studied, $10^{12}
  \, \mathrm{G}$, for which equipartition between flow and field is
  achieved, the magnetic tension favours a much earlier development of
  such long-lived high-entropy bubbles and enforces a fairly ordered
  large-scale flow pattern.  Consequently, this model, after
  exhibiting very regular shock oscillations, explodes much earlier
  than non-magnetic ones.
\end{abstract}



\section{Introduction}
\label{Sec:Intro}

Most scenarios for the explosion mechanism of core-collapse supernovae
(SNe) involve a combination of energy deposition in the matter
surrounding the nascent \pns~(PNS) and multi-dimensional hydrodynamic
flows.  Examples for means of energy transfer to the SN ejecta are the
prompt bounce shock, neutrinos, magnetic fields and acoustic waves.
In the neutrino-heating mechanism neutrinos tap the gravitational
potential energy released during collapse and deposit a part of it
behind the stalled shock. This process is enhanced and thus supported
by non-radial fluid flows triggered by hydrodynamic instabilities like
convection and the standing accretion shock instability
\citep[SASI;][]{Blondin_Mezzacappa_DeMarino__2003__apj__Stability_of_Standing_Accretion_Shocks_with_an_Eye_toward_Core-Collapse_Supernovae,Foglizzo__2001__aap__Entropic-acoustic_instability_of_shocked_Bondi_accretionI._What_does_perturbed_Bondi_accretion_sound_like?,Foglizzo__2002__aap__Non-radial_instabilities_of_isothermal_Bondi_accretion_with_a_shock:Vortical-acoustic_cycle_vs.post-shock_acceleration}

It is safe to assume that the collapsing stellar core will possess a
magnetic field of some (uncertain) strength and topology.  This
assumption is supported by observational evidence for the presence of
surface magnetic fields of up to $b_{\mathrm{WD}} \lesssim \zehn{9} \,
\mathrm{G}$ in white dwarfs, which resemble in many respects the iron
cores of evolved massive stars
\citep{Wickramasinghe_Ferrario__2000__PASP__Mag_WD}.  Furthermore,
neutron stars created in SNe may be endowed with magnetic fields
$b\sim 10^{12}\,$G (pulsars) to $\sim 10^{14}\,$G \citep[magnetars;
see,
e.g.][]{Kaspi__2010__ProceedingsoftheNationalAcademyofScience__Grandunificationofneutronstars}.
Nonetheless, the possible influence of the field on the explosion is
less clear.  The main reason for the small number of conclusive
investigations into this topic is that important effects are expected
to occur only when the magnetic field is roughly in equipartition with
the kinetic energy of the gas flow, a condition corresponding to
extremely strong fields similar to those observed in magnetars.  This
makes full MHD simulations including a treatment of the important
neutrino physics in the SN core indispensable.

Stellar evolution calculations, predicting only weak pre-collapse
magnetic fields
\citep{Heger_et_al__2005__apj__Presupernova_Evolution_of_Differentially_Rotating_Massive_Stars_Including_Magnetic_Fields},
render the prospects for magnetically affected explosions very much
dependent on the amount of field amplification happening during and
after collapse.  Rapid differential rotation may amplify a weak seed
field quite generically to dynamically relevant values, e.g.~by
winding up a poloidal field (a process linear in time) or,
exponentially in time, by the magneto-rotational instability
\citep[MRI, for a review, see][]{Balbus_Hawley__1998__RMP__MRI}.  The
potential relevance of the latter in supernova cores was first pointed
out and explored by \cite{Akiyama_etal__2003__ApJ__MRI_SN}.
Magneto-rotational explosions, theoretically discussed by
\cite{Meier_etal__1976__ApJ__MHD_SN}, have been studied in various
approximations, e.g.~by
\cite{Bisnovatyi-Kogan_Popov_Samokhin__1976__APSS__MHD_SN,Symbalisty__1984__ApJ_MHD_SN,Akiyama_etal__2003__ApJ__MRI_SN,Kotake_etal__2004__Apj__SN-magrot-neutrino-emission,Thompson_Quataert_Burrows__2004__ApJ__Vis_Rot_SN,Moiseenko_et_al__2006__mnras__A_MR_CC_model_with_jets,Obergaulinger_Aloy_Mueller__2006__AA__MR_collapse,Dessart_et_al__2007__apj__MagneticallyDrivenExplosionsofRapidlyRotatingWhiteDwarfsFollowingAccretion-InducedCollapse,Burrows_etal__2007__ApJ__MHD-SN,Cerda-Duran_et_al__2007__AA__passive-MHD-collapse,
  Obergaulinger_etal__2009__AA__Semi-global_MRI_CCSN,Masada_et_al__2012__apj__LocalSimulationsoftheMagnetorotationalInstabilityinCore-collapseSupernovae}.
Recent magnetohydrodynamic (MHD) simulations employing (detailed)
microphysics (neutrinos and a sophisticated high-density equation of
state) have been performed by \cite{Burrows_etal__2007__ApJ__MHD-SN}
and by
\cite{Scheidegger_etal__2008__aap__GW_from_3d_MHD_SN,Winteler_et_al__2012__apjl__MagnetorotationallyDrivenSupernovaeastheOriginofEarlyGalaxyr-processElements}
and
\cite{Mosta_et_al__2014__apjl__MagnetorotationalCore-collapseSupernovaeinThreeDimensions}.
While the former used a multi-group flux-limited diffusion treatment
for the neutrino transport in a Newtonian framework, the latter
approximated the effects of neutrino radiation by a parametrisation of
the pre-bounce deleptonisation of the core.  Albeit using simplified
neutrino physics, the high-resolution global simulations of
\cite{Sawai_et_al__2013__apjl__GlobalSimulationsofMagnetorotationalInstabilityintheCollapsedCoreofaMassiveStar}
addressed one of the most severe problems in numerical models of
magneto-rotational collapse, viz.~the extremely high resolution
required to resolve the fastest growing MRI modes.

However, according to stellar-evolution models, the majority of
progenitors is expected to rotate slowly.  Although many stars on the
upper main sequence show a surface rotation period close to the
critical value for mass shedding, they will most likely lose most of
their angular momentum during their subsequent evolution, e.g.~by
strong stellar winds or magnetic braking
\citep{Heger_et_al__2005__apj__Presupernova_Evolution_of_Differentially_Rotating_Massive_Stars_Including_Magnetic_Fields,Meynet_etal__2011__aap__Massive_star_models_with_magnetic_braking}.
Non-rotating cores, albeit lacking the above-mentioned efficient
channels for amplification, may experience field growth by several
effects:

\begin{enumerate}
\item Due to the extremely low resistivity of the SN core matter, the
  magnetic field lines are frozen in the flow.  Hence, the compression
  of the gas during the collapse is accompanied by an increase of the
  field strength by roughly three orders of magnitude, while the
  topology of the field does not change.
\item After bounce, convection develops in the PNS and the surrounding
  hot-bubble region.  Breaking down into three-dimensional turbulence,
  convection may provide the $\alpha$ effect responsible for a
  small-scale dynamo amplifying the field on length scales comparable
  to the scale of turbulent forcing, i.e.~the size of convective
  eddies \citep[][]{Thompson_Duncan__1993__ApJ__NS-dynamo}.  However,
  the generation of a large-scale field, e.g.~on the scale of the
  PNS, probably requires a non-vanishing kinetic helicity of the
  turbulent flow, which can most naturally be accounted for by
  (differential) rotation \citep[see,
  e.g.~][]{Brandenburg_Subramanian__2005__AN__Strong_mean_field_dynamos_require_supercritical_helicity_fluxes}.
\item
  \citet{Endeve_et_al__2010__apj__Generation_of_Magnetic_Fields_By_the_SASI}
  and subsequently
  \cite{Endeve_et_al__2012__apj__TurbulentMagneticFieldAmplificationfromSpiralSASIModes:ImplicationsforCore-collapseSupernovaeandProto-neutronStarMagnetization}
  have demonstrated that the standing-accretion-shock instability,
  growing through acoustic and advective perturbations that form a
  positive feedback cycle between the shock wave and the deceleration
  region above the PNS, has the potential to amplify the magnetic
  field by up to four orders of magnitude.  Dynamically relevant field
  strengths (in which case the field reaches at least 10\% of
  equipartition with the kinetic energy) can be reached only when the
  pre-collapse field in the stellar core is sufficiently strong.
\item Non-radial fluid motions triggered by these instabilities can
  excite perturbations of the magnetic field propagating as
  \Alfven~waves along the field lines.  The outward propagation of
  \Alfven~waves excited close to the PNS has to compete with the
  accretion of gas towards the centre.  Assuming that the accretion
  flow decelerates in this region continuously,
  \cite{Guilet_et_al__2011__apj__Dynamics_of_an_Alfven_Surface_in_Core_Collapse_Supernovae}
  argue that there must be an \emph{\Alfven~point} at which the
  \Alfven~speed equals the accretion velocity, and the propagation of
  the wave (measured in the lab frame) comes to a rest.  They show
  that \Alfven~waves are amplified exponentially at such a stagnation
  point.  For the conditions of a supernova core, the amplification
  should be most efficient for a magnetic field strength of a few
  $10^{13}$ G and could yield final fields of the order of $10^{15}$
  G.  Dissipation of the wave energy can increase the entropy of the
  gas, thus modifying the dynamics in the accretion region.  For even
  stronger fields, the \Alfven~point can be close to the shock wave.
  In this case,
  \cite{Suzuki_Sumiyoshi_Yamada__2008__ApJ__Alfven_driven_SN} find
  that the explosion can be driven solely by the energy deposited by
  the dissipation of \Alfven~waves.  This process transmitting energy
  from the (convectively active) PNS to the much less dense
  surrounding medium bears a strong similarity to the proposed
  mechanism for heating the solar corona by \Alfven~waves emerging
  from the solar surface \citep[e.g.][]{McIntosh_et_al__2011__nat__Alfv'enicwaveswithsufficientenergytopowerthequietsolarcoronaandfastsolarwind}.
\end{enumerate} 

Previous simulations of magnetohydrodynamic stellar core collapse have
used a wide variety of methods to treat the effects of neutrinos: in
the simplest models, they were either ignored
\citep[e.g.~][]{Obergaulinger_Aloy_Mueller__2006__AA__MR_collapse,Mikami_et_al__2008__apj__3d_MHD_SN}
or treated by simple parameterizations or local source terms
\citep[e.g.~][]{Cerda-Duran__2008__AA__GRMHD-code}; more complex
approaches include trapping/leakage schemes
\citep[e.g.~][]{Kotake_etal__2004__Apj__SN-magrot-neutrino-emission},
multi-dimensional, energy-dependent flux-limited diffusion
\citep[][]{Dessart_et_al__2006__apj__Multi-d_RHD_Simulations_of_PNS_Convection}
or the isotropic-diffusion source approximation, which distinguishes
between trapped and free-streaming components to avoid the treatment
of stiff source terms
\citep[][]{Liebendorfer_et_al__2009__apj__The_Isotropic_Diffusion_Source_Approximation_for_Supernova_Neutrino_Transport}.
In our approach to radiation-magnetohydrodynamics (RMHD), we employ a
new multi-dimensional and energy-dependent scheme for the neutrino
transport in supernova cores, namely a two-moment solver for the
neutrino energy (lepton number) and momentum equations with an
analytic closure relation
\citep[][]{Cernohorsky_van_Weert__1992__ApJ__Rel_2_moment_nu,Pons_Ibanez_Miralles__2000__MNRAS__hyperbol_radtrans,Audit_et_al__2002__astro-ph__hyp_RHD_closure}.
Two-moment closure schemes for neutrino transport were also applied in
relativistic simulations of black hole-torus systems by
\cite{Shibata_Sekiguchi__2012__ProgressofTheoreticalPhysics__RadiationMagnetohydrodynamicsforBlackHole-TorusSysteminFullGeneralRelativity_ASteptowardPhysicalSimulation}
\citep[see
also][]{Shibata_et_al__2011__PThPh__Truncated_Moment_Formalism_for_Radiation_Hydrodynamics_in_Numerical_Relativity}
and in stellar core collapse by
\cite{Kuroda_et_al__2012__apj__FullyGeneralRelativisticSimulationsofCore-collapseSupernovaewithanApproximateNeutrinoTransport,OConnor_Ott__2013__apj__TheProgenitorDependenceofthePre-explosionNeutrinoEmissioninCore-collapseSupernovae}.

It is the goal of this work to extend and improve our previous study
\citep{Obergaulinger_Janka__2011__ArXive-prints__Magnetic_field_amplification_in_collapsing_non-rotating_stellar_cores}.
Employing a more detailed analysis, we investigate the relevance of
field amplification mechanisms like those in points
(i)-(iv)~concerning their importance for the evolution and potential
revival of the stalled shock by core-collapse simulations including a
reasonably good treatment of the relevant microphysics.  In
particular, we will study the field growth connected to convective and
SASI activity in the post-shock layer and the role of energy transport
and dissipation by \Alfven~waves.  To this end, we perform MHD
simulations of collapse and post-bounce evolution of the core of a
star of 15 solar masses
\citep[][]{Woosley_Heger_Weaver__2002__ReviewsofModernPhysics__The_evolution_and_explosion_of_massive_stars}.
Varying the strength of the initial field, we can determine different
regimes and mechanisms of field amplification and identify the
back-reaction of the field onto the flow and the ultimate onset of an
explosion.  In contrast to current models of stellar evolution, which
predict predominantly toroidal magnetic fields
\citep{Heger_et_al__2005__apj__Presupernova_Evolution_of_Differentially_Rotating_Massive_Stars_Including_Magnetic_Fields},
we start our simulations from purely poloidal initial fields.  This
choice is in part motivated by the use of similar initial fields in
previous studies of core collapse
\cite[e.g.][]{Obergaulinger_Aloy_Mueller__2006__AA__MR_collapse,Suwa_etal__2007__pasj__Magnetorotational_Collapse_of_PopIII_Stars}.
Furthermore, in axisymmetric models neglecting rotation, there is no
topological difference between the $\theta$- and $\phi$-components of
the field, and a field containing the radial and the
$\theta$-component is qualitatively equivalent to a field consisting
of all three components.  Hence, we do not expect this choice of the
initial field geometry to have a crucial influence on our results.

This article is organised as follows: \secref{Sek:model} describes
our physical model and the numerical methods; \secref{Sek:Init}
introduces the initial conditions; \secref{Sek:Results} presents the
results of our simulations; \secref{Sek:SumCon} gives a summary of the
study and draws some conclusions.

\section{Physical model and numerical methods}
\label{Sek:model}

We assume that the evolution of the gas and the magnetic field is
described by the equations of Newtonian ideal magnetohydrodynamics
(MHD),
\begin{align}
  \label{Gl:MHD-rho}
  \partial_{t} \rho + \vec \nabla \cdot (\rho \vec v) 
  & = 0,
  \\
  \label{Gl:MHD-y_e}
  \partial_{t} (\rho Y_{\mathrm{e}}) + \vec \nabla \cdot (\rho Y_{\mathrm{e}} \vec v )
  & =  S_{\mathrm{n}}^0,
  \\
  \label{Gl:MHD-mom}
  \partial_{t} (\rho v^i )
  + \nabla_j \left( P_{\mathrm{tot}} \delta^{ij} + \rho v^i v^j - b^i b^j
  \right )
  & = 
  - \rho \nabla^i \Phi
  + S^{1;i},
  \\
  \label{Gl:MHD-erg}
  \partial_{t} e_{\mathrm{tot}} 
  + \vec \nabla \cdot 
  \left(
    (e_{\mathrm{tot}} + P_{\mathrm{tot}}) \vec v - (\vec v \cdot \vec b ) \vec b
  \right)
  & = - \rho \vec v \cdot \vec \nabla \Phi
  \\ \nonumber
  & 
  + S^0
  + \vec v \vec S^1,
  \\
  \label{Gl:MHD-ind}
  \partial_{t} \vec b 
  - 
  \vec \nabla \times
  \left(
    \vec v \times \vec b
  \right)
  & = 
  0,
\end{align}
describing the conservation of mass, electron-lepton number, gas
momentum, total energy of the matter, and magnetic flux, respectively.
In addition to these evolutionary equations, the magnetic field has to
fulfil the divergence constraint,
\begin{equation}
  \label{Gl:MHD-divb}
  \vec \nabla \cdot \vec b = 0.
\end{equation}
The symbols used in this system have standard meanings: $\rho, Y_e,
\vec v, \vec b$, $P$, and $\Phi$ are mass density, electron fraction,
velocity, magnetic field, gas pressure, and the gravitational
potential, respectively.  The total energy density is defined as the
sum of internal energy density, $\varepsilon$, kinetic energy density,
and magnetic energy density, $e_{\mathrm{tot}} = \varepsilon + \inv{2}
\rho \vec v^2 + \inv{2}\vec b^2$, and the total pressure contains the
contributions of the gas pressure and the magnetic pressure,
$P_{\mathrm{tot}} = P + \inv{2} \vec b^2$.  The source terms
$S_{\mathrm{n}}^0, S^0$, and $\vec S^{1}$, account for the exchange of
(net) electron number, energy, and momentum between the gas and the
neutrinos, respectively, and will be discussed below.

We use a finite-volume code based on the constrained-transport
formulation of the equations of Newtonian ideal magnetohydrodynamics
and employing high-resolution monotonicity-preserving reconstruction
schemes of \nth{5} order \citep{Suresh_Huynh__1997__JCP__MP-schemes}
and the HLL Riemann solver within the MUSTA framework
\citep{Toro_Titarev__2006__JCP__MUSTA}.  The stellar material is
described by the equation of state (EOS) of
\cite{Lattimer_Swesty__1991__NuclearPhysicsA__LS-EOS} with a
compressibility of $k = 220 \, \MeV$ at high densities above
$\rho_{\mathrm{thr}} = \zehnh{6}{7} \, \gccm$.  Below that threshold,
we use the treatment of the Vertex simulations as applied in
\cite{Liebendorfer_et_al__2005__apj__Supernova_Simulations_with_Boltzmann_Neutrino_Transport_A_Comparison_of_Methods}
including leptons, nuclei (Si and Ni), and radiation.  We approximate
general relativistic gravity by version A of the TOV potential of
\cite{Marek_etal__2006__AA__TOV-potential}.

We will outline the basics of our treatment of neutrino physics in the
following.  For a detailed description of the transport scheme and its
implementation, we refer to a forthcoming publication by Just,
Obergaulinger, \& Janka.  We solve the comoving-frame
energy-dependent, multi-dimensional moment equations for the energy
density, $E_{\alpha} (\omega)$, and energy flux, $F_{\alpha}^i
(\omega)$, of neutrinos of energy $\omega$ (the subscript $\alpha$
distinguishes between neutrino flavours),
\begin{align}
  \label{Gl:RT-0mom}
  \partial_{t} E_{\alpha} (\omega) 
  + \vec \nabla \cdot ( E_{\alpha} (\omega) \vec v + \vec
    F_{\alpha} (\omega))
  - \omega \nabla_j v_k \partial_{\omega} P_{\alpha}^{jk} (\omega) &
  \\ 
  \nonumber
  = S_{\alpha}^0 (\omega) &
  ,
  \\
  \label{Gl:RT-1mom}
  \partial_{t} F_{\alpha}^i (\omega) 
  + c^2 \nabla_j P_{\alpha}^{ij} (\omega)
  + \nabla_j ( F_{\alpha}^i (\omega) v^j )
  + F_{\alpha}^j (\omega) \nabla_j v^i &
  \\
  \nonumber
  - ( \nabla_j v_k)
  \partial_{\omega} (\omega Q^{ijk} (\omega)) &
  \\
  \nonumber
  =
  S_{\alpha}^{1; i} (\omega) &
  ,
\end{align}
where $P^{ij}_{\alpha} (\omega)$ (the neutrino pressure tensor) and
$Q^{ijk} (\omega)$ are the second and third angular moments of the
neutrino distribution function and $\vec v$ denotes the fluid
velocity, respectively.

In addition to energy and momentum conservation (in the absence of
source terms), the neutrino numbers need to be conserved as well, if
sources and sinks do not play a role.  This is not automatically
guaranteed by numerical schemes that solve the moment equations for
the energy density and energy flux.  However, with a suitable
discretisation scheme for Equations (\ref{Gl:RT-0mom}) and
(\ref{Gl:RT-1mom}), this problem can be overcome \citep[][Appendix
B]{Mueller_et_al__2010__apjs__A_New_Multi-dimensional_General_Relativistic_Neutrino_Hydrodynamic_Code_for_Core-collapse_Supernovae.I.Method_and_Code_Tests_in_Spherical_Symmetry}.

Conservation of energy and momentum translates into the following
relations for the source terms in the fluid equations:
\begin{eqnarray}
  \label{Gl:sourceterms-e}
  S^0 & = & 
  - \sum_{\alpha} \int \mathrm{d} \omega \, S^{0}_{\alpha} (\omega)
  ,
  \\
  \label{Gl:sourceterms-p}
  \vec S^{1}& = & 
  - \inv{c^2} \sum_{\alpha} \int \mathrm{d} \omega \, \vec
  S^{1}_{\alpha} (\omega)
  .
\end{eqnarray}
The source term for the conservation equation of net electron number,
\eqref{Gl:MHD-y_e}, follows from the source term for energy,
\begin{equation}
  \label{Gl:sourceterms-y_e}
  S^0_n = 
  - m_{\mathrm{u}}
  \int \mathrm{d} \omega \, 
  \omega^{-1} 
  \left(
    S^{0}_{\nu_e} (\omega)
    - 
    S^{0}_{\bar{\nu}_e} (\omega)
  \right)
  ,
\end{equation}
where $m_{\mathrm{u}}$ is the atomic mass unit. 

Presently only electron neutrinos and antineutrinos are considered,
i.e.~$\alpha = \nu_e, \bar{\nu}_e$.  The source terms,
$S_{\alpha}^{0,1} (\omega)$, include the interaction rates of
neutrinos with the gas by emission, absorption, and scattering
reactions; we use a reduced set of processes, viz.
\begin{enumerate}
\item emission and absorption of electron neutrinos by neutrons,
\item emission and absorption of electron anti-neutrinos by protons,
\item elastic scattering of all neutrinos off nucleons,
\item emission and absorption of electron neutrinos by heavy nuclei,
\item coherent elastic scattering of all neutrinos off heavy nuclei,
\item inelastic scattering of neutrinos off electrons and positrons.
\end{enumerate}
In our implementation of these processes, we follow
\cite{Rampp_Janka__2002__AA__Vertex}.  Due to their stiffness, we
solve these source terms implicitly and split them off from the other
parts of the conservation equations by an operator-split step.

Despite the simplified treatment of neutrino-matter interactions, in
the context of the questions focussed on in this paper, the scheme
provides a reasonably good representation of the neutrino effects that
play a role during core collapse, bounce, shock propagation and in the
accretion layer behind the stalled supernova shock. Moreover, it is a
computationally efficient treatment of the neutrino transport, and
two-dimensional simulations up to several hundred milliseconds after
bounce are well feasible.

Furthermore, fluid-velocity dependent effects such as the $P \,
\mathrm{d} V$ work associated with diverging flows appear in the
equations as well.  To close the system of moment equations, we have
to specify the tensor of the second moment, $\neutrino{P}^{ij}$, for
which we employ the maximum-entropy closure due to
\cite{Cernohorsky_Bludman__1994__ApJ__MEC-Transport}.  Using a
tensorial generalisation of a one-dimensional \emph{Eddington factor},
our method is generically multi-dimensional.

We note that our set of equations goes beyond the usual diffusion
ansatz, in which the system of moment equations is truncated at the
level of the energy equation and closed by expressing the flux in
terms of the gradient of the energy density, mostly connecting
diffusion and free-streaming limits by a flux limiter.  Retaining the
first two moment equations, our scheme leads to a hyperbolic system,
which can be solved by common methods such as high-resolution
shock-capturing methods.

In the models presented here, we include only electron neutrinos and
anti-neutrinos and neglect pair processes and muon and tau neutrinos.
This simplification affects to a certain degree the cooling of the
PNS, and hence its contraction.  Although this effect modifies the
surrounding layers where much of the field amplification takes place,
these modifications are not sufficiently strong to change our main
results concerning the processes relevant for field amplification beyond the
level of quantitative corrections.  We expect the corrections
to be of a similar magnitude as the ones due to other uncertainties
that affect the PNS cooling such as the possibility of a stiffer
nuclear equation of state.  The ``stiffness'' of the neutron-star EOS
and the corresponding faster or slower contraction of the PNS can,
however, have a decisive influence on the development of an explosion,
which is favoured or develops faster for a soft EOS with more compact
PNS
\citep{Marek_Janka__2009__apj__Delayed_Neutrino-Driven_Supernova_Explosions_Aided_by_the_Standing_Accretion-Shock_Instability,Janka__2012__ARNPS__ExplosionMechanismsofCore-CollapseSupernovae,Suwa_et_al__2013__apj__OntheImportanceoftheEquationofStatefortheNeutrino-drivenSupernovaExplosionMechanism,Couch__2013__apj__TheDependenceoftheNeutrinoMechanismofCore-collapseSupernovaeontheEquationofState}.
The goal of this paper is therefore not the determination of explosion
conditions of 2D stellar cores with magnetic fields in the most
``realistic'' manner (also the simplifications applied in our
treatment of neutrino interactions would conflict with such an aim)
but the assessment of field amplification mechanisms and of the
influence of strong fields on the development of an explosion in
comparison to the non-magnetic case.

\section{Models and initial conditions}
\label{Sek:Init}

\begin{figure}
  \centering
  \includegraphics[width=\linewidth]{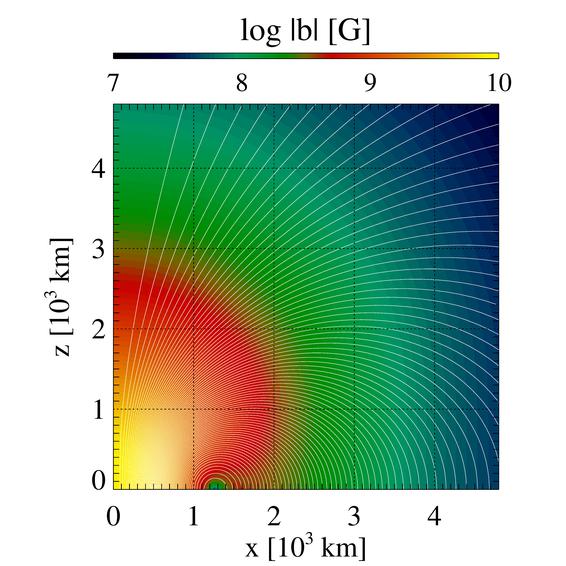}
  \caption{
    Initial (pre-collapse) purely poloidal field configuration of Model
    \modelname{B10}: field strength (colour scale) and field lines
    in the inner $\sim 5000 \, \km$.  For simplicity, only one
    quadrant is shown.
  }
  \label{Fig:B-init}
\end{figure}

The main purpose of our simulations is the differential analysis of
the effects of variations of the initial magnetic field on the
collapse and post-bounce evolution of a stellar core until explosion
when otherwise all parameters are kept equal.  We select the core of a
star of 15.0 $\msun$
\citep[][]{Woosley_Heger_Weaver__2002__ReviewsofModernPhysics__The_evolution_and_explosion_of_massive_stars}
and map the pre-collapse model to a grid of $n_r = 360$ zones.  In the
axisymmetric simulations, $n_{\theta} = 144$ lateral zones were
distributed uniformly in $\theta$ between the north and south pole.
Up to a radius of $r \approx 18 \, \km$, the radial grid was uniform
with a grid width of $\delta r = 400 \, \mathrm{m}$.  For higher
radii, the grid width was set to $\delta r = r \frac{\pi}{144}$,
resulting in an aspect ratio close to unity for grid cells.  The outer
radius of the grid was $r_{\mathrm{max}} \approx \zehnh{14.1}{3} \,
\km$.  At a radius of $r = 100 \, \km$, the grid width is $\delta r
\approx 2 \, \km$.  The resolution at the grid centre is $(\delta
r)_{\mathrm{ctr}} = 400~\mathrm{m}$.  We discretise the energy
dependence of the neutrino distribution function with $n_{\omega} =
16$ energy bins. The first zone of the energy grid covers the energy
range $[0, 5.36 \, \MeV]$, and the grid width of the other 15 zones is
given by $\delta \omega \approx 0.2915 \omega$.  The grid extends to a
maximum energy of $\omega_{\mathrm{max}} = 440 \, \MeV$.

The topology of the magnetic field at the onset of collapse is highly
uncertain.  On the main sequence, field amplification by,
e.g.,~gradual contraction of the star or convection competes with the
loss of magnetic energy in stellar winds and in work the magnetic
field does by exerting torques on the stellar matter.  In the absence
of rotation, stars lack an important ingredient of most large-scale
dynamos. In such a case, they may be dominated by a small-scale
turbulent field rather than a large-scale field.  Nevertheless, we
assume a simple initial field, viz.~a modified dipole
(\figref{Fig:B-init}).  While this may not be the most likely field
configuration in real stellar cores (though it resembles the poloidal
component of the field topology found by
\cite{Braithwaite_Nordlund__2006__aap__Stablemagneticfieldsinstellarinteriors}
simulating the magnetohydrodynamics of stellar interiors), it
represents a favourable configuration for the \Alfven-wave
amplification mechanism we are interested in because of the large,
coherent radial component allowing for the radial propagation of
\Alfven~waves.  Furthermore, we observe that the field is replaced by
a more complex small-scale field in the regions of the core affected
by hydrodynamic instabilities such as convection and the SASI.  Hence,
we deem the influence of such an artificial choice for the initial
field on our results not crucial.

The field is the same as the one used by
\cite{Suwa_etal__2007__pasj__Magnetorotational_Collapse_of_PopIII_Stars},
defined by a vector potential of the form
\begin{equation}
  \label{Gl:Init-A}
  A^{\phi} = \frac{b_0}{2} \frac{r_0^3}{r^3 + r_0^3} r \sin \theta.
\end{equation}
We set the normalisation radius $r_0$ defining the location of the
dipole to $r_0 = 1000~\mathrm{km}$ and vary the parameter $b_0$
setting the field at the centre of the core to values between $10^{8}$
and $10^{12}$ G, i.e.~in a range considerably above that found by
\citep{Heger_et_al__2005__apj__Presupernova_Evolution_of_Differentially_Rotating_Massive_Stars_Including_Magnetic_Fields}
for rotating progenitors.  In particular the strongest initial fields
can, thus, not be representative for typical pre-collapse cores.  The
models are called \modelname{B$b$}, where the number $b$ is the
decadic logarithm of $b_0$.  \figref{Fig:B-init} displays the initial
field of Model \modelname{B10}.  For comparison, we also performed
simulations of a non-magnetic model in spherical symmetry as well as
axisymmetry (Models \modelname{s15-15} and \modelname{B0},
respectively).

We assume open boundary conditions at the outer radius of our grid.
There we extrapolate the mass density and the radial velocity of the
gas in such a way that the mass accretion rate in the ghost zones,
$\partial_{t} M_{outer} \propto \rho v_r r^2$, varies linearly with
radius.

\section{Results}
\label{Sek:Results}

\subsection{Hydrodynamics of non-magnetic models}
\label{sSek:B0}

We start the presentation of the simulation results with a description
of the evolution of models without magnetic fields in spherical
symmetry and axisymmetry, Models \modelname{s15-1d} and
\modelname{B0}, respectively.  They serve us as reference cases to
which we will compare the results of the two-dimensional magnetised
models.  Furthermore, they allow for a comparison to existing
simulations that use more accurate treatments of neutrino
interactions.

We have performed a spherically symmetric (1D) simulation of the same
$15 \, \Msol$ progenitor star that we employ for the rest of the
models in our paper.  This simulation shows an overall qualitative
(and in many aspects good quantitative) agreement with results
obtained with more sophisticated neutrino-transport schemes and
including all neutrino flavours.  A detailed comparison will be
provided by Just et al.\ (in preparation).  As expected, the
spherically symmetric model does not yield a successful SN explosion.
Instead, the core collapses after $t \simeq 780\, \ms$ of post-bounce
accretion.  We note that we cannot treat the evolution of this final
collapse stage within the framework of our pseudo-relativistic
treatment of gravity, but all diagnostic parameters indicate the onset
black-hole formation, driven by continued accretion of collapsing
stellar matter.

\begin{figure*}
  \centering
  \includegraphics[width=\textwidth]{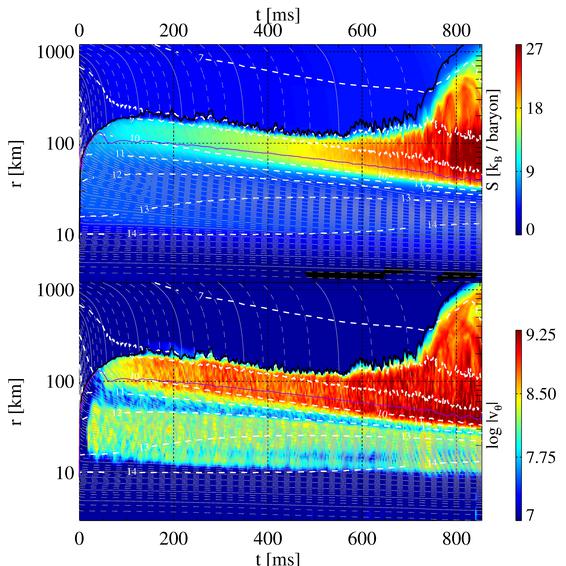}
  \caption{
    Post-bounce evolution of the non-magnetic axisymmetric model,
    \modelname{B0}.  The
    top half shows angular averages of the specific entropy of the model, while the bottom
    half displays the absolute value of the lateral velocity, also
    averaged over angles.
    In addition, trajectories of radii corresponding to chosen values
    of enclosed mass (grey), and
    iso-density contours (white) are shown, and the solid black and
    violet lines represent the average positions of the shock wave and
    the gain radius, respectively.
  }
  \label{Fig:B0-mshell}
\end{figure*}

Deviations from spherical symmetry start to show up in the
axisymmetric version of the non-magnetic model, \modelname{B0}, after
bounce.  We display the time evolution of angularly averaged specific
entropy and lateral velocity in mass-shell plots in
\figref{Fig:B0-mshell}\footnote{Although the
  definition of mass shells is not possible in multi-dimensional
  models, we will use this term to refer to shells of enclosed mass in
  angularly averaged profiles of the axisymmetric models.}.  The
angular averages of the absolute value of the non-radial velocity,
visible in the bottom half, indicate where hydrodynamic instabilities
develop.  Two potentially unstable regions appear very clearly:
\begin{enumerate}
\item Convective instability inside the PNS is limited in many
  simulations to a region below the neutrinospheres
  \citep[e.g.][]{Buras_etal__2006__AA__Mudbath_1,Dessart_et_al__2006__apj__Multi-d_RHD_Simulations_of_PNS_Convection}.
  In these simulations, convection basically develops only where the
  Ledoux-criterion for instability is fulfilled.  Although the
  Ledoux-unstable region has a very similar shape in our model,
  non-radial mass motions are not limited to this region in our case,
  but cover basically the entire outer layers of the PNS, from a
  density of $\rho = \zehn{14} \, \gccm$ to the neutrinospheres. 
  This peculiar difference is most likely a consequence of the
  omission of $\mu$ and $\tau$ neutrinos.  We take it as a matter of
  fact and will mainly discuss how the unstable region reacts to the
  presence of a magnetic field.
\item The so-called hot-bubble region behind the stalled shock wave
  exhibits unstable flows due to convection and the standing accretion
  shock instability (SASI).  The relative importance of these two
  instabilities is currently a matter of intense investigations
  \citep[see,
  e.g.][]{Burrows_et_al__2012__apj__AnInvestigationintotheCharacterofPre-explosionCore-collapseSupernovaShockMotion,Mueller_et_al__2012__apj__NewTwo-dimensionalModelsofSupernovaExplosionsbytheNeutrino-heatingMechanism_EvidenceforDifferentInstabilityRegimesinCollapsingStellarCores,Fernandez_et_al__2014__mnras__CharacterizingSASI-andconvection-dominatedcore-collapsesupernovaexplosionsintwodimensions}.
  As in the case of the instability inside the PNS, a closer
  discussion of this problem is not the topic of our work, but we
  focus on the magnetic field amplification associated with these
  non-radial flows.
\end{enumerate}

Non-radial mass motions in the gain layer cause shock deformations and
oscillations on a moderate scale.  For a long time, the minimum, mean,
and maximum shock radii follow the trend of the spherical model and
recede gradually.  The first $\sim 300 \, \ms$ after bounce are
characterised by oscillations of the shape of the shock (measured,
e.g.~by the difference between minimum and maximum shock radius).
Later, shock recession comes to a halt.  The minimum and average shock
radius never decrease below 100 km, and expand after $t \sim 550 \,
\ms$.  At this point, the post-shock instabilities and shock
oscillations start to become more violent again.  The shock adopts an
increasingly bipolar form.  Finally, it expands, and a successful
explosion sets in.  Since this reduces the mass accretion onto the
PNS, it does not undergo collapse as in the spherical version of the
model.  We note that the exact time of the onset of explosion will
certainly depend on the detailed shock oscillations that develop in
the phase of shock expansion and, thus, will have a stochastic nature.

Our results for this model should be understood in the same spirit as
the spherical model, i.e.~as a reference point for the following
discussion of the magnetic models and not as a detailed investigation
of the physics of the hydrodynamic instabilities and the neutrino
physics.  Irrespective of its cause, physical or as a result of our
approximations, we take advantage of the long delay until the
explosion sets in, because it allows us to follow the evolution of the
magnetic field for a long time before the post-shock flows are
disrupted.  Thus we may identify magnetic effects that might be
suppressed in a model exploding earlier, e.g.~due to neutrino heating.

Based on this dynamical evolution, we introduce for further reference
the following regions commonly used in CCSN theory:
\begin{description}
\item[\region{IHSP}:] The \emph{inner hydrodynamically stable PNS}
  extends from the origin to roughly the radius where the density of
  the gas drops below $10^{14} \ \gccm$.
\item[\region{PCNV}:] This inner core is surrounded by the \emph{PNS
    convection zone}, characterised by a negative gradient of the
  electron fraction and a flat or slightly negative entropy gradient.
  The outer boundary of this layer is associated with the location of
  the minimum of the $Y_e$ profile.
\item[\region{COOL}:] In the \emph{cooling layer} outside the PNS
  convection zone, the accreted matter suffers a net energy loss due
  to the production of neutrinos.  At its bottom, the flow is
  decelerated and settles onto a stable layer on the ``surface'' of
  the PNS.
\item[\region{GAIN}:] The \emph{gain region} is stirred by hot-bubble
  convection and SASI activity.  The gain radius, which defines the
  transition from neutrino cooling below to neutrino heating above,
  coincides approximately with the upper boundary of the stable layer.
  Because of neutrino heating, the gain layer develops a negative
  entropy gradient and thus postshock convection. The SASI activity,
  however, takes place in a larger volume, encompassing parts of the
  neutrino-cooling layer as well.  SASI modes are amplified between
  the shock as an outer boundary and an inner boundary layer, where
  the accretion flow gets decelerated \citep[see,
  e.g.~][]{Foglizzo_Scheck_Janka__2006__apj__Nu-driven_Convection_versus_Advection,Scheck_et_al__2008__aap__Multid_SN_sim_with_approx_nu_transport.II}.
  This happens typically between neutrinosphere and gain radius.
  Furthermore, convective overshooting may extend to regions below the
  gain radius.
\end{description}
Unless stated otherwise, we define these regions based on angularly
averaged profiles.  The resulting radii separating the four regions
are therefore only an approximate representation of the more complex
two-dimensional borders of the regions.  For instance, convective
overshooting in the \region{COOL} region leads to rather high
non-radial kinetic energies also below the average\footnote{We will
  omit this adjective in the following.} gain radius.

\subsection{Weak initial magnetic field: Model \modelname{B10}}
\label{sSek:weakfield}

Models \modelname{B08}, \modelname{B10} and
\modelname{B11} ($b_0 = 10^{8,10,11}~\mathrm{G}$, respectively)
exhibit very similar dynamics compared both to each other and to the
non-magnetised reference model.  We will focus in the discussion on
Model \modelname{B10}.

We simulated Model \modelname{B10} until a post-bounce time $t
\approx 930 \, \ms$.  The structure of the core during the
post-bounce evolution is the same as that of Model \modelname{B0} 
shown in \figref{Fig:B0-mshell}.  We will present the discussion
of the evolution of the magnetic field in the regions introduced above
in the following.

\subsubsection{The proto-neutron star}

\begin{figure*}
  \centering
  \includegraphics[width=\textwidth]{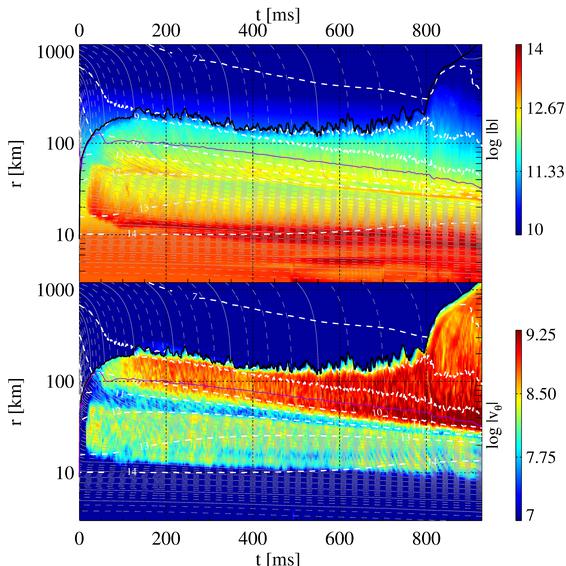}
  \caption{
    Evolution of Model
    \modelname{B10}.  This figure is similar to
    \figref{Fig:B0-mshell}, and the white, grey, violet, and black lines
    have the same meaning.  We show angular averages of the magnetic
    field strength (top) and of the absolute value of the lateral
    velocity (bottom).
  }
  \label{Fig:B10-PNS-mshell}
\end{figure*}

\begin{figure}
  \centering
  \includegraphics[width=\linewidth]{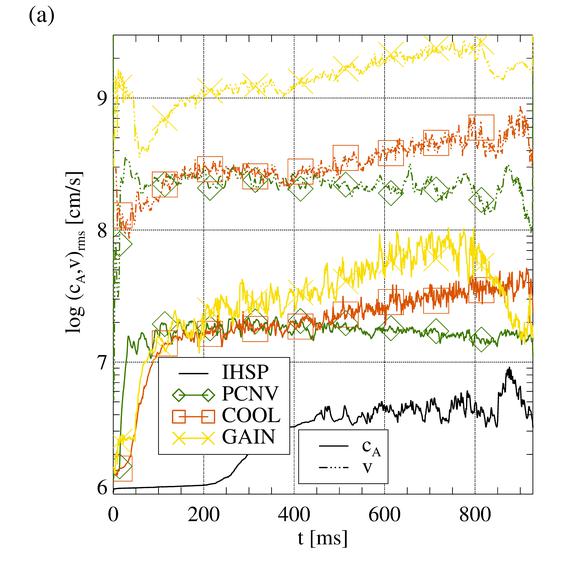}
  \includegraphics[width=\linewidth]{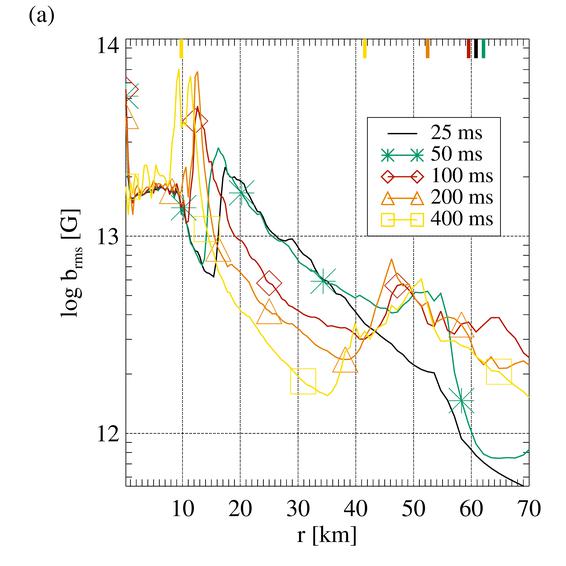}
  \caption{
    \subpanel{Top, (a)}: Volume-averaged  \Alfven~(solid lines) and
    absolute values of the flow speeds (dashed lines) in the post-shock regions of Model
    \modelname{B10} as a function of post-bounce time: the black
    line without any symbol, the green line with diamonds, the red
    line with squares and the yellow line with crosses correspond to the inner hydrodynamically stable
    core, the PNS convection zone, the cooling layer, and the gain layer, respectively.  We do not
    include the flow speed in the \region{IHSP} region; its small
    value is outside the range of the ordinate.
    \subpanel{Bottom, (b)}: profiles of the field strength (averaged
    over angles) for
    different times, as indicated in the legend.  The line for time
    $t$ is the
    result of a time average of the profiles in the time interval $[t - 5\, \ms, t
    + 5 \, \ms]$.  The small vertical tick marks  at the upper edge of
    the plot indicate the outer boundaries of the \region{IHSP}
    region (at $r \approx 10\,\km$) and of the \region{PCNV} region
    for the five times. 
  }
  \label{Fig:B10-PNS-B-tevo}
\end{figure}

The regions inside of the PNS (regions \region{IHSP}, \region{PCNV},
and \region{COOL}) are visible in \figref{Fig:B10-PNS-mshell}, and the
time evolution of the angular average values of the \Alfven~(solid
lines) and flow speeds (dashed lines) in these regions is displayed in
the \subpanel{top} panel of \figref{Fig:B10-PNS-B-tevo}.  Accretion of
matter (cf.~the grey trajectories of angularly averaged mass shells in
\figref{Fig:B10-PNS-mshell}) leads to an increasing mass of the PNS
and partially powers the neutrino emission.  The PNS (for which we
take the iso-density line $\rho = 10^{12} \, \gccm$ as a proxy)
contracts to a radius around $R_{\mathrm{PNS}} \approx 27 \, \km$ at
the end of the simulation.  The extent and intensity of the
hydrodynamically unstable flows inside the PNS is the same as in the
non-magnetic model, i.e.~we do not find a notable influence of the
magnetic field on the hydrodynamics.  Consequently, the average field
strength in the \region{IHSP} region is basically constant; see the
black line in \figref{Fig:B10-PNS-B-tevo} (\subpanel{top}) - the
increase after $t \sim 200 \, \ms$ is not caused by field
amplification in this region, but by a slow transport of magnetic
energy across the iso-density surface of $\rho=10^{14}\,$g\,cm$^3$
from the convectively active region.

\begin{figure}
  \centering
  \includegraphics[width=\linewidth]{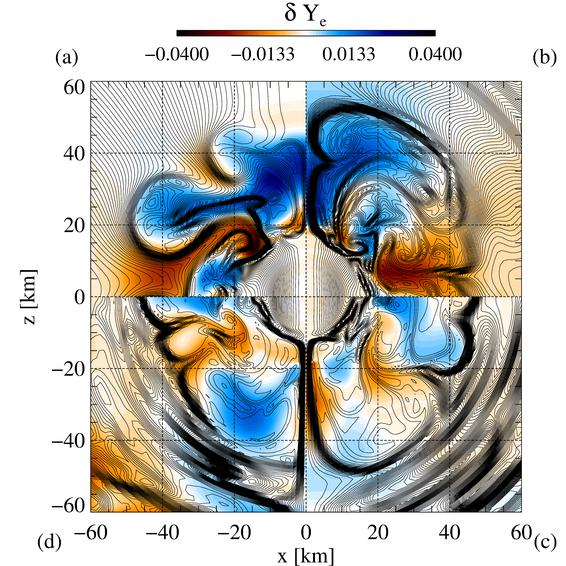}
  \caption{
    Snapshots of the PNS of Model \modelname{B10} at $t = 25, 50,
    100, 200 \, \ms$ (panels
    \subpanel{(a)}, \subpanel{(b)}, \subpanel{(c)}, \subpanel{(d)}).
    We show the deviation of the electron fraction from
    its angular average (colours) and magnetic field lines.  
    Lepton-rich matter rises in mushroom-like structures, lepton-poor
    fluid sinks inwards.
    Please note that
    each of the panels shows only one half of the snapshots, which were
    computed in a full $180 \grad$ geometry.
  }
  \label{Fig:B10-PNS-Ye-Bfield-1}
\end{figure}

Overturning motions in the convectively unstable region, on the other
hand, have a strong effect on the magnetic field.  Snapshots forming a
time series of the first $200 \, \ms$ of the evolution of the PNS are
displayed in \figref{Fig:B10-PNS-Ye-Bfield-1}.  The innermost $\approx
15 \, \km$ are stable with velocities remaining close to zero there,
and the magnetic field retains its initial topology.  Outside of this
radius, a few large convective rolls develop consisting of upflows of
high (blue) and downflows of low (red) $Y_e$ and entropy.  They twist
and stretch the magnetic field lines and amplify the field in region
\region{PCNV} until a stationary level is reached.  The solid green
line with diamonds in the \subpanel{top} panel of
\figref{Fig:B10-PNS-B-tevo} presents a volume average of the
\Alfven~speed, $c_{\mathrm{A}}$, defined as
\begin{equation}
  \label{Gl:cA}
  c_{\mathrm{A}} = \rho^{-1/2} |\vec b|.
\end{equation}
The steep early rise before $t \approx 50 \, \ms$ is a geometric
effect as the \region{PCNV} is established.  Afterwards, the mean
field increases moderately and levels off at a value around
$b_{\mathrm{\region{PCNV},rms}} \approx \zehnh{7}{13} \,
\Gauss$\footnote{We define the r.m.s. mean of a vector field as ${\vec
    u}_{\mathrm{rms}} = \langle \vec u ^2 \rangle^{1/2}$, where
  $\langle . \rangle$ represent the (volume) average.}, corresponding
to an \Alfven~speed of a few times $\zehn{7} \, \cms$.

In principle, a natural scale for the final magnetic energy achieved
by convective field amplification could be set by the kinetic energy
of the convective rolls.  In our model, however, the magnetic energy
falls short of this upper limit by about two orders of magnitude: the
\Alfven~speed is at least an order of magnitude lower than the flow
speed.  This shortfall coincides with a particular geometry of the
field: the field is rather weak in the interior of the convective
region, but strong in its upper and lower boundaries layers, as can be
seen in the profiles of the \subpanel{bottom} panel of
\figref{Fig:B10-PNS-B-tevo} and the upper part of
\figref{Fig:B10-PNS-mshell}.  Inside the convective region
\region{PCNV}, a profile of the r.m.s.~field strength roughly $\propto
r^{-2}$ is established at late times.  The power of the profile
decreases considerably, i.e.~convection does not lead to field
amplification, or, at least, is unable to compensate the loss of
magnetic flux from this region due to losses across the boundary of
the convective layer and resistive dissipation and diffusion, leading
to a field far from equipartition with the kinetic energy.  Although
our simulations are run in ideal MHD, i.e.~without an explicit
physical resistivity, inevitably the numerical discretisation leads to
errors which can behave similarly to a physical resistivity \citep[for
a thorough discussion,
see][]{Endeve_et_al__2012__apj__TurbulentMagneticFieldAmplificationfromSpiralSASIModes:ImplicationsforCore-collapseSupernovaeandProto-neutronStarMagnetization}.
We find strong indications for the influence of numerical resistivity
in the development of closed magnetic field lines in the interior of
these overturning cells (e.g.~around $(x,z) = (-22 \, \km, -22 \,
\km)$ in Panel \subpanel{(d)} of \figref{Fig:B10-PNS-Ye-Bfield-1}).
These form from magnetic flux sheets stretched by the overturning
flows until their transverse dimension is comparable to the grid
scale, at which point numerical diffusion disrupts them into separated
magnetic islands.

In two-dimensional convection, such a phenomenon, known as convective
flux expulsion, is very common
\citep[e.g.][]{Mestel__1999__Stellarmagnetism}.  Its occurrence might
be taken as indication that our simulations are under-resolved and we
should use much finer grids to reach convergence.  Besides flux
expulsion from the convection zone below, the accretion flow coming
from the shock wave is decelerated in this region, thus depositing
magnetic flux there and contributing to the magnetic energy (cf.~the
red line in the \subpanel{top} panel of \figref{Fig:B10-PNS-B-tevo}).
This part of the magnetic field exhibits a strong lateral component,
as can be seen in Panels \subpanel{(c), (d)} of
\figref{Fig:B10-PNS-Ye-Bfield-1}.

\begin{figure}
  \centering
  \includegraphics[width=\linewidth]{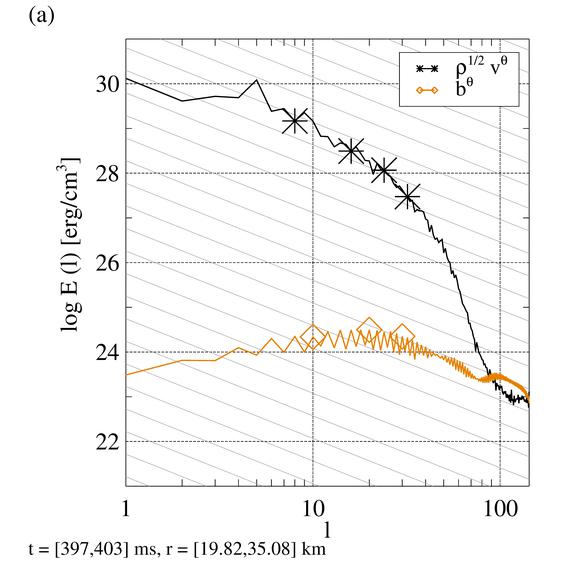}
  \includegraphics[width=\linewidth]{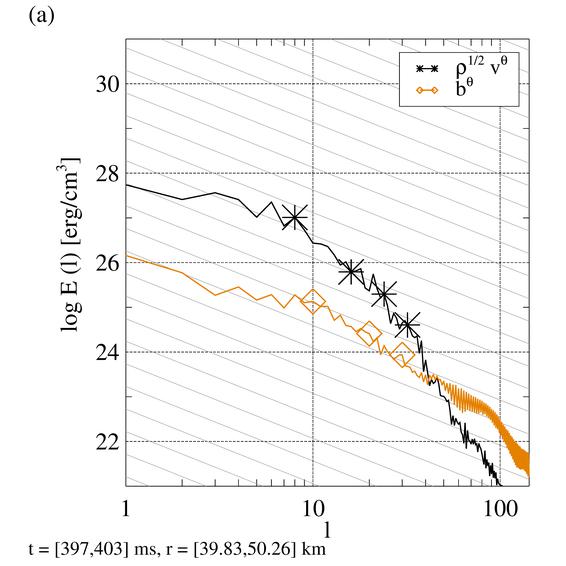}
  \caption{
    Representative lateral kinetic (black lines with stars) and magnetic energy spectra (orange lines
    with diamonds) in the PNS convection zone (Panel \subpanel{(a)}
    corresponds to $t = 400 \, \ms$) and in the stable
    layer (Panel \subpanel{(b)}) of Model \modelname{B10}.  The
    spectra are taken as temporal 
    and radial averages over the ranges indicated in the panels.  For
    comparison, the grey lines show the turbulent Kolmogorov scaling.
  }
  \label{Fig:B10-PNS-ellspec}
\end{figure}

This evolution is reflected in the energy spectra of the angular
components of the velocity and magnetic field.  We decompose a
variable $f (t,r,\theta)$ into spherical harmonics of degree $l$:
\begin{equation}
  \label{Gl:spherharmdeco}
  a_{l;f} = \int \mathrm{d} \Omega f ( t,r,\theta ) Y^0_l (\theta),
\end{equation}
where the spherical harmonics $Y^0_l ( \theta ) = \sqrt{\frac{2 l +
    1}{4\pi}} P_l^0 (\cos \theta)$ are defined in terms of the
associated Legendre polynomials $P^0_l$.  From the coefficients
$a_{l;f}$, we compute an energy spectrum, $E_{l;f} = a_{l;f}^2$.
Setting $f = \sqrt{\rho v_{\theta}^2}$ and $f = |b_{\theta}|$,
$E_{l;f}$ gives the lateral kinetic and magnetic energy of a spectral
mode of degree $l$ .  We show selected kinetic and magnetic energy
spectra of the $\theta$-components at a position in the lower
\region{PCNV} region for times $t = 400 \, \ms$ in Panel
\subpanel{(a)} of \figref{Fig:B10-PNS-ellspec}.  The curves can be
compared to the Kolmogorov scaling, $l^{-5/3}$ (grey lines)
characterising the enstrophy cascade at low to intermediate $l$ and
the steeper scaling, typically $l ^ {-3}$, of the energy cascade at
high $l$
\citep[][]{Hanke_et_al__2013__apj__SASIActivityinThree-dimensionalNeutrino-hydrodynamicsSimulationsofSupernovaCores}.
Lower modes of the velocity spectra ($l \lesssim 10$) have a mostly flat
spectrum, and a Kolmogorov scaling is a good approximation in a
rather limited intermediate range ($l \lesssim 30$), while higher
modes are characterised by a steeper decline towards the dissipation
scales close to the grid resolution.  The normalisation of the
low-mode part of the spectrum remains roughly constant, hinting at a
constant energy in the convective motions in the PNS.  The magnetic
field possesses a mostly flat spectrum with much less spectral energy
throughout the well-resolved modes.  Over the course of the evolution,
we observe a slow tendency of decreasing spectral power of the
magnetic field on the lowest angular scales, while high-order modes
develop a pronounced bump as overturns break down the large-scale
structures of the magnetic field in the PNS convection layer into
small-scale modes and, finally, dissipation converts those into
thermal energy.

The structure of the field lines as shown in
\figref{Fig:B10-PNS-Ye-Bfield-1} hints towards another shortcoming of
our models: the magnetic field is strongest along the symmetry axis
where the geometry does not allow for the development of lateral
velocities.  This effect can also be understood in terms of the
expulsion of magnetic flux into a region where convection is
suppressed.  However, the reason for the suppression in this case is
not physical as the effect happens at the angular boundaries of the
\region{PCNV} region; instead, it is an artifact of the assumption of
axisymmetry.

The magnetic flux expelled from the interior of the convection zone
tends to accumulate at the upper and lower boundaries where the
velocities are close to zero, see the maxima in the \subpanel{bottom}
panel of \figref{Fig:B10-PNS-B-tevo} and the green-yellow and red
bands in the top half of \figref{Fig:B10-PNS-mshell}.  The fluid
velocities close to the radial and angular boundaries of the unstable
layer are strongly aligned with the magnetic field and both are
tangential to the region where overturning velocities operate: lateral
at the top and bottom, and predominantly radial along the axis.  As a
consequence, very little magnetic flux is advected into the fluid
vortices.  In Panels \subpanel{(c)} and \subpanel{(d)} of
\figref{Fig:B10-PNS-Ye-Bfield-1}, a few radial flows between the
large convective cells can be identified, which drag along field
lines.  Injection of magnetic flux into the unstable region by these
narrow, finger-like flows and its amplification by the convective
motions is apparently insufficient to balance the loss of flux by
diffusion and dissipation.  In addition to the small amount of flux
advected into the convective zone, their shape makes these magnetic
fingers very vulnerable against dissipation or resistive
instabilities.  Therefore, they can be destroyed very easily, which
further contributes to limiting the potential of field amplification.

The thin layers of non-radial field on top of the \region{PCNV} layer
are not as prone to resistive disruption as the filaments in the
unstable region.  This highlights the importance of the small-scale
dynamics for the field evolution.  Numerical resistivity has only an
important effect on structures with a size close to the grid
resolution.  The layers of lateral fields consist of far wider flux
sheets, and also the radial streams injecting magnetic flux into the
\region{PCNV} layer are relatively thick.  The latter, however, are
quickly stretched by the convective motions and in this process become
increasingly thin until their transverse dimension approaches the grid
width.

The energy spectra of the lateral flow in the stable layer (panel
\subpanel{(b)} of \figref{Fig:B10-PNS-ellspec}) feature a
Kolmogorov-like power law in a short range of intermediate $10
\lesssim l \lesssim 30$, and flatter and steeper scalings at low and
high $l$, respectively.  The magnetic spectra are similar to the
velocity spectra, though at $l \lesssim 10$ they possess less power by
about two orders of magnitude.  In this part of the spectrum, the
spectral magnetic energy, though larger than in the \region{PCNV}
region, is still less than the spectral kinetic energy.  Both energies
are at a level much lower than that of the velocity field in the
convective region.

If the resulting field structure remains frozen into the matter
throughout the further evolution, the newly-born neutron star would
possess a fairly peculiar field structure, where the observed surface
and crustal field is enhanced w.r.t.~the layers underneath and has a
strong non-radial component.  If this geometry does not change during
the subsequent evolution, it might lead to a neutron star whose
surface is shielded from external flows by a strong magnetic field.
Furthermore, the magnetic energy stored in the layers of a non-radial
surface field of alternating polarity (for Model \modelname{B10} of
the order of $\zehn{44} \, \erg$) might be released by reconnection
and trigger explosive events in the magnetosphere.

Whether or not this field geometry develops in a more realistic model,
in which the three main restrictions of our simulations--neglect of
$\mu/\tau$ neutrinos, axisymmetry, and low resolution--are removed,
remains to be seen in future models.

\subsubsection{The post-shock region}

\begin{figure}
  \centering
  \includegraphics[width=\linewidth]{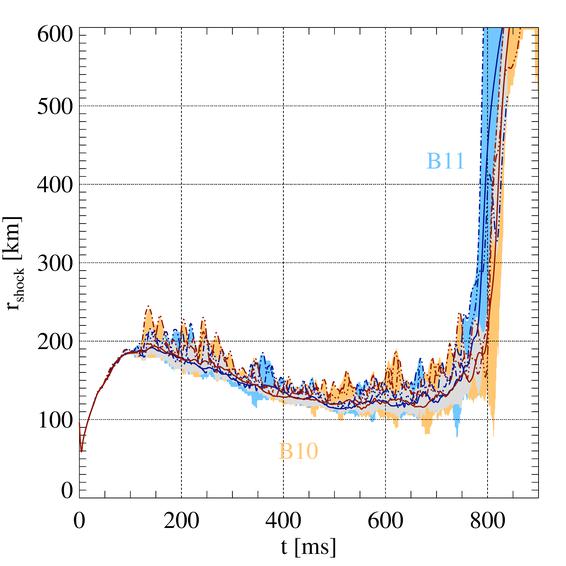}
  \caption{
    Shock positions
    of Models \modelname{B10} and \modelname{B11} as functions of time.  For each model, all
    shock radii at a given time are inside the orange/blue bands (the
    overlap of both bands is coloured grey).  In addition, the average
    shock radii and the shock position along the north and south pole
    are shown with solid and dashed lines, respectively.  
  }
  \label{Fig:B10-shockmodes}
\end{figure}

The magnetic field of Models \modelname{B08}, \modelname{B10}, and
\modelname{B11} is too weak to lead to significant changes in the
behaviour of the post-shock region (see \figref{Fig:B10-PNS-mshell}
for the evolution of Model \modelname{B10}).  In
\figref{Fig:B10-shockmodes}, we present the evolution of the minimum,
maximum, average and polar shock positions of the two models, and the
right panels show the decomposition of the shock radius of Model
\modelname{B10} into spherical harmonics during the post-bounce
evolution.  Similarly to
\cite{Burrows_et_al__2012__apj__AnInvestigationintotheCharacterofPre-explosionCore-collapseSupernovaShockMotion},
we compute the expansion coefficients, $a_l(t)$, from the shock
position as a function of angle, $R_{\mathrm{sh}} (t, \theta)$, as
\begin{equation}
  \label{Gl:shockspherharm}
  a_l (t) =\inv{\sqrt{4 \pi (2 l +1)}} 
  \int_{\theta = 0}^{\theta = \pi}
  \mathrm{d} (-\cos \theta) 
  Y_l^0 (\theta) R^0_{\mathrm{sh}} (t, \theta).
\end{equation}

Again, the shock stalls early and recedes, while showing a varying
degree of non-spherical oscillations.  After $t \sim 400 \, \ms$, the
shock radius stabilises at values just above $100 \, \km$.  The early
shock phase is fairly spherical with small amplitudes of low-order
modes.  The dipole mode, $l = 1$, grows until reaching saturation
around $t \sim 150 \, \ms$.  Around $t \sim 300 \, \ms$, the dipole
mode shows a minimum.  Afterwards, shock oscillations become more
violent, and, most prominently, large dipole modes with a significant
quadrupolar contribution develop.  Eventually, these oscillations
evolve into a coherent expansion of the shock wave across all
latitudes, signifying the onset of the explosion.  For both models,
this happens at almost the same time, insignificantly later than in
the non-magnetic model.  We do not attribute this short delay to the
presence of the magnetic field; instead, it is the result of the
stochastic nature of the non-radial flows governing the oscillatory
motion of the shock wave.

The post-shock magnetic field is amplified by flux-freezing accretion
and by the turbulent flows behind the shock.  For the latter, two
primary sources exist, viz.~convection and the SASI.  Though a
detailed analysis of the relative importance of these two
instabilities would be interesting and might be crucial for
understanding the explosion mechanism, we do not focus on this here
and limit ourselves to those aspects most relevant for the magnetic
field.  We note that
\cite{Fernandez_et_al__2014__mnras__CharacterizingSASI-andconvection-dominatedcore-collapsesupernovaexplosionsintwodimensions}
have emphasised that a SN explosion is intimately linked to the
presence of large-scale bubbles of high entropy \citep[see
also][]{Hanke__2012__apj__IsStrongSASIActivitytheKeytoSuccessfulNeutrino-drivenSupernovaExplosions,Couch__2013__apj__TheDependenceoftheNeutrinoMechanismofCore-collapseSupernovaeontheEquationofState,Dolence_et_al__2013__apj__DimensionalDependenceoftheHydrodynamicsofCore-collapseSupernovae}.
Depending on the prevalence of convection or the SASI, different
possible formation mechanisms exist for these, allowing for a
multitude of explosion scenarios.

\begin{figure}
  \centering
  \includegraphics[width=0.45\textwidth]{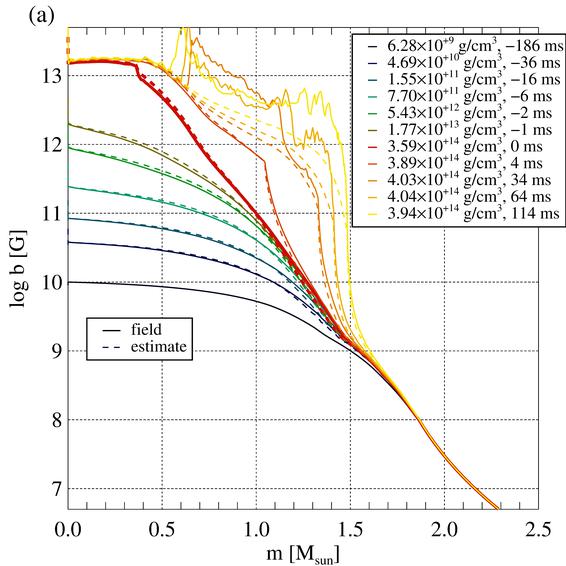}
  \includegraphics[width=0.45\textwidth]{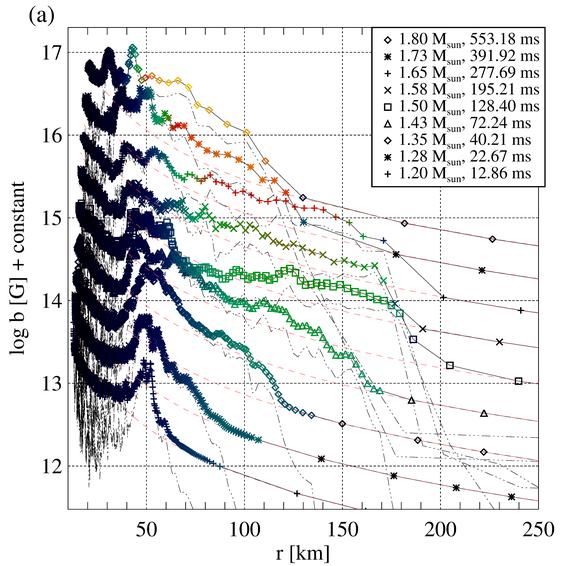}
  \caption{
    Panel \subpanel{(a)} compares the angularly averaged field strength
    of the core as a function of enclosed mass for various times
    during collapse and shortly after bounce.  The times and the
    corresponding central densities are listed in the legend.
    Panel \subpanel{(b)} panel
    shows the evolution of the magnetic field after bounce.  Each
    solid line is the trajectory of the radius corresponding to a
    chosen value of enclosed mass
    as it is accreted from outside the shock onto the PNS, i.e.~one
    can follow its evolution from the right to the left until it
    reaches its final radius at the end of the simulation.  To
    avoid/reduce confusion, the lines are scaled by a constant factor;
    without this scaling, they would lie very close to each other.
    Different symbols distinguish between different Lagrangian mass
    coordinates (from top to bottom as indicated in the legend,
    together with the corresponding time at which the mass value
    passes the shock wave).  The solid lines and symbols represent the
    total angularly averaged magnetic field strength, and the black
    dash-dot-dot-dotted lines are the average $\theta$-components.  The
    pink dashed lines show the field strength
    that would result from radial compression alone, and the colour of
    the symbols encodes the specific entropy of the mass shell in
    order to allow for an identification of, e.g.~the shock wave with
    pre-shock gas in black and post-shock showing up in colours.
  }
  \label{Fig:B10-SAS-tevos}
\end{figure}

\begin{figure}
  \centering
  \includegraphics[width=\linewidth]{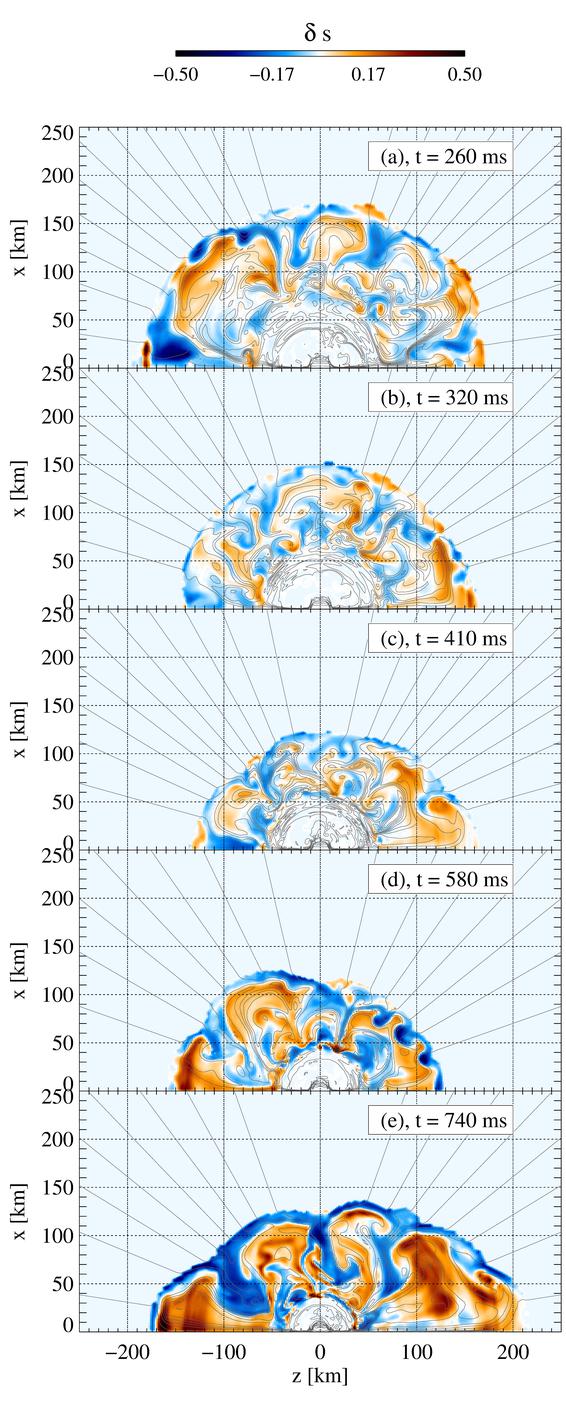} 
  \caption{
    Relative deviation of the entropy from its angular
    average, $\delta s$ (\eqref{Gl:delta_s}), and magnetic field lines of Model
    \modelname{B10} for five times after bounce.
  }
  \label{Fig:B10-SAS-ent-2}
\end{figure}

First, we will discuss the amplification of the field in the phases
before the explosion develops.  We can identify several effects
contributing to the amplification of the magnetic field.  We simplify
the discussion by assuming a magnetic field that is either completely
radial or completely non-radial; the former is a good approximation
outside the shock (cf.~\figref{Fig:B10-SAS-ent-2}).  The radial
dependence of the two types of fields follows from two different
constraints.  The divergence constraint fixes the form of a purely
radial field to $b^r (r) \propto r^{-2}$, which can hold only for $r
\ne 0$.  For a non-radial field, the magnetic flux frozen into the
surface of a fluid element, i.e.~the surface integral of $b^{\theta}$,
will be conserved during the collapse.  Based on this, we can obtain
two estimates for the field amplification by compression:
\begin{itemize}
\item From flux freezing, we find that the magnetic field of a fluid
  element at mass coordinate $m$ scales as $b (t) = b (t =0) \left(
    r (m,t) / r (m,0) \right)^{-2}$.
\item The density scales as $\propto r^{-3}$, which we can combine
  with the above estimate to a scaling $b (m) \propto \rho (m)
  ^{2/3}$.
\end{itemize}
Both scaling laws are fulfilled very well at the centre of the core,
but only approximately valid at intermediate mass coordinates ($m
\approx 1 \, \msol$).  A better description would require us to take
into account the dynamics of the collapse.  Though this task would be
simplified by the simple homologous velocity profile, we do not
perform it here, but use simple empirical findings for the scaling
field profile.  At all mass coordinates, an average of the two is a
good estimate for amplification by collapse for the specific field
profile developing in our models (dashed lines in Panel \subpanel{(a)}
of \figref{Fig:B10-SAS-tevos}).  By comparing the actual profile to
this estimate, we can assess how much extra amplification is provided
by other mechanisms.

When a fluid element falls through the shock wave, a non-zero
$\theta$-component of the field is formed (see
\figref{Fig:B10-SAS-ent-2}).  Across the shock, the field lines are
bent way from the shock normal, and the $\theta$-component of the
field is amplified strongly at the shock, although it remains on
average weak compared to the radial one, leading to an overall small
amplification factor for the total magnetic field.  At late times, the
pre-shock field, though still almost perfectly radial, locally
possesses a significant component parallel to the shock wave because
the shock deviates considerably from spherical symmetry (cf.~Panel
\subpanel{(e)} of \figref{Fig:B10-SAS-ent-2}), leading to a more
pronounced shock amplification.  Averaged over all latitudes, the jump
of the field strength across the shock wave is only moderate and
amounts to an amplification factor of the order of a few $10\,\%$.
Thus, the shock wave does not prominently show up in the profiles of
the magnetic field shown in the profiles of the field strength
displayed in Panel \subpanel{(a)} of \figref{Fig:B10-SAS-tevos}.

\begin{figure}
  \centering
  \includegraphics[width=0.5\textwidth]{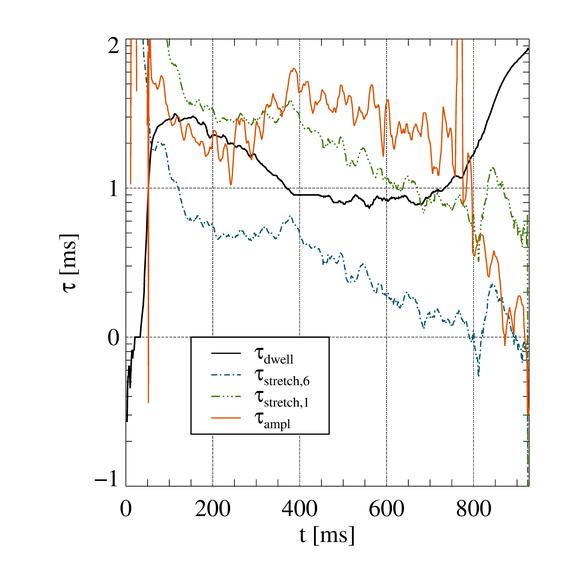}
  \caption{
    Comparison of the dwell time of matter in the gain layer (solid line)
    to two estimates for the time scale for turnover of
    eddies of the unstable flow (dashed-dotted, dash-dot-dot-dotted lines).
    For those, we estimated the typical velocity from the total
    kinetic energy and the mass in the gain layer and used two
    estimates for the length scale, $L_{1} = R_{\mathrm{gain}} \pi $
    and $L_{6} = R_{\mathrm{gain}} \pi/6 $, respectively.  The solid red
    line presents the net amplification time measured for fluid
    elements passing through the gain radius as a function of time.
  }
  \label{Fig:B10-SAS-tevos-2}
\end{figure}

The amplification by the post-shock flow is much more pronounced.  The
yellow lines in the \subpanel{top} panel of
\figref{Fig:B10-PNS-B-tevo} show the mean values of the flow velocity
and the \Alfven~speed in the \region{GAIN} layer.  Due to the
hydrodynamic instabilities, this region exhibits the highest flow
speeds of all post-bounce regions.  The magnetic energy, indicated by
the \Alfven~velocity, is about four orders of magnitude lower.  After
an initial rise in the \Alfven~speed before $t \sim 100 \, \ms$, both
lines are roughly parallel.  This suggests that the field
amplification by the flow is strong initially, while afterwards the
amplification factor does not vary strongly.

Soon after bounce, we find field strengths clearly in excess of our
simple estimate for compressive amplification (see the orange and
yellow lines in Panel \subpanel{(a)} of \figref{Fig:B10-SAS-tevos}).
Since the hot-bubble region appears very compressed in profiles
presented as a function of mass coordinate, we continue the discussion
with Panel \subpanel{(b)} of \figref{Fig:B10-SAS-tevos}: The
enhancement of the field strength of a fluid element w.r.t.~the
compression estimate (dashed lines) remains high for all mass shells
until the onset of the explosion.  We cannot state the definite reason
for the difference between the decay of the field in the PNS and the
sustained strong field in the hot-bubble region, but we tentatively
argue for the importance of the advection of magnetic flux through the
shock wave.

Above, we noted that very little injection of magnetic flux into the
\region{PCNV} region of the PNS occurs, and what little there is,
happens in a geometry that limits the efficiency of field
amplification quite severely.  Across the shock, the situation is
completely different.  The radial infall provides a continuous inflow
of magnetic flux into the unstable region.  Furthermore, a slow but
steady accretion flow superposes the vortical motion behind the
shock wave, in contrast to the essentially vanishing radial velocity
in the \region{PCNV} region.  The presence of the radial velocity has a
number of important consequences for the development of hydrodynamic
instabilities as well as for the magnetic field:
\begin{itemize}
\item The development of convective structures has to compete with the
  accretion of matter through the unstable region.  As pointed out by
  \cite{Foglizzo_Scheck_Janka__2006__apj__Nu-driven_Convection_versus_Advection},
  convective cells will only grow from seed perturbations if their
  growth time is smaller than the advection time by a factor of $\sim 3$.
\item Radial accretion forms a crucial part of the advective-accoustic
  cycle responsible for the development of the SASI
  \citep{Foglizzo_et_al__2007__apj__Instability_of_a_Stalled_Accretion_Shock_AAC,Guilet_Foglizzo__2012__mnras__On_the_linear_growth_mechanism_driving_the_SASI}.
  The superposition of two instabilities, convection and the SASI,
  leads to more complex flows than in the \region{PCNV} layer with far
  less stationary, unstable cells.  This is also reflected in the shape
  of the outer boundary: spherical for region \region{PCNV}, deformed
  for the \region{GAIN} layer.
\item We already noted the large-scale advection of radial magnetic
  flux into the unstable region across the shock wave.  Even if all
  other conditions were the same as in the PNS, this might eventually
  lead to a steady state characterised by a balance between convective
  flux expulsion and advective flux injection.
\item The geometry of the field is rather different from that in the
  PNS.  In the latter, fluid elements spend a very long time trapped
  in the vortical motion, and the field frozen into the flow is wound
  up into increasingly thin structures, which finally are disrupted by
  (numerical) resistivity.  In the gain layer, on the other hand, a
  fluid element spends only a limited time, and the attached magnetic
  field lines are not stretched in as many turnovers as in the
  \region{PCNV} layer.  This has two opposing effects:  on the one
  hand, it limits the maximum amplification, but, on the other hand,
  it prevents the creation of structures near the grid resolution
  limit susceptible to resistive disruption.
\end{itemize}

\begin{figure}
  \centering
  \includegraphics[width=0.42\textwidth]{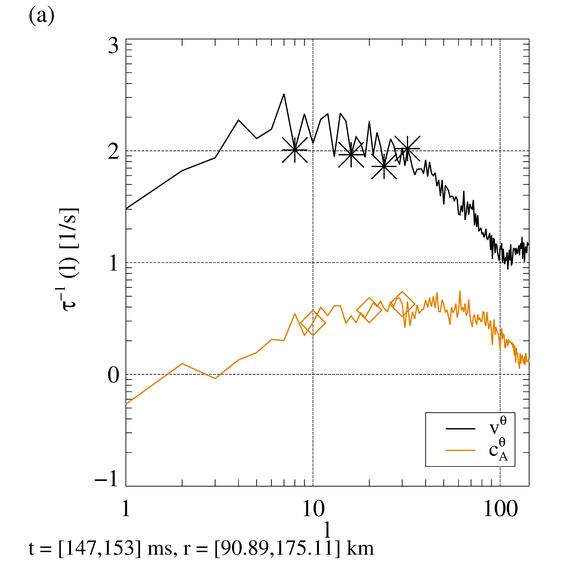}
  \includegraphics[width=0.42\textwidth]{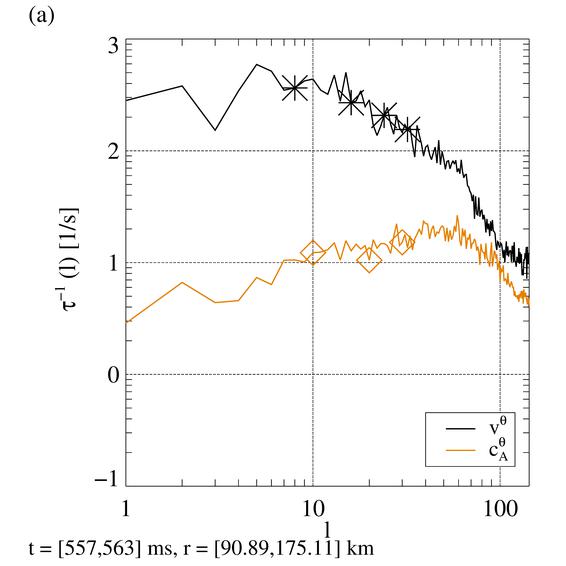}
  \includegraphics[width=0.42\textwidth]{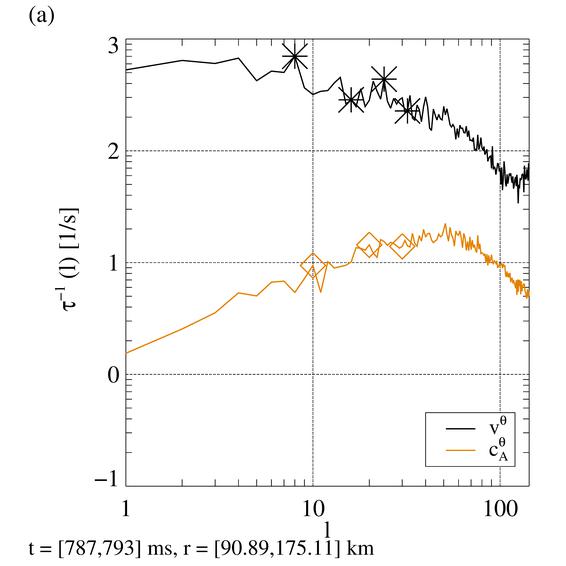}
  \caption{
    The inverse of the turnover time and the \Alfven~travel time as
    a function of mode number for time for $t = 150, 560, 790 \, \ms$.   The spectra are computed as averages over times
    $[t -3, t+3] \, \ms$
     and over radii between the lower boundary of the unstable region
     and the minimum shock radius.
   }
  \label{Fig:B10-GAI-ellspec}
\end{figure}

All of these factors contribute to the specific evolution of the field
in the hot-bubble region.  Conditions for the development of the
convection in the post-bounce accretion flow and typical properties of
the convective flows were given by
\cite{Foglizzo_Scheck_Janka__2006__apj__Nu-driven_Convection_versus_Advection}
For marginally unstable flows $H/r_{\mathrm{sh}} \sim 1/2$, which is
approximately fulfilled in our model ($H$ is the entropy scale height),
they estimate that modes of degree $l \approx 6$ should grow.  In our
case, this translates into convective structures extending over a
length scale roughly around $L_{\mathrm{conv}} \sim 40 \, \km$, in
decent agreement with our simulation (see Panels \subpanel{(a), (b),
  (c)} of \figref{Fig:B10-SAS-ent-2}).

In the unstable region, we can determine estimates for the source term
in the evolution equation for the magnetic energy density,
\begin{equation}
  \label{Gl:emag-evo}
  \partial_{t} \frac{\vec b^2}{2} + \nabla_i \frac{\vec b^2}{2} v^i
  = 
  - \frac{\vec b^2}{2} \nabla_i v^i + b^i b^j \nabla_j v^i.
\end{equation}
The r.h.s.~of this equation describes the growth of the magnetic energy
density by work the flow does against the Lorentz force.  It consists
of a compression term (work against the magnetic pressure) and an
anisotropic stretching term (work against the magnetic tension).  We
are mostly interested in the latter, because this is the term behind
the excess of the field strength w.r.t.~the compression profiles
(dashed lines in \figref{Fig:B10-SAS-tevos}, Panel \subpanel{(b)}).  

To get an order-of-magnitude estimate, we can approximate the growth
of the magnetic energy locally as $\inv{2} \vec b^2 V/L$, where $V$
and $L$ are typical velocities and length scales of the unstable flow.
Thus, the field grows on a time scale $\tau_{\mathrm{stretch}} \sim
L/V$.  The growth of the field in a fluid element has to compete with
the advection of the fluid element through the gain layer, since
amplification can only occur as long as the fluid element is inside
the unstable region.  Thus, we can get an estimate of the amount of
field amplification by comparing $\tau_{\mathrm{stretch}}$ to the
dwell time of mass in the gain layer, $\tau_{\mathrm{dwell}}$.  We can
relate the dwell time to the mean accretion velocity fairly
straightforwardly, $ \tau_\mathrm{dwell} = \int_{\mathrm{gain~layer}}
\mathrm{d} r / \bar{v}_{r}$, but the estimate of the stretch time is
hampered by the uncertainty of $L$ and $V$.  If we could get estimates
for them, we could translate them into a simple formula for the
amplification factor of the field for a given fluid element, i.e.~the
ratio between the field strength after amplification by the turbulent
flow and the field strength in the absence of turbulence (but
including compression), $\vec b_{\mathrm{nt}}$:
\begin{equation}
  \label{Gl:amplif-stretch-simple}
  f = | \vec b | / | \vec b_{\mathrm{nt}} |
  \sim 
  \int_{\mathrm{gain~layer}} \frac{\mathrm{d} r}{- \bar{v}_r}
  \frac{V}{L}.
\end{equation}

To use typical values of our model, we estimate $V$ from the mean
velocity of convective eddies from the kinetic energy of the
\region{GAIN} region and from the mean $\theta$-velocity in this
region.  The length scale, $L$, is even less well constrained.  For
this reason, we compare the dwell time to two estimates for the
amplification time, one using $L = R_{\mathrm{gain}} \pi/6$,
i.e.~relatively small modes typical for convection, and $L =
R_{\mathrm{gain}} \pi$, i.e.~large-scale modes which might be more
characteristic for the SASI.  The two curves for
$\tau_{\mathrm{stretch}}$ (\figref{Fig:B10-SAS-tevos-2} resulting from
the two approximations) should limit the true value of the stretching
time scale from below and above.  Since the flows in the model show a
small-scale pattern first (see Panels \subpanel{(a,b,c)} of
\figref{Fig:B10-SAS-ent-2}) and large-scale flows later (Panel
\subpanel{(d)}), the dwell time (solid line) should be compared to the
lower, dash-dot-dot-dotted line for early $t$, and to the upper,
dash-dotted line at late $t$.  We find that the amplification time is
always of the same order as, and usually less than the dwell time,
which would translate into a moderate amplification of the field in
the \region{GAIN} layer, as we observe in the simulations.  To test
the quality of our simple considerations, we show the actual
amplification time scale, $\tau_{\mathrm{ampl}}$, in the same figure
(red line of \figref{Fig:B10-SAS-tevos-2}).  For each angularly
averaged mass shell in the gain layer, we divide the magnetic energy
by its time derivative and average the resulting amplification time
over the time the fluid element spends in the gain layer.

Though the agreement is not overwhelming, we find that the time-scale
criterion reproduces the order of magnitude and the trends reasonably
well.  The amplification time is longer than the estimates
$\tau_{\mathrm{stretch}}$ (dashed lines), because the actual
efficiency of amplification is reduced by the advection through the
gain layer.  During the entire post-bounce evolution, we find an
amplification factor varying around a value $f \sim 5...10$.  At
first, when the advection is slow and high-order convective modes
dominate, the real amplification time is short.  Later, when the
advection is comparably fast and the unstable flow is dominated mostly
by low-order modes, the amplification time grows before starting a
slow secular evolution to shorter time scales again as the stretching
time decreases.

The time evolution of the spectra corroborates the assumption of a
shift from modes of intermediate $l$ to the dipole mode which we used
in the discussion of the field amplification time scale above.  We can
check this more quantitatively by converting the spectral energy
density of the velocity field into an estimate for the typical
turnover time scale for modes of given $l$, $\tau_l(t,r,l) \sim
r/(l+1) / \sqrt{E(t,r,l) / \bar{\rho} (t,r)}$ ($\bar{\rho}$ is the
angularly averaged mass density), which is in decent agreement with
the above estimates for $\tau_{\mathrm{stretch}}$.  The inverse of
this quantity is shown in \figref{Fig:B10-GAI-ellspec}.  At $t = 150
\, \ms$, the overturn time for dipole and quadrupole modes is greater
than the minimum value of $\tau_l$ in the \region{GAIN} layer, which
we find at modes around $l \gtrsim 8$.  Later, $\tau_l$ decreases most
strongly for the lowest-order modes, leading to overturning times for
the dipole mode that are similar to those at intermediate $l$.

Hence, stretching of field lines should become more efficient at
generating large-scale fields.  Very strong power in the dipole mode
can be encountered mostly in the lower radial range of the unstable
region.  At higher radii, the spectrum of the magnetic field for the
largest modes is below the Kolmogorov scaling.

We summarise the most important points: 
\begin{itemize}
\item The largest contribution to field amplification comes from the
  radial infall.  We can characterise the resulting field, dominated
  by the radial component, by simple, smooth functions of radius (see
  dashed lines in both panels of \figref{Fig:B10-SAS-tevos}).
\item After the development of strong non-spherical deformations of
  the shock wave, the (mostly radial) pre-shock field is amplified at
  the shock wave, contributing on average a few $10 \%$ to the total
  amplification.
\item On top of this, the post-shock instabilities amplify the field
  and generate a strong $\theta$-component.  The factor of
  amplification of the field is the result of a competition between
  vortical flows and the advection of magnetic flux out of the
  unstable region.  Thus, we can quantify the efficiency of this
  process roughly by the ratio of the time scales of the turnover
  motion and the accretion.  In our case, this competition leads to an
  amplification factor around 5...10.
\item The magnetic energy remains much smaller than the kinetic energy
  in the gain layer since the termination of the amplification is set
  by the aforementioned competition of time scales and not by
  back-reaction of the field onto the flow.
\item The magnetic field, amplified during its passage through the
  gain layer, accumulates in the stable layer outside the PNS and
  forms sheets of mostly lateral field.  Here, the magnetic field has
  an energy comparable to the kinetic energy of the matter, which is
  almost at rest; however, it is still weak compared to the internal
  energy and, thus, has little effect on the structure of the layer.
\item The spectrum of the velocity and the magnetic field in the
  unstable region follow a Kolmogorov scaling for a limited range of
  mode numbers between $l \gtrsim 10$ and $l \lesssim 30$ but are
  separated by three orders of magnitude.  At no mode number $l$
  (except in the dissipation range) do we find energetic equipartition
  between the flow and magnetic field.
\end{itemize}

\subsubsection{Dependence on resolution}

\begin{figure}
  \centering
  \includegraphics[width=\linewidth]{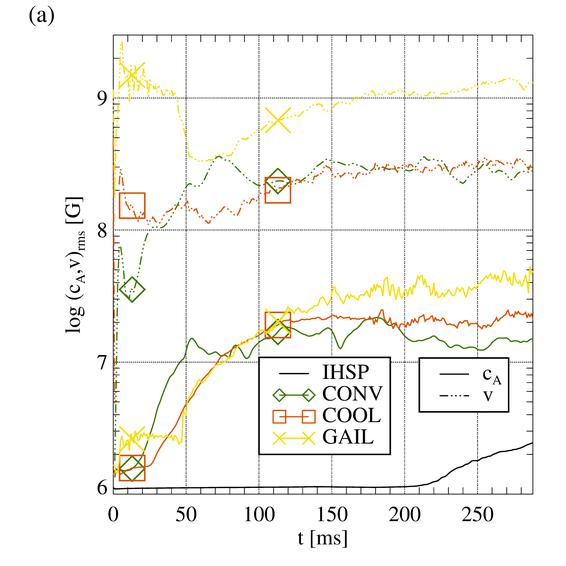}
  \caption{
    Same as Panel \subpanel{(a)} of \figref{Fig:B10-PNS-B-tevo}, but
    for the high-resolution resimulation of model \modelname{B10}.
  }
  \label{Fig:B10HR}
\end{figure}

In order to assess the sensitivity of our results to numerical
resistivity, which depends crucially on grid resolution, we
resimulated model \modelname{B10} on a grid of $n_r \times n_\theta=
720 \times 288$ zones.  Like in the simulation described above, the
uniform grid width in the centre was $\delta r = 400 \, \mathrm{m}$,
but outside of this region the grid is finer by a factor of 2 in
radius and in angle than in model \modelname{B10}.  The increased
computational costs of the high-resolution simulation restricted us to
run only up to a post-bounce time of $t_{\mathrm{fin}} = 280 \, \ms$.
The change in resolution leaves the dynamics mostly unchanged.  We
show the time evolution of the average values of the flow and
\Alfven~speeds in \figref{Fig:B10HR}.  Compared to model
\modelname{B10} (\figref{Fig:B10-PNS-B-tevo}, Panel \subpanel{(a)}),
the modifications are only minor.  The most prominent difference
concerns the \Alfven~velocity in the \region{GAIN} region: comparing
the two simulations at $t \approx 250 \, \ms$, the volume average of
the high-resolution version is larger by a factor of about $1.4$.
This result can indicate either a higher saturation level of the
amplification or a more rapid growth of the magnetic energy (note that
it grows during the entire post-bounce phase of the simulation for our
standard resolution).

\subsubsection{Other weak-field models}

Neither a decrease of the initial field strength by two orders of
magnitude ($b_0 = \zehn{8} \, \mathrm{G}$) nor an increase by an order
of magnitude to $b_0 = \zehn{11} \, \mathrm{G}$ leads to significant
changes of the evolution.  The mechanism of field amplification as
well as the dynamic feedback is almost the same, although, of course,
the details of the stochastic flows show differences.  Consequently,
the onset of the explosion and the time at which the shock starts to
expand are very similar.  The factors by which the magnetic field is
amplified are essentially the same as discussed above.  The only
exception to the resulting scaling with $b_0$ is due to an artifact of
our axisymmetric models, viz.~the strong radial field along the polar
axis.

\subsection{Strong initial magnetic field}
\label{sSek:B12}

A very strong initial field of $b_0 = 10^{12} \, \mathrm{G}$ leads to
a very different evolution of Model \modelname{B12}.  Most
prominently, shock runaway occurs about 400 ms earlier than in
Model \modelname{B10}.  We will contrast the evolution of the two
models in the following.

\subsubsection{The proto-neutron star}

\begin{figure}
  \centering
  \includegraphics[width=0.48\textwidth]{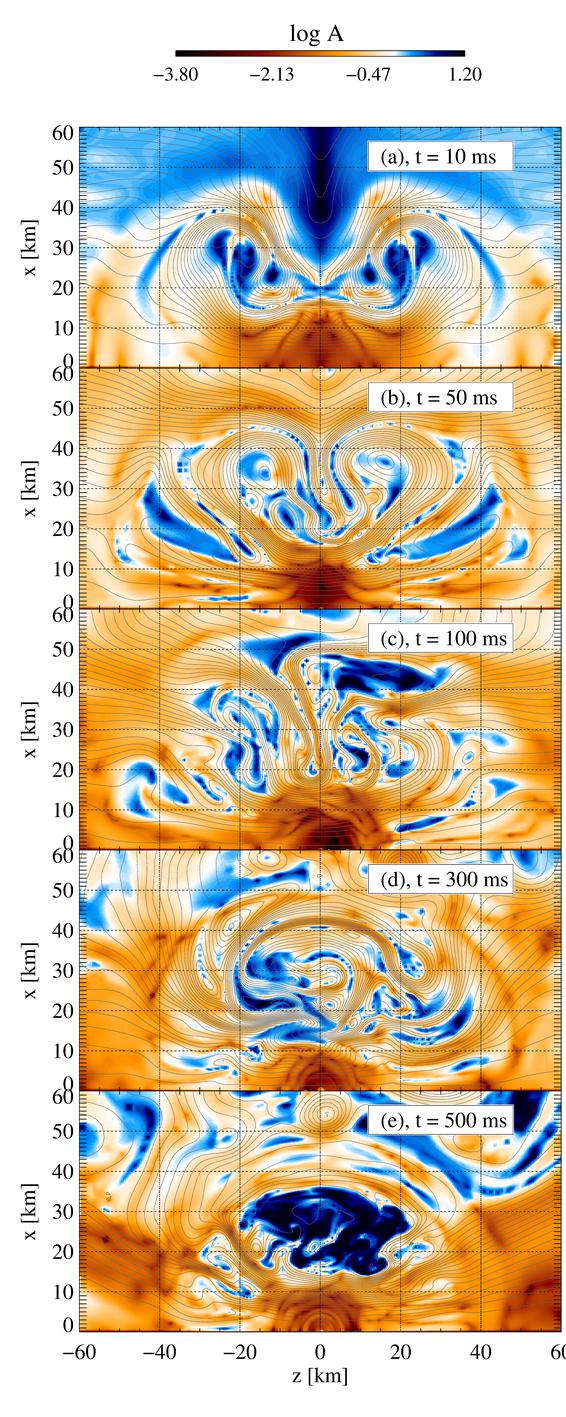}
  \caption{
    Logarithm of the \Alfven~number, $\mathbf{A} = |\vec v| /
    c_{\mathrm{A}}$, and magnetic field lines of
    the flow in the innermost regions of Model \modelname{B12} for
    five times after bounce.  Sub- and super-\Alfvenic~regions are
    shown in red and blue colours, respectively.
  }
  \label{Fig:B12-PNS-Ye-Bfield-1}
\end{figure}

\begin{figure*}
  \centering
  \includegraphics[width=\textwidth]{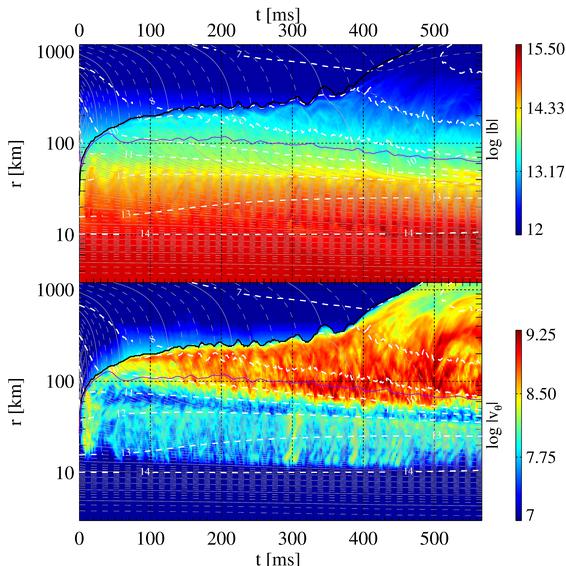}
  \caption{
    Same as \figref{Fig:B10-PNS-mshell}, but for Model \modelname{B12}.
  }
  \label{Fig:B12-mshells-1}
\end{figure*}

A large quadrupolar vortex develops outside the stable core at radii
between 15 and 50 km immediately after bounce.  We show this pattern
in Panels \subpanel{(a)} and \subpanel{(b)} of
\figref{Fig:B12-PNS-Ye-Bfield-1}, displaying the \Alfven~number of the
flow and the magnetic field lines at $t = 10 \, \ms$ and $50 \, \ms$,
respectively.  While this transient still lasts, a sustained
convective flow develops.  During the entire evolution, the outer
boundary of the convection zone contracts at a rate very similar to
Model \modelname{B10} (see \figref{Fig:B12-mshells-1}).  Because the
magnetic energy is much less than the thermal energy, the thermal
structure of the core (density, entropy, electron fraction) is hardly
modified by the presence of the field.

\begin{figure}
  \centering
  \includegraphics[width=0.4\textwidth]{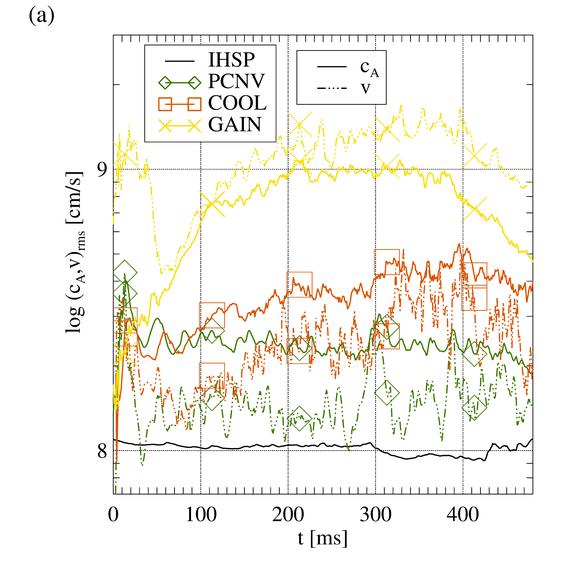}
  \includegraphics[width=0.4\textwidth]{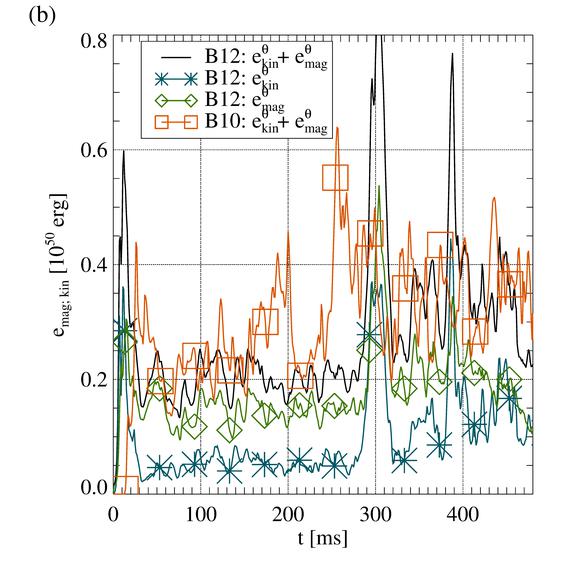}
  \includegraphics[width=0.4\textwidth]{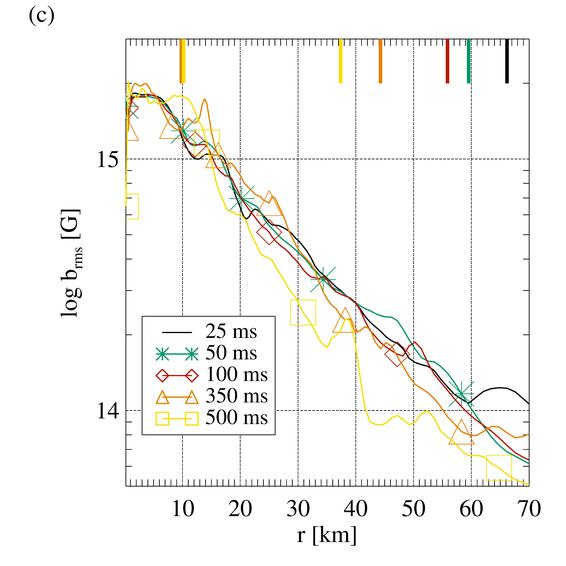}
  \caption{
    Panel \subpanel{(a)}: angular averages of \Alfven~and flow speed in the
    four post-shock regions of Model \modelname{B12}.
    Panel \subpanel{(b)}: 
    $\theta$-components of
    the kinetic and
    magnetic energies of Model \modelname{B12} in the PNS
    convection zone in comparison to the sum of the kinetic and
    magnetic energy in the PNS convection zone of Model
    \modelname{B10}.
    Panel \subpanel{(c)}: radial profile of the average field
    strength in the innermost 70 km of Model \modelname{B12} for five
    times after bounce.
  }
  \label{Fig:B12-PNS-emag}
\end{figure}

The velocity and the magnetic field, on the other hand, differ
strongly from those in Model \modelname{B10}.  We show the evolution
of the angular averages of the \Alfven~and the flow speed in the
post-shock regions of the core in Panel \subpanel{(a)} of
\figref{Fig:B12-PNS-emag}.  Except for the innermost stable core,
where the velocity is essentially zero, the kinetic and magnetic
energies are near equipartition.  In the \region{PCNV} region (green
lines), the flow of the strong-field model is slower than the
\Alfven~speed by a factor of around 2.  Furthermore, it is also slower
than the convective flow in the PNS of Model \modelname{B10}, which
can also be seen in the lower halves of \figref{Fig:B10-PNS-mshell}
and \figref{Fig:B12-mshells-1}.

In Model \modelname{B12}, where the magnetic field cannot be
neglected, we expect to find a reduction of the kinetic energy
w.r.t.~Model \modelname{B10}, for which very little of the kinetic
energy of this model is used to amplify the seed magnetic field,
allowing convection to run at the maximum kinetic energy that can be
attained given the thermal structure of the core.  The reduction is
caused by the conversion of kinetic to magnetic energy and due to the
suppression of turnover motions by the tension of a magnetic field
close to equipartition.  The sum of the kinetic and magnetic energies
generated by the instability, on the other hand, should be roughly the
same.  Panel \subpanel{(b)} of \figref{Fig:B12-PNS-emag} compares the
time evolution of the energies of the $\theta$-components of the
velocity and the magnetic field in the \region{PCNV} region of Model
\modelname{B12} to Model \modelname{B10}.  The total convective
energies (black line and red line with squares) are approximately
equal, though with a wide range of fluctuations, but for the
strong-field model, the convective region is magnetically rather than
kinetically dominated.  We note that we get somewhat different values
if we include the radial components.  Even without additional
convective amplification, the radial field is very strong, in
particular along the $z$-axis, and thus the total magnetic field in
the \region{PCNV} region is considerably stronger than the total
convective energy of Model \modelname{B10}.

\begin{figure}
  \centering
  \includegraphics[width=0.48\textwidth]{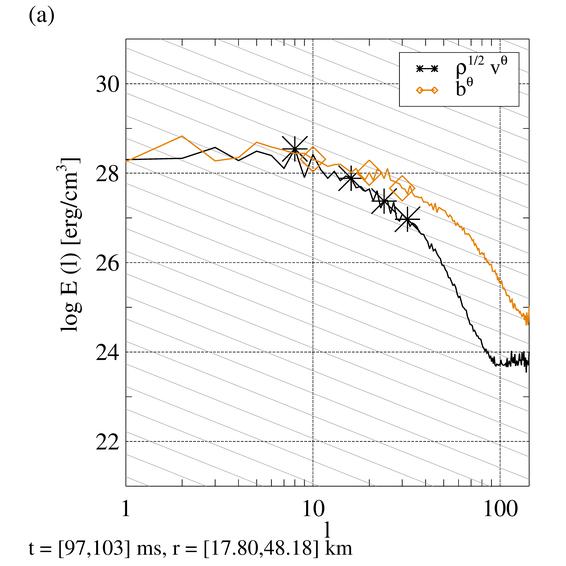}
  \caption{
    Energy spectra of the angular components of the velocity (black lines with stars)
    and magnetic field (orange lines with diamonds) in the PNS of
    Model \modelname{B12} at
    $t = 100 \, \ms$.  
    The energy spectra, $E(l)$, are
    shown in comparison to the Kolmogorov scaling, $l^{-5/3}$ (grey lines).
  }
  \label{Fig:B12-PNS-spectra}
\end{figure}

The ratio between kinetic and magnetic energies is reflected in the
large sub-\Alfvenic~regions in the \region{PCNV} region at all times
(red colours in \figref{Fig:B12-PNS-Ye-Bfield-1}).  In particular,
the magnetic field dominates over the flow in a $\approx 10 \, \km$
wide column parallel to the symmetry axis.  Magnetic tension is
sufficiently strong to suppress turnover motions in this region to
a high degree.  Closer to the equator, the magnetic field is weaker,
and a certain amount of convective activity is possible.  At $t = 50
\, \ms$, we still find the imprints of the early quadrupolar vortex,
but later, it is replaced by smaller overturns.  

Later ($t = 100 \, \ms$), the magnetic field is concentrated into
several narrow radial flows, between which the field is rather weak
and the flow is super-\Alfvenic.  Angular energy spectra of the
velocity and the magnetic field show that the lateral magnetic field
is close to equipartition throughout the entire spectrum and exceeds
the lateral kinetic energy near the dissipation range.  Both spectra
are flat up to $l \sim 8$. Up to $l \lesssim 30$, the velocity spectra
show a power-law scaling close to Kolmogorov, and the magnetic field
is slightly less steep.

During the subsequent evolution, larger weakly magnetised bubbles
develop transiently, e.g.~around $t = 300 \, \ms$ (Panel
\subpanel{(d)} of \figref{Fig:B12-PNS-Ye-Bfield-1}).  At later times,
they tend to last for longer times than in the early post-bounce
phase.  Since the strong radial field suppresses them at the pole,
they appear in the equatorial region only.  The bubbles are filled by
gas of an electron fraction exceeding the angular average and
surrounded by sheets of strong magnetic fields (\subpanel{bottom}
panel of \figref{Fig:B12-PNS-Ye-Bfield-1}).  In the angular averages
of the lateral velocity (lower half of \figref{Fig:B12-mshells-1}),
the bubbles appear as phases during which $\langle v_{\theta} \rangle$
is enhanced to the same level as in Model \modelname{B10} (the yellow
features starting to become visible for $t \gtrsim 300 \, \ms$).  The
most intense of these features show up in the evolution of the
magnetic field strength since they lead to a temporary growth of the
magnetic energy and a redistribution of the field towards larger
radii.

Because of the weak turnover motions, radial profiles of the magnetic
field strength (Panel \subpanel{(c)} of \figref{Fig:B12-PNS-emag})
show relatively little variation during the post-bounce evolution.  In
contrast to Model \modelname{B10}, only little field amplification
beyond that provided by radial compression takes place, as can be seen
in the comparison of the angular averages of the field strength as a
function of enclosed mass in Panel \subpanel{(b)} of
\figref{Fig:B12-coll-bampl}.  As in Model \modelname{B10}, overturning
eddies can expel the magnetic flux from the convective cells at later
times, as can be seen in the last time step shown in
\figref{Fig:B12-PNS-Ye-Bfield-1}, Panel \subpanel{(e)}.  However,
because of the smaller convective velocities and because these eddies
are only present intermittently, the expulsion is less pronounced: the
last time step displayed in Panel \subpanel{(c)} of
\figref{Fig:B12-PNS-emag} (yellow line) is only moderately below the
profile generated by pure field compression in the infall.

With less flux expulsion, the increased field strength in the stable
layers below and above the convection zone is less pronounced than in
Model \modelname{B10}--though not entirely absent (in Panel
\subpanel{(c)} of \figref{Fig:B12-PNS-emag}, see the maximum at $r
\approx 38 \, \km$ at $t = 500 \, \ms$).  The angular average of the
magnetic field strength (top half of \figref{Fig:B12-mshells-1}) does
not show the enhancement of the field strength along the upper
boundary of the \region{PCNV} region that is present in
\figref{Fig:B10-PNS-mshell}.  Radial fields connect the inner core to
the hot-bubble region in a wide cone surrounding the polar axis.
Closer to the equator, the field has a stronger $\theta$-component and
forms a sheet around the PNS (Panels \subpanel{(d,e)} of
\figref{Fig:B12-PNS-Ye-Bfield-1}).  In contrast to the weak-field
model (Panel \subpanel{(d)} of \figref{Fig:B10-PNS-Ye-Bfield-1}), for
which this layer is composed of a stack of thin sheets of opposite
polarity, it contains only a lateral field of positive sign.

\subsubsection{The post-shock layer}

\begin{figure}
  \centering
  \includegraphics[width=0.48\textwidth]{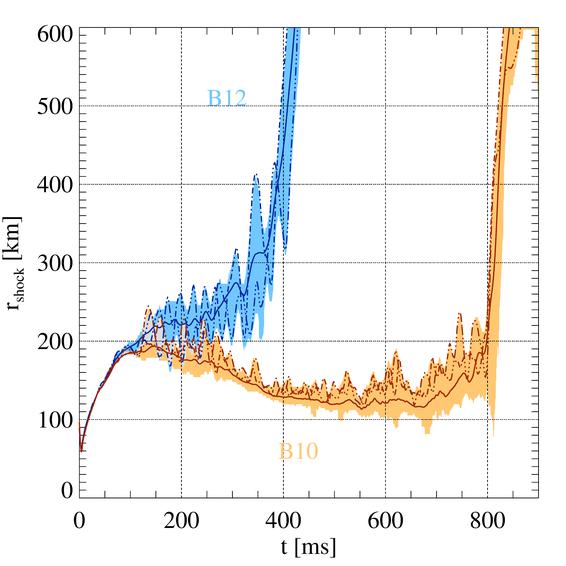}
  \caption{
    Comparison of the shock radii of Models
    \modelname{B12} and \modelname{B10}.  Colours have the
    same meaning as in 
    \figref{Fig:B10-shockmodes}.
  }
  \label{Fig:B12-shockvgl}
\end{figure}

\begin{figure}
  \centering
  \includegraphics[width=0.48\textwidth]{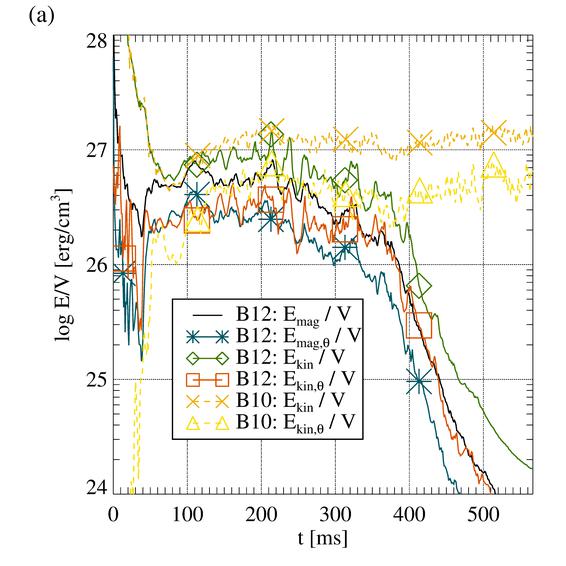}
  \includegraphics[width=0.48\textwidth]{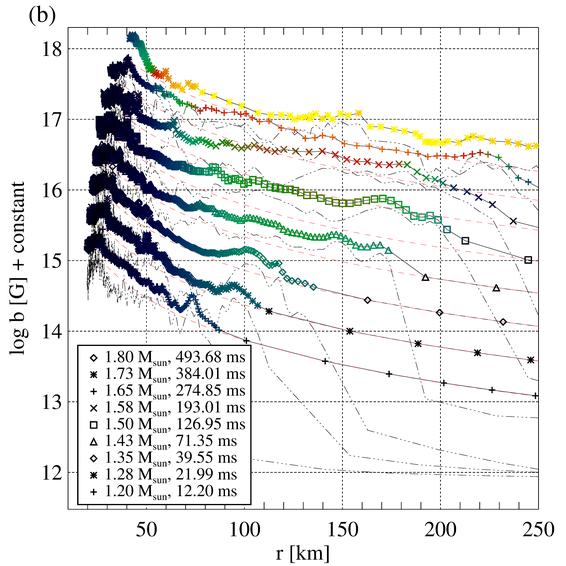}
  \caption{
    Panel \subpanel{(a)}:  time evolution of the total kinetic
    and magnetic energy densities in the gain layer of Model
    \modelname{B12} and corresponding $\theta$-components 
    compared them to the total and lateral kinetic energy in the gain layer of
    Model \modelname{B10}.
    Panel \subpanel{(b)}: same as Panel \subpanel{(b)} of \figref{Fig:B10-SAS-tevos}.
  }
  \label{Fig:B12-coll-bampl}
\end{figure}

\begin{figure*}
  \centering
  \includegraphics[width=0.48\textwidth]{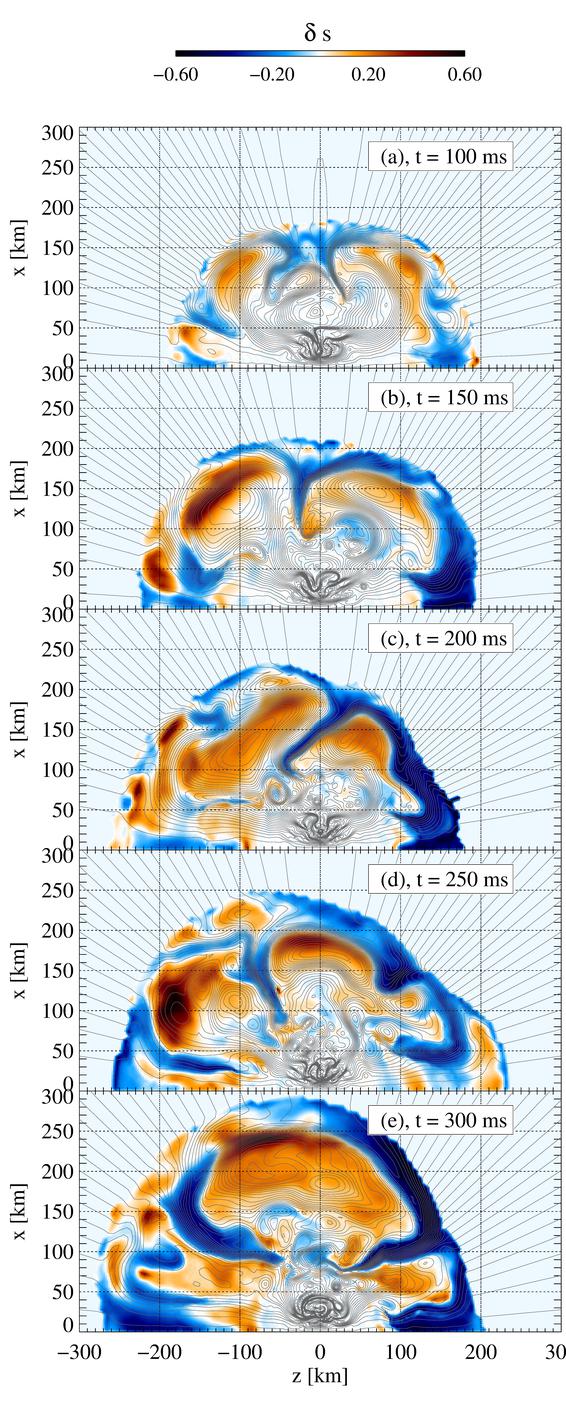}
  \hspace{0.03\textwidth}
  \includegraphics[width=0.48\textwidth]{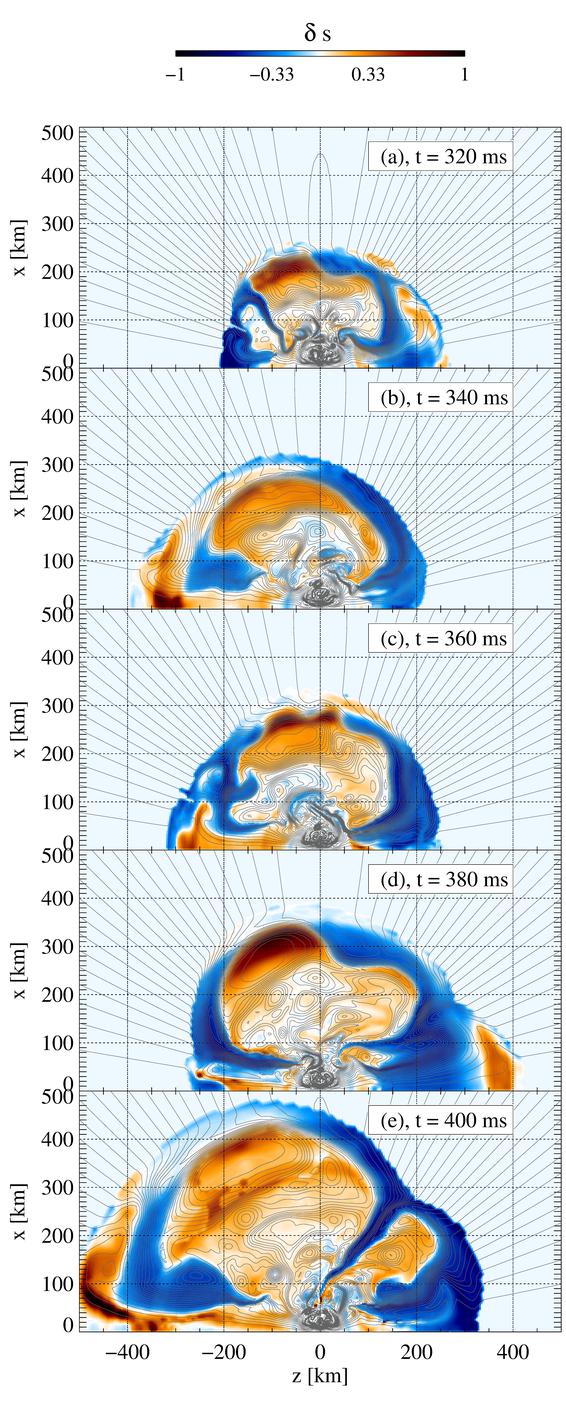}
  \caption{
    Snapshots of the post-shock region of Model \modelname{B12} at
    ten times between $t = 100 \, \ms$ and $t = 400 \, \ms$, displaying the
    entropy contrast, i.e.~the relative deviation of the entropy from its
    angular average (colour scale), and field lines (black solid lines).
  }
  \label{Fig:B12-sas-ent-Bfield-1}
\end{figure*}

The shock wave of Model \modelname{B12} begins to deviate from the
evolution of Model \modelname{B10} already earlier than $t \sim 100 \,
\ms$, as shown in \figref{Fig:B12-shockvgl}.  Instead of slowly
contracting, the average radius passing an intermediate plateau of
$r_{\mathrm{sh}} \approx 230 \, \km$ before it starts to expand,
reaching $310 \, \km$ at $t \approx 340 \, \ms$.  Afterwards, the
expansion quickly gains speed, and the average shock radius exceeds
$600 \, \km$ after $t \approx 420 \, \km$.  After the brief period of
shock stagnation at $t \sim 100 \, \ms$, the shock surface becomes
highly aspherical, highlighted by the wide blue band displaying the
range of shock radii in the model.

The non-radial components of the magnetic field and the velocity have
almost the same energy until the shock starts to expand, whereas the
mean accretion flow leads to an excess of the radial kinetic energy
over the radial magnetic energy, albeit only by a small factor (see
Panel \subpanel{(a)} of \figref{Fig:B12-coll-bampl}).  Compared to
Model \modelname{B10}, the lateral energy densities in the gain layer
are greater until $t \sim 120 \, \ms$ and on the same level
afterwards, and the total (kinetic plus magnetic) energy densities are
similar.  We note, however, that the post-shock volume of Model
\modelname{B12} is greater than that of Model \modelname{B10} after $t
\sim 150 \, \ms$.  Accounting for this effect, the energies of flow
and field in Model \modelname{B12} are considerably greater.

These data demonstrate that the magnetic field in the gain layer of
Model \modelname{B12} reaches equipartition with the kinetic energy
globally (and locally can exceed this level by far).  Furthermore, the
fact that both energies combined are on the same level as in the
weakly magnetised case indicates that the basic mechanisms behind the
generation of turbulent velocity and magnetic fields are the same for
both models, and that the strong field does not lead to, e.g.~an
alternative instability tapping into additional reservoirs of energy.
Similarly, values of the total stress tensor components are similar in
both models, but the distribution between (hydrodynamic) Reynolds
stress and (magnetic) Maxwell stress is different.  Compared to the
internal energy, the magnetic energy is globally weak (a few $\%$),
though in the most intense flux sheets the magnetic energy density is
closer to equipartition with the internal energy density.

Since the seed magnetic field is already close to equipartition with
the flow, the post-shock flow contributes only little to further field
amplification.  Following the evolution of an angularly averaged mass
element as it falls through the shock onto the PNS (Panel
\subpanel{(b)} of \figref{Fig:B12-coll-bampl}), we find only a modest
growth above the estimate for a purely radial infall and a less
efficient creation of a $\theta$-component by the shock and the
post-shock flow than in Model \modelname{B10}.

For a detailed discussion of the evolution of the model, we refer to
\figref{Fig:B12-mshells-1} (time evolution of angularly averaged
profiles of magnetic field strength, lateral velocity, and isodensity
contours and radii of enclosed mass of the angularly averaged model)
and to \figref{Fig:B12-sas-ent-Bfield-1} presenting a time series of maps of the
entropy contrast, i.e.~the relative
deviation of the entropy from the angular mean, 
\begin{equation}
  \delta s (r,\theta)=(
  s (r,\theta)- \langle s \rangle (r) ) / \langle s \rangle (r),
  \label{Gl:delta_s}
\end{equation}
where the angular average is computed only from post-shock zones.

Similarly to the PNS, the \region{GAIN} region right after bounce is
dominated by a large quadrupolar vortex resulting from the small
imbalance between the pressure profiles at the poles and the equator
generated by the magnetic tension during collapse.  These flows decay
later, and convection and the SASI establish a new pattern of
non-spherical flows after $t \approx 100 \, \ms$.  While caused by a
different mechanism, these flows show a similar geometry,
viz.~predominantly large-scale bubbles characterised by an entropy
exceeding the angular average.  In a visualisation of the entropy
contrast, $\delta s$, (\figref{Fig:B12-sas-ent-Bfield-1}) these
regions show up in red.  The gas inside these bubbles has a small
mean radial velocity and, thus, remains trapped for a long time,
whereas accretion proceeds mainly through a few narrow downflows at
the poles and close to the equator.  Sheets of a strong magnetic field
with a magnetic energy density exceeding the kinetic energy density of
the flow surround the hot bubbles.  Directing the matter that falls
through the shock wave towards the downflows, these flux sheets shield
the bubbles against the infall and suppress their disruption by
small-scale turbulence.  Consequently, they are responsible for the
persistence of the bubbles over long times and the slow, regular shock
oscillation.

The accretion flows can be the site of additional field amplification
by a mechanism not discussed so far.  Along the polar axis, an
\Alfven~point forms separating super-\Alfven ic ($|\vec v| >
c_{\mathrm{A}}$) accretion at large radii from sub-\Alfven ic ($|\vec
v| < c_{\mathrm{A}}$) accretion at smaller radii.  Thus, Alfven waves
generated closer to the PNS can travel against the accretion flow up
to this point, where they accumulate.
\cite{Guilet_et_al__2011__apj__Dynamics_of_an_Alfven_Surface_in_Core_Collapse_Supernovae}
have suggested that an instability will amplify the waves there, and
their dissipation contributes to the transport of thermal energy into
the gain layer.  We note that our models may only show this effect for
an already strong initial field because only then the \Alfven~point is
located sufficiently far outside the PNS.  We find indications for the
accumulation of \Alfven~waves at this point in the evolution of
patterns of the curvature radius of magnetic field lines.
Perturbations propagate from the PNS convection zone along the radial
field lines close to the poles, until they accumulate at the
\Alfven~point, leading to an enhanced field strength.  We are,
however, not able to clearly identify the contribution of this
mechanism to field amplification or heating because the field is
already very strong there.  Thus, we leave this issue open for further
study.

\begin{figure}
  \centering
  \includegraphics[width=\linewidth]{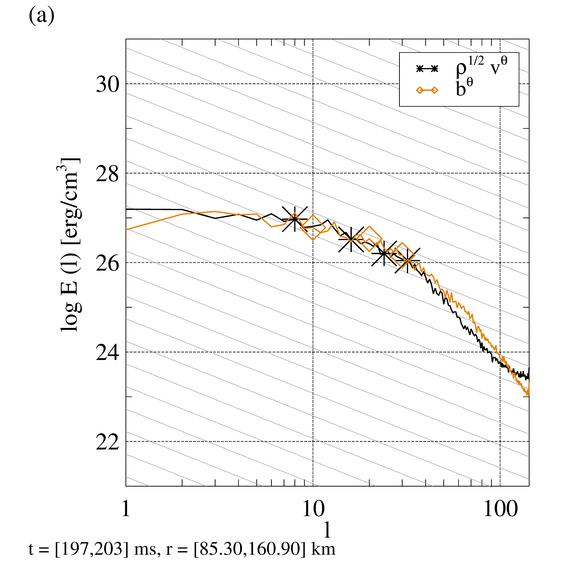}
  \caption{
    Angular energy spectra of the velocity and
    the magnetic field in the \region{GAIN} region of Model
    \modelname{B12} at  $t =
    200 \, \ms$.  For
    orientation, we show the Kolmogorov scaling $E(l) \propto
    l^{-5/3}$ in thin grey lines.  The spectra are computed as averages over times
    $[t -3, t+3] \, \ms$
    and over radii between the lower boundary of the unstable region
    and the minimum shock radius.
  }
  \label{Fig:B12-SAS-spectra}
\end{figure}

To explore the dynamics of the gain layer of Model \modelname{B12} in
spectral space, we compare the coefficients of the expansion of the
transverse components of the magnetic field and of the flow velocity
into spherical harmonics in the lower \region{GAIN} region in
\figref{Fig:B12-SAS-spectra}.  These variables show an extended flat
spectrum up to $l \sim 10$ and a Kolmogorov scaling up to a higher
mode number $l \sim 40$.  The energy spectra of the lateral components
of velocity and magnetic field are close to equipartition across the
entire range of wave numbers.  Besides a roughly equal energy between
field and flow, this indicates that the components of the stress
tensor due to advection and the Lorentz force acting on structures
corresponding to a spectral mode of degree $l$ balance each other.  In
particular, the magnetic tension force resisting the bending of field
lines is of the same order as the hydrodynamic forces and limits the
kinetic energy of the turbulence.

\subsection{Intermediate magnetic fields}
\label{sSek:B11.5}

\begin{figure*}
  \centering
  \includegraphics[width=\textwidth]{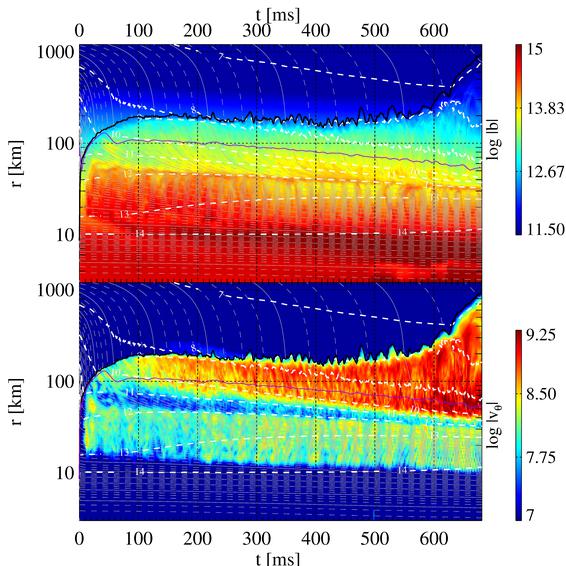}
  \caption{
    Same as \figref{Fig:B10-PNS-mshell}, but for Model \modelname{B11.5}.
  }
  \label{Fig:B115-mshell}
\end{figure*}

\begin{figure}
  \centering
  \includegraphics[width=0.48\textwidth]{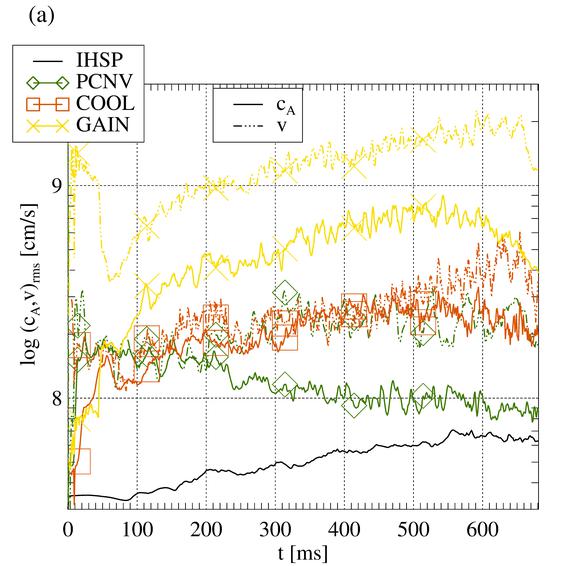}
  \includegraphics[width=0.48\textwidth]{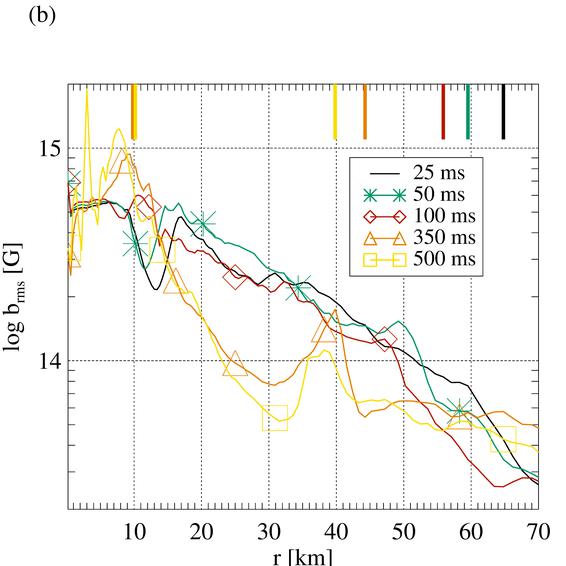}
  \caption{
    Panel \subpanel{(a)}: angularly averaged values of \Alfven~and flow speed in the
    four post-shock regions of Model \modelname{B11.5}.
    Panel \subpanel{(b)}: radial profile of the mean field
    strength in the innermost 70 km of Model \modelname{B11.5} for 5
    times after bounce.
  }
  \label{Fig:B115-PNS-emag}
\end{figure}

An initial field strength of $b_0 = \zehnh{3.16}{11} \, \mathrm{G}$
(Model \modelname{B11.5}; space-time plot shown in
\figref{Fig:B115-mshell}) does not modify the dynamics of the PNS
significantly.  We refer to Panel \subpanel{(a)} of
\figref{Fig:B115-PNS-emag} for the time evolution of the mean flow
speed and \Alfven~velocity across the different regions of the core.
Convection develops in the same region as in Model \modelname{B10}.
During the first $\sim 200 \, \ms$ of post-bounce time, the flow and
the magnetic field are close to equipartition in the \region{PCNV}
layer (green lines), but afterwards expulsion of the magnetic flux
leads to a decrease of the magnetic energy, whereas the kinetic energy
grows and achieves an energy comparable to the weakly magnetised case.
As a consequence of flux expulsion, the PNS at late times exhibits two
layers of enhanced field strength at the bottom and top of the
\region{PCNV}; see the profiles for $t = 350, 500 \, \ms$ in Panel
\subpanel{(b)} of \figref{Fig:B115-PNS-emag} and the upper half of
\figref{Fig:B115-mshell}.  The layer at the outer boundary of the PNS
contains a mostly lateral field, except for high latitudes, where it
is dominated by an intense radial field strong enough to locally
suppress convection.

\begin{figure}
  \centering
  \includegraphics[width=0.48\textwidth]{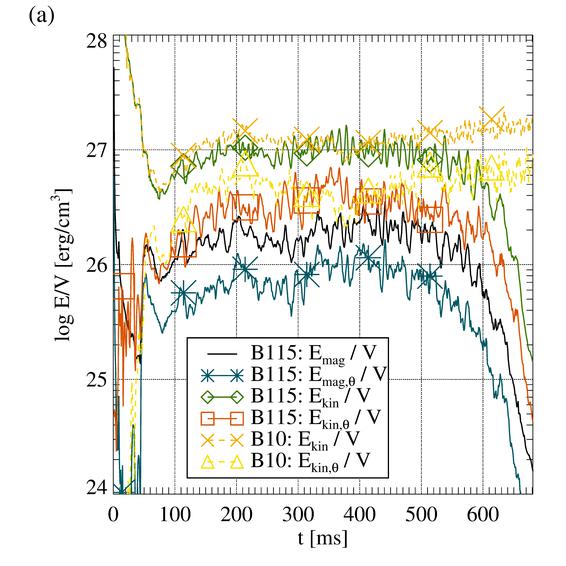}
  \includegraphics[width=0.48\textwidth]{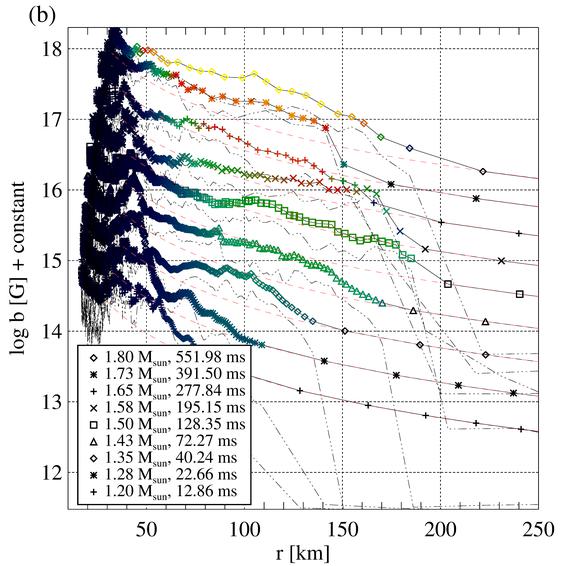}
  \caption{
    Panel \subpanel{(a)}: volume averages of the kinetic and
    magnetic energy densities (total and $\theta$-components) over the \region{GAIN} layer of Model
    \modelname{B11.5} compared to those of  Model \modelname{B10}.
    Panel \subpanel{(b)}: same as Panel \subpanel{(b)} of
    \figref{Fig:B12-coll-bampl}, but for Model \modelname{B11.5}.
  }
  \label{Fig:B115-SAS-emag}
\end{figure}

The evolution of the post-shock region combines elements from models
with weak and models with strong fields.  Unlike in Model
\modelname{B12}, the magnetic field is below equipartition during
the entire evolution, in terms of both the modulus and the
$\theta-$component, though only by a factor of roughly four, as we
show in the overview of the time evolution of the partial energies in
the \region{GAIN} layer in Panel \subpanel{(a)} of
\figref{Fig:B115-SAS-emag}.  Thus, the field interferes less with
the development of hydrodynamic instabilities than in Model
\modelname{B12}, and the kinetic energy reaches, on average, the
same level as in Model \modelname{B10} (dashed lines in the
figure).  We examine the resulting field amplification in Panel
\subpanel{(b)} of \figref{Fig:B115-SAS-emag}, which follows the
mean field strength of mass elements falling at different times
through the shock onto the PNS.  While the shock wave enhances the
$\theta$-component, but does little to the total field strength, both
components of the field grow in the overturning vortices behind the
shock.  The properties of these flows are similar to the weakly
magnetised case, leading to a similar amplification factor.

\begin{figure}
  \centering
  \includegraphics[width=0.48\textwidth]{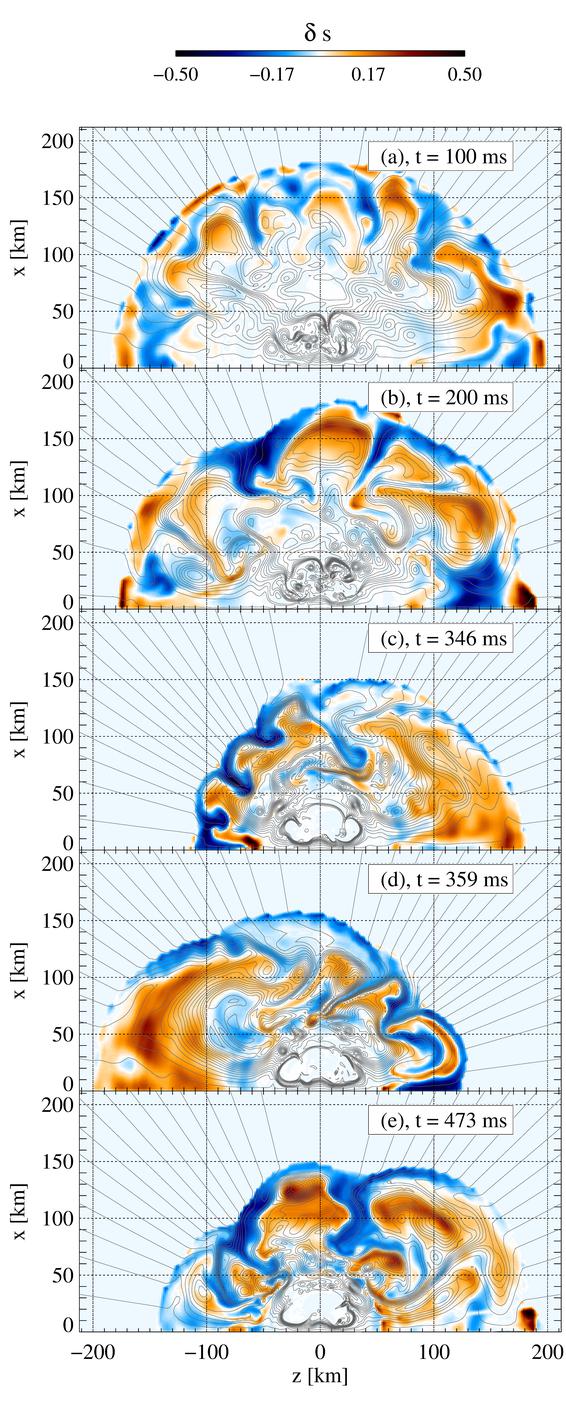}
  \caption{
    Snapshots of the evolution of the entropy contrast, $\delta s$
    (\eqref{Gl:delta_s}), for Model \modelname{B11.5} showing the relative
    entropy contrast and magnetic field lines.
  }
  \label{Fig:B115-2dstruct}
\end{figure}

Snapshots of the 2D structure of the model
(\figref{Fig:B115-2dstruct}) show elements both from Models
\modelname{B10} and \modelname{B12}.  The early flow is characterised
by small-scale features (Panels \subpanel{(a)} and \subpanel{(b)}) in
the flow as well as in the field, similar to the weak-field case.
Later (Panels \subpanel{(d)} and \subpanel{(e)}), larger bubbles
reminiscent of the ones in Model \modelname{B12}
(cf.~\figref{Fig:B12-sas-ent-Bfield-1}) develop, surrounded by narrow
downflows and shielded by strongly magnetised sheets, and the shock
assumes a strongly non-spherical shape.

\subsection{Comparison between models}
\label{sSek:CompGlob}

\subsubsection{Magnetic properties}

We list important parameters of our models in \tabref{Tab:models}:
apart from the initial field strength, $b_0$, we present time averages
(over $t \in [290 ~ \ms, 300 \, \ms]$) of the mean field strength in the
four regions defined above.  Moreover, time-averaged (over $t \in [290
\, \ms,300 \, \ms]$) ratios of magnetic energy to internal energy and
to kinetic energy in the same regions, $\beta^{\mathrm{i}}$ and
$\beta^{\mathrm{k}}$, respectively, are listed.  We have selected this
interval since none of the models is close to an explosion at
that time.

The distribution of the magnetic field strength summarises the patterns
discussed in the previous sections.  The innermost stable region has a
dynamically negligible magnetisation with $\beta^{\mathrm{i}}$ reaching
$\zehnh{2.61}{-5}$ for Model \modelname{B12}, corresponding to an
average field strength of $\zehnh{1.64}{15} \, \mathrm{G}$.  In this
region, the velocity is essentially zero after bounce, suppressing all
mechanisms of field amplification except for compression during
collapse.  Hence, the field has almost the same geometry as in its
initial state, and the mean field strength is roughly proportional to
$b_0$.

Models with weak initial field (\modelname{B08, B10, B11}) develop a
PNS convection zone (\region{PCNV}) in which the field is at first
amplified passively (Model \modelname{B11} reaches only a magnetic
energy of $5 \, \%$ of the convective kinetic energy), but later
expelled from the \region{PCNV} layer and accumulated at its lower and
upper radial boundaries as well as at the symmetry axis.  In the limit
of a passive field, which does not react back onto the flow, the
evolution of the field strength depends only on the properties of the
flow such as eddy sizes and turnover times.  Hence, the factor by
which these processes change the magnetic field strength should be
independent of the seed field.  The proportionality is violated in our
models: stronger fields (\modelname{B11}) are amplified less than
weaker ones.  This violation occurs mostly at the polar axis, where
the assumed axisymmetry leads to the formation of an artificially
strongly magnetised column.  The stronger the initial field, the more
the field amplification is limited by the suppression of convection by
magnetic tension.

If we take the values in the \region{COOL} region as approximations to
the surface field of the PNS, we get values that are compatible with
most pulsars (weak-field models) or on the upper end of the field
distribution with magnetar fields of order $\zehn{14}\, \mathrm{G}$
(strong-field models).

In the \region{GAIN} layer, we find similar trends as in the
\region{PCNV} region.  Weak fields are amplified by a constant factor,
while strong fields reach energies close to the kinetic energy and,
thus, are limited by dynamic feedback.  The artificial effect of the
axis on the field amplification affecting the results in the PNS is
less of an issue here (the mean field and the $\beta$ parameters are
closer to proportionality to $b_0$) since the combination of
advection, SASI, and convection does not favour the development of
stationary convective cells similar to those found in the convective
layer of the PNS.

\begin{table*}
  \centering
  \caption{
    Compilation of parameter values that characterise the field strength
    and evolution in our computed models with magnetic field.  The
    initial central
    field strength is $b_0$, and $b_{\mathrm{IHSP}}$, $b_{\mathrm{PCNV}}$, $b_{\mathrm{COOL}}$, and
    $b_{\mathrm{GAIN}}$ are the average field strengths in the PNS at
    densities exceeding $10^{14} ~ \gccm$, in the PNS
    convection zone, in the stable layer surrounding
    the PNS, and in the gain layer, respectively.
    Furthermore, $\beta^{\mathrm{i}}_{...}$ and $\beta^{\mathrm{k}}_{...}$
    are the average ratios of magnetic to internal
    energy in the same regions, and the
    average ratio of magnetic to kinetic energy in the same regions,
    respectively.  These quantities are averaged over the time
    interval of 290--300 ms after core bounce, i.e.~before any of the
    models explodes.
  }
  \begin{tabular}{l|ccccc}
    \hline
    Model 
    & \modelname{B08}
    & \modelname{B10}
    & \modelname{B11}
    & \modelname{B11.5}
    & \modelname{B12}
    \\ 

    \hline
    \hline
    $b_0$
    & $10^8$
    & $10^{10}$
    & $10^{11}$
    & $3.16 \times 10^{11}$
    & $10^{12}$
    \\ 
    \hline
    $b_{\region{IHSP}}$ [G] 
    & $4.02\times10^{11}$
    & $3.02\times10^{13}$
    & $2.66\times10^{14}$
    & $7.51\times10^{14}$
    & $1.64\times10^{15}$
    \\ 
    $\beta^{\mathrm{i}}_{\region{IHSP}}$
    & $1.61\times10^{-12}$
    & $9.12\times10^{-9}$
    & $7.07\times10^{-7}$
    & $5.54\times10^{-6}$
    & $2.61\times10^{-5}$
    \\ 
    $\beta^{\mathrm{k}}_{\region{IHSP}}$
    & 0.000225
    & 1.08
    & 9105.42
    & 1149.13
    & 677.40
    \\ 

    \hline
    $b_{\region{PCNV}}$ [G] 
    & $8.23\times10^{11}$
    & $4.71\times10^{13}$
    & $1.25\times10^{14}$
    & $2.52\times10^{14}$
    & $6.01\times10^{14}$
    \\ 
    $\beta^{\mathrm{i}}_{\region{PCNV}}$
    & $5.96\times10^{-10}$
    & $1.98\times10^{-6}$
    & $1.40\times10^{-5}$
    & $5.64\times10^{-5}$
    & $3.14 \times 10^{-4}$
    \\ 
    $\beta^{\mathrm{k}}_{\region{PCNV}}$
    & $2.40\times10^{-6}$
    & 0.00614
    & 0.0476
    & 0.32
    & 1.42
    \\ 

    \hline
    $b_{\region{COOL}}$ [G] 
    & $6.62\times10^{10}$
    & $4.95\times10^{12}$
    & $2.15\times10^{13}$
    & $4.62\times10^{13}$
    & $8.20\times10^{13}$
    \\ 
    $\beta^{\mathrm{i}}_{\region{COOL}}$
    & $6.73\times10^{-10}$
    & $3.80\times10^{-6}$
    & $7.82\times10^{-5}$
    & $3.95 \times 10^{-4}$
    & $1.51 \times 10^{-3}$
    \\ 
    $\beta^{\mathrm{k}}_{\region{COOL}}$
    & $8.90\times10^{-7}$
    & 0.00435
    & 0.15
    & 0.64
    & 2.97
    \\ 

    \hline
    $b_{\mathrm{GAIN}}$ [G] 
    & $1.42\times10^{10}$
    & $1.19\times10^{12}$
    & $8.36\times10^{13}$
    & $1.68\times10^{13}$
    & $2.28\times10^{13}$
    \\ 
    $\beta^{\mathrm{i}}_{\mathrm{GAIN}}$
    & $2.99\times10^{-9}$
    & $2.19\times10^{-5}$
    & $0.00103$
    & $0.00511$
    & $0.0319$
    \\ 
    $\beta^{\mathrm{k}}_{\mathrm{GAIN}}$
    & $1.00\times10^{-7}$
    & 0.000626
    & 0.0411
    & 0.15
    & 0.59
    \\

    \hline
    \hline

  \end{tabular}

  \label{Tab:models}
\end{table*}

\subsubsection{Neutrino emission}
\label{sSek:Neutrino}

\begin{figure}
  \centering
  \includegraphics[width=0.48\textwidth]{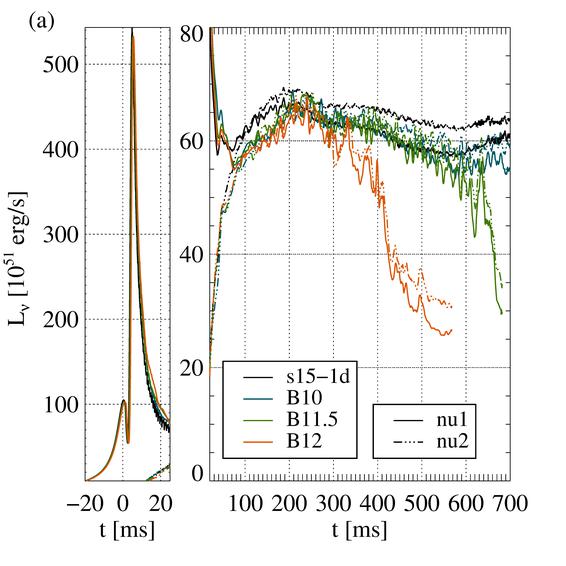}
  \includegraphics[width=0.48\textwidth]{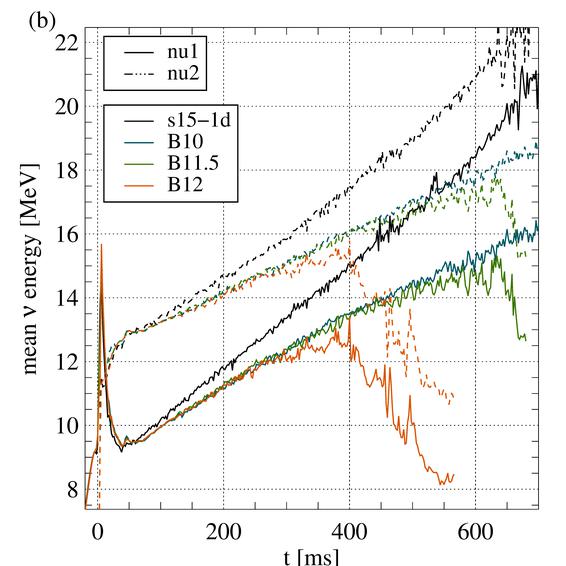}
  \caption{
    Panel \subpanel{(a)}: comparison of the time evolution of the neutrino luminosities of the
    magnetised models (\modelname{B10}, \modelname{B11.5}, and
    \modelname{B12}) to the
    spherically symmetric model.  Solid and  
    dash-dot-dot-dotted lines display the luminosities of
    the electron neutrinos and the antineutrinos, respectively.
    Panel \subpanel{(b)}: mean energies of electron neutrinos (solid) and
    antineutrinos (dash-dot-dot-dotted) of the same models as functions of time.
    The curves labelled by s15-1d display the results of the corresponding
    spherically symmetric simulation briefly described in \secref{Sek:Results}.
  }
  \label{Fig:Neutrinoemission}
\end{figure}

The total neutrino luminosities emitted by the core of three of the
magnetised and the spherically symmetric models are presented in
Panel \subpanel{(a)} of \figref{Fig:Neutrinoemission}.  For all
models, $L_{\nu_e}$ increases gradually during collapse.  At prompt
shock breakout, we observe the well-known prominent burst of the
electron neutrino luminosity.  The burst in the figure is located not
right at bounce but at a somewhat later time, $t \approx 12 ~ \ms$;
this is the consequence of the finite propagation time from the inner
core where the burst is produced to the radius where we measure the
luminosity, $r = 500 \, \km$.  After the short burst, lasting for
about five milliseconds (full width at half maximum), the total
($\nu_e$ plus $\bar{\nu}_e$) neutrino luminosity settles to a roughly
constant value of the order of $1.3 \times 10^{53} ~ \mathrm{erg} /
\mathrm{s}$.  In this phase, electron anti-neutrinos contribute about
half of the total luminosity.

All axisymmetric models emit slightly lower luminosities than Model
\modelname{s15-1d} during the first $\sim 200 \, \ms$ after the
neutrino burst.  This reduction, representing the contribution of PNS
convection to the total energy transport out of the PNS
\citep{Buras_etal__2006__AA__Mudbath_1}, does not depend on the
magnetic field strength.  The curves representing the different
initial fields lie almost on top of each other for long parts of the
evolution until the onset of an explosion causes a decrease of the
mass accretion rate onto the PNS and hence a drop of the luminosity,
which occurs earlier for models with stronger initial fields since, as
we will describe below, the explosion time is correlated to the
magnetic field strength.

The average energies (Panel \subpanel{(b)} of
\figref{Fig:Neutrinoemission}) of electron neutrinos and
anti-neutrinos increase in this phase to values of $14$ and $15$ MeV
at $t = 300 ~ \ms$, respectively.  Because of their weaker interaction
with the surrounding matter, the anti-neutrinos have the tendency to
decouple from the gas slightly deeper on average than the electron
neutrinos and, therefore, have a mean energy exceeding that of the
electron neutrinos by about $25 \%$ early on.  The mean energies of
both flavours increase strongly in spherical symmetry as the neutron
star becomes more massive to evolve to collapse
\citep{Sumiyoshi_et_al__2007__apj__DynamicsandNeutrinoSignalofBlackHoleFormationinNonrotatingFailedSupernovae.I.EquationofStateDependence,Sumiyoshi_et_al__2008__apj__DynamicsandNeutrinoSignalofBlackHoleFormationinNonrotatingFailedSupernovae.II.ProgenitorDependence,Fischer_et_al__2009__aap__Theneutrinosignalfromprotoneutronstaraccretionandblackholeformation}.
Similarly to the luminosities, the mean energies of the axisymmetric
models deviate from the spherical case, but show little difference
among each other, except different times of explosion, which are
marked by a steep drop of the mean energies.

\subsubsection{Explosion in dependence of the magnetic field}
\label{sSek:Expl}

\paragraph{Model \modelname{B10}.}

The previous discussion of the field amplification was focused mostly
on the accretion phase characterised by a steady or slowly receding shock
wave.  In the following, we will briefly outline the most important
features of the final transition to shock expansion and, eventually,
explosion.

Discussions of the physics of the onset of explosions very often focus
on global quantities such as the total energy in the gain layer and
the competition of the heating time scale of matter by neutrinos and
its advection onto the PNS
\citep[e.g.~][]{Thompson__2000__apj__AccretionalHeatingofAsymmetricSupernovaCores,Scheck_et_al__2008__aap__Multid_SN_sim_with_approx_nu_transport.II,Thompson_Murray__2001__ApJ__MHD_conv_SN,Buras_etal__2006__AA__Mudbath_1}
or local conditions such as the antesonic condition
\citep{Pejcha_Thompson__2012__apj__ThePhysicsoftheNeutrinoMechanismofCore-collapseSupernovae}
connecting sound and escape velocity in the gain layer.  The resulting
explosion criteria can be translated into a critical curve in the
space of the luminosity of neutrinos and the accretion rate of mass
onto the PNS, above which explosions occur
\citep{Burrows_Goshy_1993__apjl__ATheoryofSupernovaExplosions}.  A
steady-state solution for the accretion flow is possible only for a
luminosity below the critical one, and the lack of such a solution for
higher luminosity leads to shock expansion and explosion
\citep[see][for a detailed
study]{Fernandez__2012__apj__HydrodynamicsofCore-collapseSupernovaeattheTransitiontoExplosion.I.SphericalSymmetry}.
These models work best in spherical symmetry or for the spherical ($l=
0$) mode of the shock surface in multi-dimensional cores.
Non-spherical mass motions due to e.g.~convection and the SASI lead to
a reduction of the critical luminosity \cite[see,
e.g.][]{Janka_Mueller__1996__aap__NeutrinoheatingconvectionandthemechanismofType-IIsupernovaexplosions,Herant_Benz_Colgate__1992__apj__HD_SN1987A-2d_early_evolution,Nordhaus_et_al__2010__apj__Dimension_as_a_Key_to_the_Neutrino_Mechanism_of_CCSNe,Hanke__2012__apj__IsStrongSASIActivitytheKeytoSuccessfulNeutrino-drivenSupernovaExplosions,Couch__2013__apj__TheDependenceoftheNeutrinoMechanismofCore-collapseSupernovaeontheEquationofState}.

Recently,
\cite{Dolence_et_al__2013__apj__DimensionalDependenceoftheHydrodynamicsofCore-collapseSupernovae},
and subsequently
\cite{Fernandez_et_al__2014__mnras__CharacterizingSASI-andconvection-dominatedcore-collapsesupernovaexplosionsintwodimensions},
have studied multi-dimensional explosions dominated either by
convection or the SASI in the post-shock region.  High-entropy bubbles
can be generated and destroyed by both instabilities on a wide range
of eddy sizes.  These bubbles, in turn, can affect the evolution of
the SASI and the oscillations of the shock wave.  The onset of the
explosion is characterised by large bubbles persisting for many eddy
turnover times.

We will characterise the explosion of Model \modelname{B10} along
these lines.  This will allow us to find features that distinguish
this model from the ones with stronger initial fields to be discussed
afterwards.

\begin{figure*}
  \centering
  \includegraphics[width=0.48\textwidth]{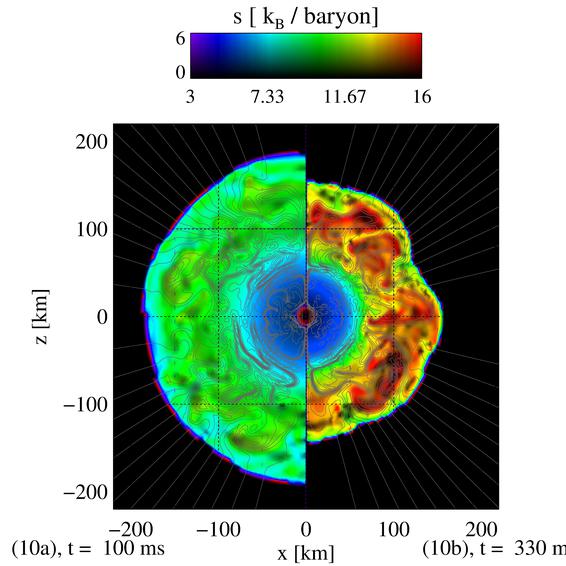}
  \includegraphics[width=0.48\textwidth]{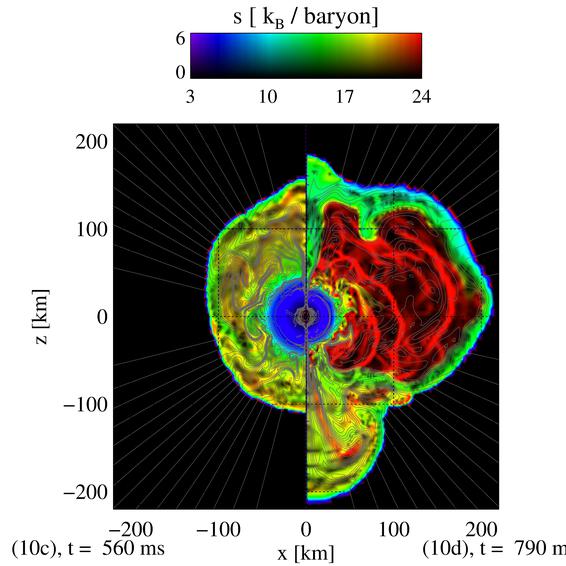}
  \includegraphics[width=0.48\textwidth]{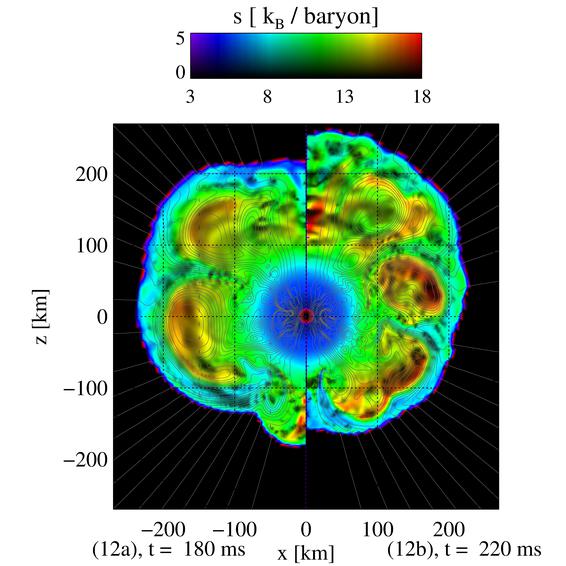}
  \includegraphics[width=0.48\textwidth]{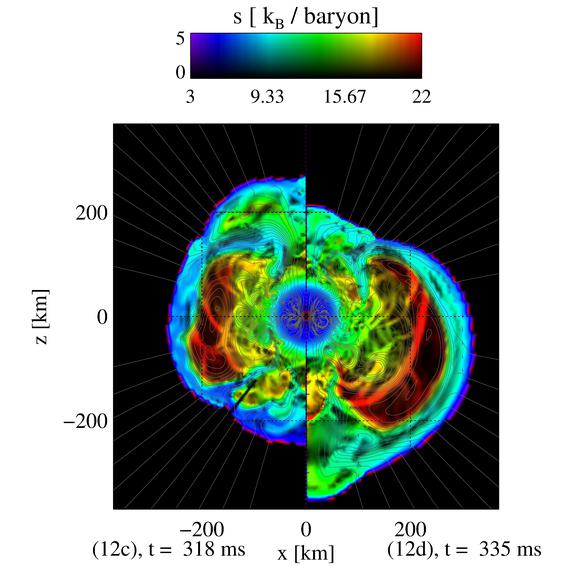}
  \caption{
    Visualisation of magnetic field lines, entropy and magnetosonic
    waves of Models \modelname{B10} (\subpanel{top} panels
    \subpanel{(10a,10b,10c,10d)}) and \modelname{B12}
    (\subpanel{bottom} panels \subpanel{(12a,12b,12c,12d)}) at four
    times after bounce.  The hue of the colour scale shows the
    specific entropy and its lightness represents the normalised
    length scale of the variation of the total pressure, $r |\vec
    \nabla \log P_{\mathrm{tot}}|$, in which magnetosonic wave fronts
    show up as light and dark patterns.  Video versions of these
    visualisations are available in the electronic edition of the
    article.
  }
  \label{Fig:B10-SAS-ent-waves-1}
\end{figure*}

\begin{figure*}
  \centering
  \includegraphics[width=0.48\textwidth]{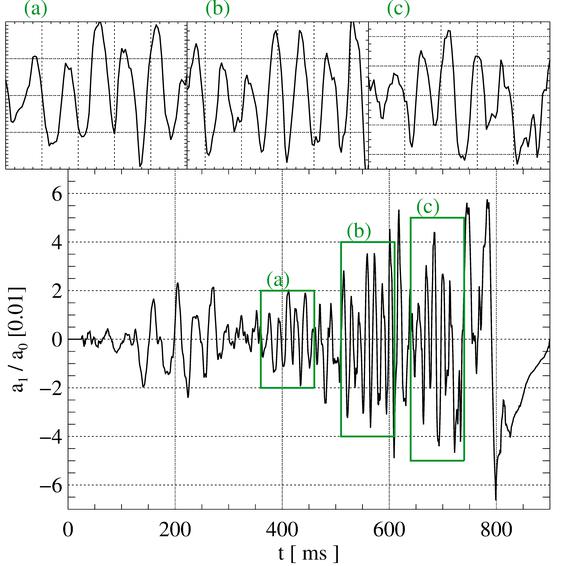}
  \includegraphics[width=0.48\textwidth]{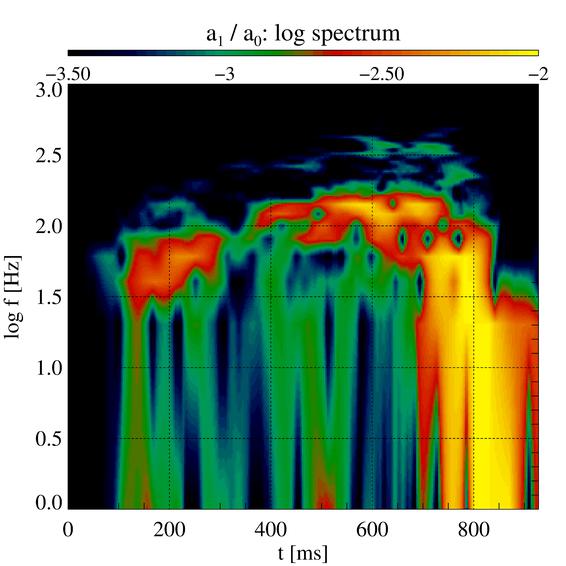}
  \includegraphics[width=0.48\textwidth]{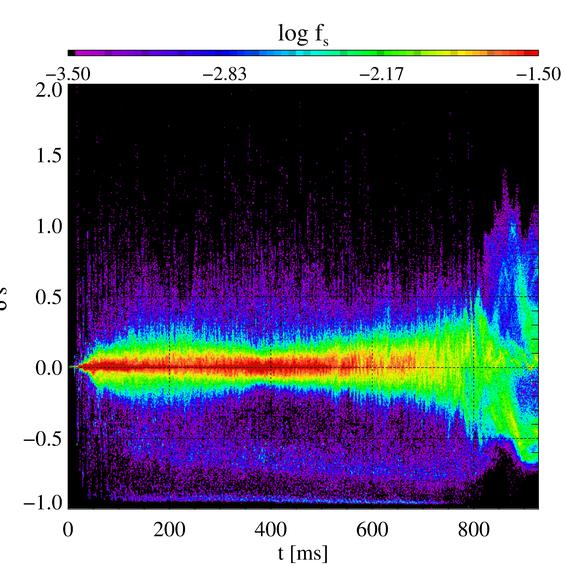}
  \includegraphics[width=0.48\textwidth]{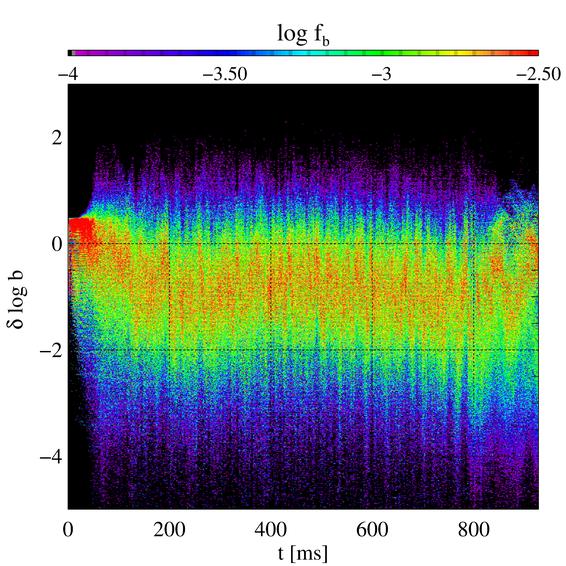}
  \caption{
    Overview of the evolution of the shock radius and the post-shock
    conditions of Model \modelname{B10}: 
    \subpanel{Top left} panel: time evolution of the first coefficient
    of the expansion of the shock radius in spherical harmonics, $a_1
    / a_0$.  
    \subpanel{Top right} panel: spectrogram of the same variable.  
    \subpanel{Bottom left} panel: distribution of the volume filling
    factor of the relative deviation of the entropy from its angular
    average, $\delta s$.  The histogram was constructed by binning
    $\delta s$ into 1600 equidistant bins between -1 and 4.  
    \subpanel{Bottom right} panel: same as the \subpanel{bottom left}
    panel, but for the logarithm of the magnetic field strength.
  }
  \label{Fig:B10-shock-l1modes}
\end{figure*}

As shown in \figref{Fig:B10-shockmodes}, the shock wave is mostly
spherical as it expands to a radius of 180 km at $t \approx 100 \,
\ms$, although aspherical flows are already present (see Panel
\subpanel{(a)} of \figref{Fig:B10-SAS-ent-2} and Panels
\subpanel{10a,10b,10c,10d} of \figref{Fig:B10-SAS-ent-waves-1}).
During the next $\sim 400 \, \ms$, strong asymmetries develop while
the shock contracts.  To quantify these asymmetries, we present the
time evolution of the normalised amplitude of the $l = 1$ coefficient,
$a_1 / a_0$, of the spherical harmonics expansion of the shock radius
in the \subpanel{top left} panel of \figref{Fig:B10-shock-l1modes}.
The \subpanel{top right} panel contains a spectrogram of this
variable, i.e.~the spectral amplitude $f$ at time $t$ computed from a
Fourier transform of $a_1(T) / a_0(T), T \in [t - 50 \, \ms, t + 50 \,
\ms]$.

We try to connect the deformations of the shock surface to the
structure of the post-shock flow.  The \subpanel{bottom left} panel
of \figref{Fig:B10-shock-l1modes} shows histograms of the entropy
contrast $\delta s$ (see \eqref{Gl:delta_s}), i.e.~the variable for
which two-dimensional maps are shown in \figref{Fig:B10-SAS-ent-2},
for all post-bounce times.  We constructed a linear grid in $\delta s$
covering the range between -1 and 4 with 1600 equidistant bins, each
attributed with the volume filling factor $f_s$ of all zones in the
region between the PNS and the shock wave that possess a certain value
of $\delta s$.  Typically, the distribution is peaked around $\delta s
= 0$ because most zones have an entropy close to the mean value.  It
gradually broadens as bubbles of increasing size develop.  Very
large bubbles of high entropy surrounded by cold downflows are
reflected in a broader distribution with a less pronounced peak.

After $t \approx 700 \, \ms$, $|\delta s|$ increases, and the bubbles
persist for a longer time.  The oscillations of $a_1/a_0$ break down,
and the spectrum exhibits a strong low-frequency component.  At this
point, the distribution of $f_s$ is very broad, no longer exhibiting a
strong peak at $\delta s = 0$.  The weakening of this peak correlates
with the onset of an increasingly rapid outward motion of the shock
radii (\figref{Fig:B10-shockmodes}).

The dynamics observed agrees well with those described by
\cite{Dolence_et_al__2013__apj__DimensionalDependenceoftheHydrodynamicsofCore-collapseSupernovae}
and
\cite{Fernandez_et_al__2014__mnras__CharacterizingSASI-andconvection-dominatedcore-collapsesupernovaexplosionsintwodimensions}.
Compared to the models presented there, Model \modelname{B10} shows a
very extended post-bounce phase and the shock motion as well as the
generation and disappearance of large-scale structures undergoes
several phases during which the $l=1$ shock mode oscillates either in
a more regular, SASI-like, or a more complex, convection-like,
pattern.

The magnetic field is mostly organised in fairly thin filaments.
Thus, the volume filling factor of the magnetic field is small, and
most zones have a low magnetisation.  This can be seen in the volume
filling factor of the logarithm of the magnetic field strength, $f_b$,
obtained, similarly to $f_s$, based on a logarithmic grid in the
deviation of $|\vec b|$ from its angular average, $\delta b = b -
\langle b \rangle$.  We display $f_b$ in
the \subpanel{bottom right} panel of \figref{Fig:B10-shock-l1modes}.
The predominance of thin structures of strong field lead to a negative
displacement of the maximum of the histograms from $\delta b =
0$.  The location of this maximum exhibits a variability on time
scales of a few tens of ms, i.e.~similar to the most prominent
frequencies of the oscillations of $a_1/a_0$ Besides this variability,
we find little imprint of the overall dynamics, in particular of the
transition to explosion.

\begin{figure}
  \centering
  \includegraphics[width=0.5\textwidth]{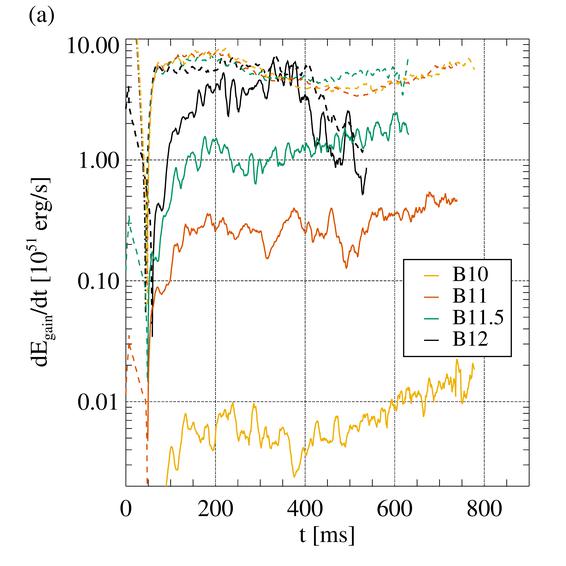}
  \includegraphics[width=0.5\textwidth]{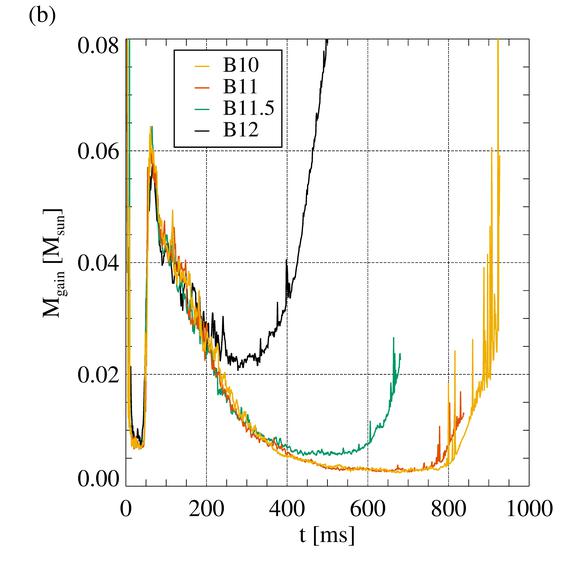}
  \caption{
    Panel \subpanel{(a)}: evolution of the transfer of energy between
    matter and neutrinos, i.e.~neutrino heating, (dashed lines) and between matter and
    magnetic fields due to field amplification and
    dynamic feedback (solid lines) in the gain layer of the magnetised models.
    Panel \subpanel{(b)}: evolution of the mass of the gas in the
    \region{GAIN} region of the magnetised models.
  }
  \label{Fig:Gain-mass-q}
\end{figure}

Neutrino heating is crucial for the dynamics of the \region{GAIN}
layer.  The dashed lines in Panel \subpanel{(a)} of
\figref{Fig:Gain-mass-q} display the volume-integrated heating rates,
$Q_{\mathrm{\nu}}$, of all magnetised models as functions of time
(dashed lines).  They exceed a level of $Q_{\mathrm{\nu}} \gtrsim
\zehnh{6}{51} \, \erg / \mathrm{s}$ for 200 ms around $t = 200 \, \ms$
after bounce and decrease afterwards by a factor of 2.  In the same
panel, we show for comparison the volume-integrated source terms of
the magnetic energy (r.h.s.~of \eqref{Gl:emag-evo}) (solid lines).
Their positive sign during most of the evolution indicates the
generation of magnetic energy by the flow due to compression and
stretching of field lines.  Since this quantity scales with the square
of the field strength, it is several orders of magnitude below
$Q_{\nu}$ in Model \modelname{B10}, a further sign of the negligible
level of dynamic feedback of the field onto the flow in this model.

Finally, we discuss the evolution of the mass in the \region{GAIN}
layer, $M_{\mathrm{gain}}$, presented for the magnetised models in
Panel \subpanel{(b)} of \figref{Fig:Gain-mass-q}.  After reaching a
maximum value of $M_{\mathrm{gain}} \approx 0.065 \Msol$ after bounce,
the mass decreases to a few thousandths of a solar mass, until the
onset of explosion leads to its extremely steep increase at late times.

\paragraph{Model \modelname{B12}.}

Similar to our analysis of Model \modelname{B10}, we can convert the
spectral energy into turnover times for eddies and \Alfven~waves.  For
weak fields (cf.~\figref{Fig:B10-GAI-ellspec}), the eddy turnover time
is comparable to the advection time of fluid elements in the
\region{GAIN} layer, allowing for dynamically relevant effects of the
non-spherical motions.  For Model \modelname{B12}, both eddy turnover
and lateral \Alfven~crossing times are comparable to the advection
time.  Thus, the post-shock structure is dominated in equal parts by
the hydrodynamic instabilities and the field.  In particular, the
largest structures and bubbles are in force balance between Reynolds
and Maxwell stresses.

These findings point towards an important role of the magnetic field
in establishing and maintaining large-scale flow patterns.  To further
explore the magnetic effects on the development of the SASI, we
discuss the pattern of magnetosonic waves, which can be identified in
the normalised length scale of variations of the total pressure,
$L_{P} = r/P_{\mathrm{tot}} |\vec \nabla P_{\mathrm{tot}}|$, shown as
variations of the lightness superimposed on the colour scale
representing the specific entropy in the two-dimensional snapshots of
\figref{Fig:B10-SAS-ent-waves-1} (Panels \subpanel{12a,12b,12c,12d}).  These
variations reveal a geometry characterised by only few wave fronts
extending over large ranges of latitude.  Since the magnetic field
restricts the advection to narrow downflows, magnetosonic waves are
emitted from the deceleration region only at a few locations, which
evolve only slowly in time.  Emitted into all directions in the lower
section of a downflow, the waves travel mostly radially in the
accretion channels, in particular at the poles, and in an oblique
direction across the bubbles in between.  Arriving at the shock wave,
they can couple back to the infall and close the advective-acoustic
cycle fostering the SASI.

The magnetic field suppresses the disruption of the large bubbles by
convective turbulence observed in Model \modelname{B10}.  The
quadrupolar topology of the field, constituted by a large field
vortex in each hemisphere (see left panels of
\figref{Fig:B12-sas-ent-Bfield-1} and Panel \subpanel{(12a)} of
\figref{Fig:B10-SAS-ent-waves-1}), acts as an additional
stabilising factor.  The formation of one large bubble, which could
interrupt the SASI cycle, requires that one of these vortices be
disrupted or pushed aside towards the axis.  This process is
inhibited by the super-equipartition magnetic tension.

\begin{figure}
  \centering
  \includegraphics[width=0.48\textwidth]{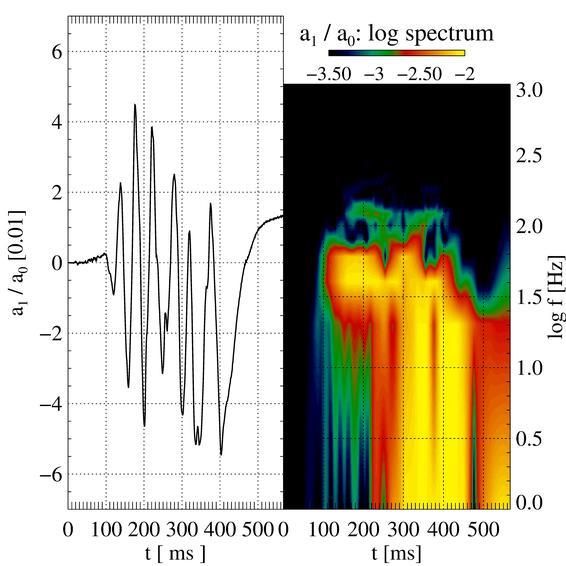}
  \caption{
    Time evolution and spectrogram of the
    normalised dipole mode $a_1 / a_0$ of Model \modelname{B12}.
  }
  \label{Fig:B12-shockmodes}
\end{figure}

Consequently, the SASI generates a fairly regular, pattern of slow
shock oscillations for a long period.  The dipole mode of the shock
deformation (\subpanel{bottom} panel of \figref{Fig:B12-shockmodes})
shows fairly regular oscillations of a high amplitude in this stage.
The period of these oscillations, around $\tau \approx 40 \, \ms$, is
greater than in Model \modelname{B10} and remains rather stable until
the onset of rapid shock expansion.  The evolution of the shock is a
response to the post-shock structure.

Shock sloshings tend to occur via the
development of new large bubbles at one pole, the displacement of the
existing (usually two) equatorial bubbles towards the other pole, and,
finally, mergers between bubbles, and less via the disruption of
bubbles by secondary instabilities.  Panel \subpanel{(12b)} of
\figref{Fig:B10-SAS-ent-waves-1} shows an intermediate stage during
the oscillations.  Three vortices of the field are present, and the
entropy is rising in the downflow at the north pole.  The bubbles will
later merge, and the shock will oscillate towards the south pole.
Panels \subpanel{(c)} and \subpanel{(d)} show the model at a later
phase.  They correspond to subsequent positive and negative peaks in
$a_1 / a_0$.  By this time, we find only one predominant hot bubble
near the equator and two polar downflows (see also
\figref{Fig:B12-sas-ent-Bfield-1}, right column of panels).  This
bubble is not disrupted as the shock oscillates between the two
hemispheres, but retains its coherence.  The oscillations of the shock
are mostly due to alternating intensities of the polar downflows.

Eventually, a runaway of the shock radius and an explosion develop
from these large bubbles.  In this last phase of the evolution, the
model remains dominated by the equatorial bubble that persists until
the end of the simulation while smaller ones may develop transiently
near the poles (\figref{Fig:B12-sas-ent-Bfield-1}, right column of panels).

\begin{figure}
  \centering
  \includegraphics[width=0.48\textwidth]{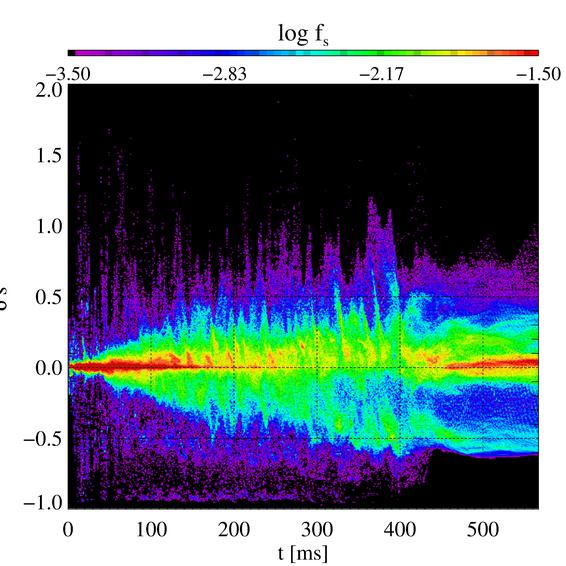}
  \includegraphics[width=0.48\textwidth]{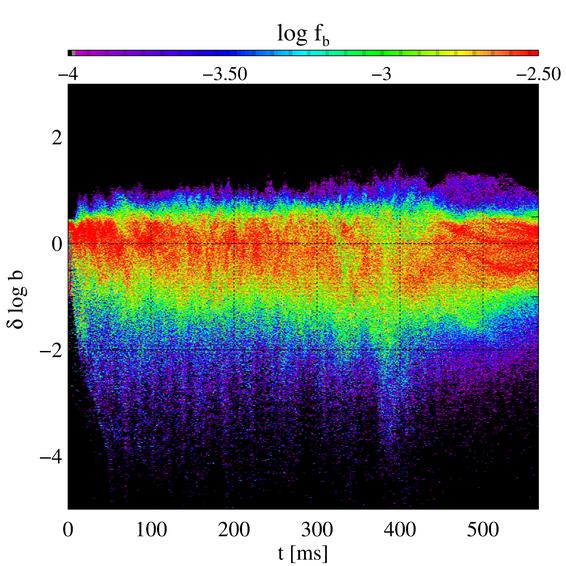}
  \caption{
    Distributions of the volume filling factors of the
    relative entropy contrast, $\delta s$, and of the logarithm of the
    magnetic field, $\delta \log b$, of Model \modelname{B12}.
  }
  \label{Fig:B12-shock-1}
\end{figure}

The transition to the explosion is heralded by the broadening of the
distribution of the volume filling factor of the entropy contrast
(\figref{Fig:B12-shock-1}).  After $t \sim 140 \, \ms$, the red band
at $f_s = 0$ weakens, and significant fractions of the gas deviate
from the mean entropy by at least a third.  During the subsequent
evolution, the pulsations of the large, high-entropy bubbles appear
as a series of more pronounced maxima at $\delta s \sim 0.1$ (red
knots in the figure).  After $t \sim 230 \, \ms$, the maximum of the
distribution is very wide, pointing towards a dichotomy between
increasingly hot bubbles (high $\delta s > 0$) and cool downflows (low
$\delta s < 0$).  This broadening precedes the actual shock expansion
slightly and, thus serves as an early indicator for the subsequent
explosion.

Not only the evolution of the distribution of the entropy is different
from that of Model \modelname{B10} (in particular w.r.t.~the mechanism
behind the development of large bubbles), but also the volume filling
factor of the magnetic field strength is rather different.  The maxima
of the histograms are at $\delta b = 0$, not at a lower value as in
Model \modelname{B10}.  The flux sheets comprising most of the
magnetic energy are thicker.  Thus, a larger volume fraction of the gas
possesses a high magnetisation.  We note that the important role of
the field for the explosion is not imprinted into these distributions in
a way similar to the ones of $\delta s$.  This is, on the other hand,
not too surprising, since the magnetic field does not undergo strong
amplification in the post-bounce flow and, thus, changes to its
structure are limited only.

We compare our model to the results of
\cite{Fernandez_et_al__2014__mnras__CharacterizingSASI-andconvection-dominatedcore-collapsesupernovaexplosionsintwodimensions}
in the regime dominated by large bubbles.  They point out that the
SASI cycle can be interrupted as large-scale bubbles form and force
the accretion flow into (typically) one narrow downflow.  In contrast,
regular shock oscillations coexist with large, magnetically stabilised
bubbles in our model.  This difference might be due to the presence of
wide polar accretion channels, which can close the advective-acoustic
cycle.  These channels lead to a rather spherical explosion geometry
instead of the bipolar geometry of the model of
\cite{Fernandez_et_al__2014__mnras__CharacterizingSASI-andconvection-dominatedcore-collapsesupernovaexplosionsintwodimensions}
(cf.~e.g.~our \figref{Fig:B12-sas-ent-Bfield-1} to their Fig.~3).

The total rate of neutrino heating, $Q_{\nu}$, differs somewhat from
that of Model \modelname{B10} (Panel \subpanel{(a)} of
\figref{Fig:Gain-mass-q}) despite little differences in the neutrino
luminosity and the mean neutrino energies between the two models (see
\figref{Fig:Neutrinoemission}).  It is lower early on and does not
exhibit the extended plateau up to $Q_{\nu} \sim \zehnh{8}{51} \, \erg
/ \mathrm{s}$.  This difference is caused by the different structures
of the \region{GAIN} layer.  In a comparison of the 2D structure of
the models, we find that the lower boundary of the regions where
neutrino heating exceeds neutrino cooling lies at larger radius, hence
lower density, for the model with stronger field, leading to a
slightly reduced neutrino optical depth in the heating region.  We
note that the rate at which magnetic energy is generated (solid line)
has the same order of magnitude as $Q_{\nu}$, and at late times even
comes close to it.  Around $t \sim 350 \, \ms$, both terms balance
each other approximately, before they drop roughly in parallel when
the explosion develops.

The total mass of the \region{GAIN} layer (Panel \subpanel{(b)} of
\figref{Fig:Gain-mass-q}) is hardly affected by the strong field early
on, but due to the large shock radius after $t \sim 200 \, \ms$ and the
early explosion, it does not drop to the same low values as in Model
\modelname{B10} later, but, instead, remains above $\sim 0.022 \Msol$
at all times.

\paragraph{Model \modelname{B11.5}.}

The radius of the shock wave begins to stop receding and deviate from
that of Model \modelname{B10} after $t \sim 300 \, \ms$, i.e.~about
200 ms later than in Model \modelname{B12} (see
\figref{Fig:B115-mshell}).  After $t \sim 400 \, \ms$, the shock
expands, gradually at first, then at a very fast rate once its mean
radius exceeds 200 km at $t \sim 600 \, \ms$.

The dipole mode of the shock surface shows weak and irregular
oscillations until $t \sim 250 \, \ms$.  The amplitude is of the same
order as in Model \modelname{B10} early on.  This stage is succeeded
by a rather long phase of regular, large-amplitude oscillations,
indicative of a strong SASI mode.  Their frequency is slightly lower
(periods around 12 ms) than during similar epochs in Model
\modelname{B10}, which is consistent with its slightly larger shock
radius.  Finally, the power shifts to a broader range of lower
frequencies as the average shock radius starts to increase.

Despite the high energy of the non-spherical flow, the magnetic field
strength reaches local kinetic equipartition, in particular in a wide
column along the polar axis.  Similar to the situation in Model
\modelname{B12}, this allows for the development of very persistent
polar accretion flows in addition to the more variable ones at lower
latitudes.  In the maps of the entropy contrast given in
\figref{Fig:B115-2dstruct}, these show up as blue, i.e.~cool regions, 
threaded by mostly radial field lines.

High-entropy bubbles surrounded by strong magnetic flux sheets develop
between the accretion channels.  Though their number, position, and
size vary with time, the strong field acts to preserve, on average,
the balance between the two hemispheres for a long time, analogous to
Model \modelname{B12}.  Besides the presence of the polar accretion
flows, this leads to the long phase of regular SASI oscillations with
a dipole mode that corresponds to the alternating expansion and
compression of the bubbles on either side of the equator. Panels
\subpanel{(c)} and \subpanel{(d)} of \figref{Fig:B115-2dstruct} show
the model during subsequent peak amplitudes of $a_1 / a_0$.  The
large-scale topology of the field is conserved, and the two time steps
show an approximate mirror symmetry with reversed positions of large
high-entropy bubbles and cool (polar) downflows.

This pattern, stabilised by the magnetic field, is repeated until,
finally, the symmetry is lost and one bubble expands strongly at the
cost of the others.  Afterwards, the shock starts to expand rapidly,
leading to an explosion.

For this model, both the energy generation rates (Panel \subpanel{(a)}
of \figref{Fig:Gain-mass-q}) and the mass in the \region{GAIN} (Panel
\subpanel{(b)} of \figref{Fig:Gain-mass-q}) are intermediate between
Models \modelname{B10} and \modelname{B12}.  Though $Q_{\nu}$ is less
than that of Model \modelname{B10}, it significantly exceeds the
source term of the magnetic energy at all times.  $M_{\mathrm{gain}}$
drops to almost the same values as for Model \modelname{B10}, but the
earlier onset of the explosion translates into a minimum higher value
around $t \sim 500 \, \ms$.

\paragraph{Onset of explosions.}

The extended presentation of our results demonstrates that the
transition to explosions can be well described within the framework of
\cite{Dolence_et_al__2013__apj__DimensionalDependenceoftheHydrodynamicsofCore-collapseSupernovae}
and
\cite{Fernandez_et_al__2014__mnras__CharacterizingSASI-andconvection-dominatedcore-collapsesupernovaexplosionsintwodimensions}.
High-entropy bubbles are formed and disrupted in the turbulent flows
dominated to varying extent by convection or the SASI.  During most of
the post-bounce time, bubbles have small spatial scales and short
coherence time scales, whereas the onset of the explosion is
characterised by the development of large bubbles that persist for
several dynamical time scales.  For Models \modelname{B08, B10}, and
\modelname{B11}, the influence of the magnetic field on the evolution
is negligible.

Similarly, large-scale bubbles initiate the explosion of Model
\modelname{B12}.  The strong magnetic fields, however, affect the
development of an explosion indirectly by modifying the dynamics of
high-entropy bubbles in the gain layer rather than in a more direct
manner, e.g.~by adding magnetic pressure to the thermal pressure.
Strong magnetic sheets couple large regions as the \Alfven~speed is
comparable to the flow speed, thus favouring the development of large
structures, and magnetic tension resists their disruption, making the
bubbles very persistent.  Such an evolution leads to conditions
favouring an explosion much earlier than in a model with a weak
magnetic field.  Model \modelname{B11.5} presents an intermediate
case, where the flow is affected by magnetic fields less and
large-scale bubbles develop somewhat later than in Model
\modelname{B12}.

\section{Summary and conclusions}
\label{Sek:SumCon}

We have studied (some of) the processes leading to the amplification
of the magnetic field in a non-rotating stellar core during the
collapse and post-bounce accretion phases of a supernova.  In
non-rotating stars, a variety of amplification mechanisms considered
in previous works, e.g.~winding of the field by differential rotation
or the MRI, are not viable.  Instead, compression amplified the field
during infall and convection and the SASI may constitute (small-scale)
dynamos.  Additionally, \Alfven~waves travelling against the gas flow
may be amplified once they reach an \Alfven~point where the
(co-moving) \Alfven~velocity equals the gas velocity, a condition that
could be fulfilled in the accretion flow onto the PNS.  If efficient
amplification occurs, the field may be able to affect the dynamics of
the core, e.g.~by altering the geometry of SASI flows or by energy
dissipation through \Alfven~waves in the upper layers of the
hot-bubble region.

Reducing the complexity of the problem, simplified models,
e.g.~one-dimensional simulations with assumptions for the excitation,
propagation and dissipation of \Alfven~waves
\citep{Suzuki_Sumiyoshi_Yamada__2008__ApJ__Alfven_driven_SN}, toy
models of \Alfven~waves in decelerating flows
\citep{Guilet_et_al__2011__apj__Dynamics_of_an_Alfven_Surface_in_Core_Collapse_Supernovae},
and 2D and 3D MHD simulations without neutrino transport
\citep[][]{Endeve_et_al__2010__apj__Generation_of_Magnetic_Fields_By_the_SASI,Endeve_et_al__2012__apj__TurbulentMagneticFieldAmplificationfromSpiralSASIModes:ImplicationsforCore-collapseSupernovaeandProto-neutronStarMagnetization}
have demonstrated that these effects could, in principle, be relevant.
On the other hand, to study their evolution under less idealised
conditions, in particular their interplay with a highly dynamical
background, more calculations of self-consistent models are required,
i.e.~multi-dimensional MHD including a treatment of neutrino transfer
through the stellar core.

To this end, we have performed axisymmetric simulations of the
collapse and the post-bounce evolution of the core of a non-rotating
star of 15 solar masses possessing a purely poloidal initial field.
Using a new code for neutrino-magnetohydrodynamics, we have solved the
MHD equations coupled to the system of two-moment equations for the
neutrino transport; the closure for the moment equations was provided
by an analytic variable Eddington factor.  We included descriptions
for the most important reactions between electron neutrinos and
antineutrinos and the stellar matter, viz.~nucleonic, nuclear, and
leptonic emission and absorption and scattering off nucleons and
nuclei.  Our current approach ignores muon and tau neutrinos and is
not as accurate in treating neutrino-matter interactions as the most
sophisticated existing transport schemes
\cite[e.g.~][]{Buras_etal__2006__AA__Mudbath_1,Lentz_et_al__2012__apj__InterplayofNeutrinoOpacitiesinCore-collapseSupernovaSimulations},
but possesses full two-dimensionality including velocity effects.
Hence, our models allow for a fairly reliable assessment of the main
MHD effects in self-consistent SN core models in the presence of all
basic dynamical features found in supernova simulations, e.g.~the
stagnation of the prompt shock wave, PNS and hot-bubble convection,
and the SASI activity.  We will discuss remaining major limitations
below.

The principal results of our simulations and the main conclusions to
be drawn from them are:
\begin{enumerate}
\item Non-magnetic and weakly magnetised models were followed for a
  post-bounce evolution of about 800 ms, during which the shock
  exhibits different phases of more or less regular oscillations of
  its dipole mode, corresponding to post-shock flows dominated by the
  SASI or convection.  The mean shock radius contracts to values
  slightly above 100 km, until it starts to expand again and, finally,
  reaches more than 1000 km, at which point we stopped the
  simulations.  Similar to the results of
  \cite{Dolence_et_al__2013__apj__DimensionalDependenceoftheHydrodynamicsofCore-collapseSupernovae}
  and
  \cite{Fernandez_et_al__2014__mnras__CharacterizingSASI-andconvection-dominatedcore-collapsesupernovaexplosionsintwodimensions},
  this transition to explosion can be characterised by the development
  of one large bubble of high entropy.  In agreement with
  \cite{Fernandez_et_al__2014__mnras__CharacterizingSASI-andconvection-dominatedcore-collapsesupernovaexplosionsintwodimensions},
  we find that this corresponds to a pronounced widening of the
  distribution of volume-filling factors of the relative entropy
  contrast.
\item The magnetic field is amplified kinematically by the turbulent
  flows developing due to convection and the SASI.  The amplification
  factor does not depend on the initial field unless it reaches,
  starting from a high initial value, equipartition with the kinetic
  energy and thus the maximum possible energy that can be attained.
  The maximum field we observe locally is a few times $10^{15} ~
  \mathrm{G}$ at the base of the PNS convection layer.  In the PNS
  convection zone, the magnetic field is transiently amplified by the
  overturning flow, but suffers losses due to the expulsion of
  magnetic flux from convection cells.  As a consequence, the PNS is
  surrounded by a layer of strong field mostly parallel to its
  surface.  If no changes of the magnetic topology occur as the PNS
  cools and our 2D results can be confirmed by 3D simulations, the
  neutron star formed in the explosion will be magnetically shielded,
  in contrast to what is usually assumed in models of structure,
  cooling, and evolution of neutron stars
  \citep[e.g.][]{Geppert_Rheinhardt__2006__aap__Magnetars_versus_radio_pulsars.MHD_stability_in_newborn_highly_magnetized_NSs,Ciolfi_et_al__2009__mnras__Relativisticmodelsofmagnetars:thetwistedtorusmagneticfieldconfiguration}.
  Such a geometry might have important consequences for accretion on
  the PNS and it might open up the possibility of powering explosive
  events in the magnetosphere triggered by impulsive reconnection.
  Furthermore, it is similar to the field configurations used by by
  \cite{Vigano_et_al__2013__mnras__Unifyingtheobservationaldiversityofisolatedneutronstarsviamagneto-thermalevolutionmodels}
  to explain the thermal luminosity of isolated neutron stars.
\item When falling through the non-spherical accretion shock, the
  magnetic field is bent by lateral flows, creating a component
  parallel to the shock.  These structures are advected towards the
  PNS convection zone.  The field accumulates there, leading to a
  layer of strong magnetic fields.  Since the structures are
  associated with a strong lateral component of the field, this layer
  is dominated by the $\theta$-component of the field.  We find that
  the field amplification achieved by the turbulent flows is the
  result of the competition between the radial advection and the
  overturning flows and, hence, the growth of the field is connected
  to the ratio of the advection time scale to the time scale of eddy
  turnover.  The field carried by a fluid element can only be
  amplified as long as the fluid parcel is inside the region of vortex
  motion.  Hence, fast turbulent flows in a slowly accreting layer are
  most conducive to field amplification.  This effect, rather than the
  feedback of a magnetic field amplified to equipartition with the
  flow, sets the maximum to the field strength that our models with
  weak initial fields can reach.  In fact, the final fields remain far
  below the kinetic energy across the entire spectrum of modes.
\item Due to the non-radial field geometry, \Alfven~waves propagate at
  constant radius rather than upwards.  Therefore, we do not find an
  \Alfven~surface in the accretion flow, although there are sub- as
  well as superalfv\'enic regions in this layer.  This limits the
  efficiency of the amplification mechanism proposed by
  \cite{Guilet_et_al__2011__apj__Dynamics_of_an_Alfven_Surface_in_Core_Collapse_Supernovae}.
  Conditions for the latter effect are more favourable if accretion
  occurs through a sufficiently steady columns.  In our axisymmetric
  models, this is the case mostly along the polar axis where the
  geometry enforces radial fields and flows.  An \Alfven~surface forms
  in the radial field of the accretion column at a radius depending on
  the initial field strength.  For fields of $b_0 \gtrsim 10^{11} ~
  \mathrm{G}$ initially, we observe that perturbations created in the
  PNS convection zone propagate along the field upwards into the
  accretion column, but a clear identification of their contribution
  to the amplification of the field and the energy transport was not
  possible.
\item For the strongest initial field, $b_0 = 10^{12} ~ \mathrm{G}$,
  the combined kinetic and magnetic energies and stresses in the
  post-shock layer are similar to the ones for weak initial fields,
  but the contributions of magnetic and kinetic terms are now roughly
  equal.  Consequently, the field is able to shape the post-shock
  flow.  The \Alfven~speed equals the flow speed, and thus magnetic
  forces couple large volumes and favour the formation of a flow
  dominated by low-order multipole modes, viz.~a quadrupolar pattern
  of accretion columns at the poles and near the equator.  Furthermore,
  magnetic tension suppresses the disruption of these flow structures.
  Consequently, the shock oscillations of this model show an
  extraordinary regularity, large amplitudes, and a low frequency.  In
  this model with its special flow topology, we find the most
  pronounced shock expansion of all models, setting in $\sim 400 \,
  \ms$ earlier than for weak fields.  Similar to non-magnetic models,
  the explosion geometry is dominated by large high-entropy bubbles,
  but in contrast to the weak-field case, these bubbles are seeded by
  the low-order modes enforced by the magnetic field rather than
  slowly developing from the convective/SASI modes.  Strong initial
  fields in the progenitor core of $b_0 \sim \zehn{12} \, \mathrm{G}$
  thus favour an earlier onset of the explosion.
\end{enumerate}

Though limited to non-rotating models in axisymmetry, our study is
similar in scope to that of
\cite{Endeve_et_al__2010__apj__Generation_of_Magnetic_Fields_By_the_SASI,Endeve_et_al__2012__apj__TurbulentMagneticFieldAmplificationfromSpiralSASIModes:ImplicationsforCore-collapseSupernovaeandProto-neutronStarMagnetization},
and our main results are in agreement with theirs.  We find,
consistent with their results, efficient amplification of the magnetic
field in the unstable regions of the hot bubble and the PNS, and, in
particular, along the symmetry axis, where the flow is forced into
stable, narrow accretion columns.  Furthermore, all pre-collapse
initial fields weaker than $b_0 \sim 10^{11} ~ \mathrm{G}$ are
amplified by roughly the same average factor in our models, while
dynamic backreaction limits the amplification of stronger initial
fields to smaller average factors.  This threshold of the pre-collapse
field strength above which the amplification is limited by feedback
corresponds to field strength at the surface of the PNS similar to the
one reported by
\cite{Endeve_et_al__2010__apj__Generation_of_Magnetic_Fields_By_the_SASI}
(cf.~their Fig.~5).  Having in mind the dependence of the field growth
rate on the ratio between advection and eddy-turnover times, we may
speculate that the different factors of magnetic field growth in their
and our models are caused by strongly different sizes of the gain layers
and different accretion profiles.

Besides many similar aspects, we have to note a striking difference
between our results and the ones by
\cite{Endeve_et_al__2010__apj__Generation_of_Magnetic_Fields_By_the_SASI}:
they find a complete suppression of the SASI by their strongest
magnetic field, while in our models, the SASI is able to operate even
for the strongest initial fields, albeit with a modified geometry.
Among the differences in the setup of the two sets of simulations, the
one most likely cause of such a strong discrepancy may be the chosen
initial geometry of the magnetic field.  While
\cite{Endeve_et_al__2010__apj__Generation_of_Magnetic_Fields_By_the_SASI}
start with a strictly radial field (split monopole), our initial
fields are generated by an off-centre dipole and, thus, have strong
non-radial components.  Studies of the evolution of the SASI in a
magnetised medium by
\cite{Guilet_Foglizzo__2010__apj__Toward_a_MHD_Theory_of_the_SASI:Toy_Model_of_the_AAC_in_a_Magnetized_Flow}
and
\cite{Guilet_et_al__2010__French_Soc_AA__Effects_of_a_moderately_strong_magnetic_field_in_core_collapse_supernovae}
show that the interaction between the magnetic field and the SASI
modes entails stabilising as well as destabilising effects, but their
relative importance depends on the field geometry.  The destabilising
effects are strongest for a non-radial field and absent for a radial
field.  This indicates a possible origin of the differences between
our simulations and those by
\cite{Endeve_et_al__2008__JPCS__Magnetic_field_generation_by_SASI},
although this issue requires a closer investigation.

In summary, our results suggest that magnetic field amplification to
interesting strengths can efficiently take place during stellar core
collapse even in the absence of rotation.  In addition to an
enhancement due to compression by the radial collapse, we find that
non-radial fluid flows associated with convection and SASI activity,
and the interaction of \Alfven~waves in the accretion funnels can
amplify the initial iron-core fields.  Present stellar evolution
models
\citep[][]{Heger_et_al__2005__apj__Presupernova_Evolution_of_Differentially_Rotating_Massive_Stars_Including_Magnetic_Fields}
predict a field strength in slowly rotating pre-collapse iron cores of
the order of $10^{9}$--$10^{10}$\,G and a predominantly toroidal field
geometry.  Because a large-scale dynamo is less likely to operate
inside a non-rotating core, cores should generically possess even
weaker and less ordered fields in the limit of very slow rotation than
for rapid rotation.  Starting with fields of the mentioned strength as
an upper limit, field strengths of typical pulsars
($10^{12}$--$10^{13}$\,G) can be reached.  Magnetar fields of
$10^{14}$--$10^{15}$\,G result when the progenitor core is assumed to
posses a pre-collapse field between a few $10^{11}$\,G and
$10^{12}$\,G.  Only in the latter case, the fields around the nascent
neutron star obtain dynamical importance and might have an influence
on the supernova explosion mechanism.

Though offering some insight into the magnetic-field evolution in
non-rotating magnetised cores, our study has several important
shortcomings:
\begin{enumerate}
\item We have used a new solver for the neutrino transport with
  simplified neutrino-matter interactions, constraining ourselves to
  electron neutrinos and antineutrinos.  While the accuracy of such a
  simplification in comparison to more complete neutrino treatments
  will have to be assessed in a separate study, we do not deem this a
  problem for the presented investigation because we are able to
  capture the most important dynamical effects in a supernova core.
\item To save computational costs, we have restricted ourselves to
  axisymmetric simulations.  In the light of the anti-dynamo theorems,
  this is a severe limitation leading to a wrong, possibly too low,
  level of amplification of the field in the turbulent regions.
  Furthermore, our models do not allow for the development of shear
  layers associated with non-axisymmetric spiral modes of the SASI,
  which may also be a site of efficient amplification of the magnetic
  field \citep[see
  also][]{Endeve_et_al__2010__apj__Generation_of_Magnetic_Fields_By_the_SASI}.
\item Axisymmetry may also affect the amount to which the magnetic
  flux is expelled from the PNS convection zone.
\item Moreover, axisymmetry restricts the dynamics of the accretion
  flows, favouring the development of very stable accretion columns
  along the poles.  As we have described, field amplification shows
  very distinct features in and below these columns.  We presume that
  the dynamics in three dimensions is more similar to what we have
  seen in off-axis accretion flows, i.e.~less stable \Alfven~surfaces
  and less efficient amplification of \Alfven~waves.  Very strong
  initial fields of the order of $b_0 = 10^{12} ~ \mathrm{G}$ lead to
  high field strengths that are able to dominate the post-shock
  accretion flow.  This might establish coherent, stable accretion
  columns even in 3D.  For such fields, our axisymmetric results may
  hence be a reasonable approximation.
\item Turbulent field amplification may be very sensitive to
  dissipation effects, physical and numerical.  Our models, based on
  \emph{ideal} MHD, neglect dissipation by physical viscosity and
  resistivity, but are computed on relatively coarse numerical grids,
  corresponding to excessive numerical dissipation.  This is at least
  partially confirmed by a resimulation of one of our weak-field
  models on a grid with twice the standard resolution, which shows a
  magnetic energy in the gain layer twice as large as in the standard
  model, whereas most other variables change only weakly.
\item Therefore, we are not able to fully resolve the turbulent
  (inverse) cascades of magnetic and kinetic energy and helicity
  covering many orders of magnitude in wave number in a supernova
  core.  The effect of insufficient resolution on \Alfven~waves is
  probably less serious although their wave number should increase as
  they approach the \Alfven~point, requiring enhanced resolution.
  Simulations of cores at a resolution corresponding to numerical
  viscosity and resistivity below the physical scales of these
  dissipative effects are by far too expensive today, and will remain
  so for a long time.  To tackle this difficulty, a combination of
  different approaches would be desirable, viz.~global direct
  numerical simulations of the core with drastically enhanced physical
  transport coefficients, and simple sub-grid models for MHD
  turbulence based on idealised local simulations neglecting most
  aspects of, e.g.~neutrino physics.  We are, however, aware of the
  lack of reliable sub-grid models for MHD at present, obstructing
  further progress in this direction.
\end{enumerate}

Apart from these methodological shortcomings, open physical questions
are, e.g.~the influence of the progenitor on the establishment of
certain patterns in the accretion flow, effects of different initial
field geometry, and the influence of slow rotation of the core on our
findings.  In particular the last issue may prove interesting as it
would enable a large-scale dynamo.  We defer these questions as well
as the more technical problems listed above to future investigations.

\section{Acknowledgements}
\label{Sek:Ackno}

We thank J{\'e}r{\^o}me Guilet, Thierry Foglizzo, Jeremiah Murphy,
Adam Burrows, and Oliver Just for interesting and educative
discussions and Andreas Marek for providing us with the EOS table.  We
thank the anonymous referee for their careful revision of the
manuscript and their helpful comments.  M.O.~and
M.{\'A}.A.~acknowledge support by the European Research Council (grant
CAMAP-259276), and from the Spanish Ministerio de Ciencia e
Innovaci{\'o}n (grant AYA2010-21097-C03-01 \emph{Astrof{\'i}sica
  Relativista Computacional}).  Furthermore, at Garching this work was
supported by the European Research Council through grant ERC-AdG
No.~341157-COCO2CASA and by the Deutsche Forschungsgemeinschaft
through the Transregional Collaborative Research Center SFB/TR~7
``Gravitational Wave Astronomy'', the Cluster of Excellence EXC~153
``Origin and Structure of the Universe''
(\texttt{http://www.universe-cluster.de}), and the
Max-Planck-Princeton Center for Fusion and Astro Plasma Physics.  The
simulations were performed using the clusters \emph{Lluisvives} and
\emph{Tirant} of the Universitat de Val{\`e}ncia.

\bibliographystyle{aa}

\end{document}